\documentclass[]{aa} % for the long lists of affiliations 
\usepackage{graphicx}
\usepackage{txfonts}
\usepackage{comment}
\usepackage{lscape}
\usepackage{longtable}
\usepackage{natbib,twoopt}
\bibpunct{(}{)}{;}{a}{}{,}             %% natbib format for A&A and ApJ
\usepackage[usenames]{color}                                   % Comment this if compiled with [referee] option

\newcommand\starlight{{\sc starlight}}          	% STARLIGHT 
\begin{document}

\title{The cosmic evolution of the spatially-resolved star formation rate  and stellar mass of the CALIFA survey  }

\authorrunning{L\'opez Fern\'andez et al.}
\titlerunning{Cosmic evolution of SFR, sSFR, and stellar mass density}

\author{
R. L\'opez Fern\'andez\inst{1},
R. M. Gonz\'alez Delgado\inst{1},
E. P\'erez\inst{1},
R. Garc\'{\i}a-Benito\inst{1}, 
R. Cid Fernandes\inst{2},
W. Schoenell\inst{3},
S. F. S\'anchez\inst{4},
A. Gallazzi\inst{5},
P. S\'anchez-Bl\'azquez\inst{6},
N. Vale Asari\inst{2},
C. J. Walcher\inst{7}
}

\institute{
Instituto de Astrof\'{\i}sica de Andaluc\'{\i}a (CSIC), P.O. Box 3004, 18080 Granada, Spain. (\email{rosa@iaa.es})
\and
Departamento de F\'{\i}sica, Universidade Federal de Santa Catarina, P.O. Box 476, 88040-900, Florian\'opolis, SC, Brazil
\and
Universidade de Sao Paulo, Instituto de Astronom\'{\i}a, Geof\'{\i}sica e Ciencias Atmosf\'ericas, Rua do Matao 1226, 05508-090, Sao Paulo, Brazil.
\and
Instituto de Astronom\'\i a,Universidad Nacional Auton\'oma de M\'exico, A.P. 70-264, 04510 M\'exico D.F., Mexico
\and
INAF-Osservatorio Astrofisico di Arcetri, Largo Enrico Fermi 5, 50125 Firenze, Italy
\and
Depto. de F\'\i sica Te\'orica, Universidad Aut\'onoma de Madrid, 28049 Madrid, Spain
\and
Leibniz-Institut f\"ur Astrophysik Potsdam (AIP), An der Sternwarte 16, D-14482 Potsdam, Germany
}

\date{\today}

% \abstract{}{}{}{}{} 
% 5 {} token are mandatory

\abstract{
We investigate the cosmic evolution of the absolute and specific star formation rate (SFR, sSFR) of galaxies as derived from a spatially-resolved study of the stellar populations in a set of 366 nearby galaxies from the CALIFA survey. 
The sample spans stellar masses from $M_\star \sim 10^9$ to $10^{12} M_\odot$ and a wide range of Hubble types. 
The analysis combines GALEX (FUV and NUV) and  SDSS (u, g, r, i, z) images with the 4000 \AA\ break, H$\beta$, and [MgFe]$^{'}$ indices measured from the CALIFA data-cubes to constrain parametric models for the star formation history (SFH), which are then used to study the cosmic evolution of the star formation rate density ($\rho_{\rm SFR}$), the sSFR,  the main sequence of star formation (MSSF),  and the stellar mass density ($\rho_\star$). 
Several SFH laws are used to fit the observational constrains. A delayed-$\tau$ model, SFR  $\propto ( t_0 - t) \exp(-( t_0 - t)/\tau)$, provides the best results, in good agreement with those obtained from cosmological surveys.
Our main results from this model are: 
{\em (a)} The mass currently in the inner ($\leq 0.5$ half light radius, HLR) regions formed at earlier epochs than that in the outer (1--2 HLR) regions of galaxies. 
The time since the onset of the star formation is larger in the inner regions ($ t_0 \sim 13$ to 10 Gyr) than in the outer ones ($ t_0 \sim 11$ to 9 Gyr), for all the morphologies,  while the e-folding time-scale $\tau$ is similar or smaller in the inner than in the outer regions. These results confirm that galaxies of any Hubble type grow inside-out. 
{\em (b)} The sSFR declines rapidly as the Universe evolves, and faster for early than for late type galaxies, and for the inner than for the outer regions of galaxies.   
{\em (c)} The evolution of $\rho_{\rm SFR}$ and $\rho_\star$ agrees well with results from cosmological surveys, particularly with the recent results from the GAMA/G10-COSMOS/3D-HST survey. 
At low redshift, $z\leq 0.5$, most star formation takes place in the outer regions of late spiral galaxies, while at $z > 2$  the inner regions of the progenitors of the current E and S0 are the major contributors to $\rho_{\rm SFR}$. 
{\em (d)}
Similarly, the  inner regions of galaxies are the major contributor to $\rho_\star$  at $z > 0.5$, growing their mass faster than the outer regions, with  a lookback time at $50\%$
$\rho_\star$ of $t_{50}\sim 9$ and 6 Gyr for the inner and outer regions.
{\em (e)} The MSSF follows a power-law at high redshift, with the slope evolving with time,  but always being sub-linear,  in good agreement with the $Illustris$ simulation. 
{\em (f)}  In agreement with galaxy  surveys at different redshifts, the average SFH of CALIFA galaxies indicates that galaxies grow their mass mainly in a mode that is well represented by a delayed-$\tau$ model, with the  peak at $z \sim 2$ and an e-folding time  of $\sim 3.9$ Gyr.  
}

%{x}{x}{x}{} 

% aims heading (mandatory){aaa}
% methods heading (mandatory){aaa}
% conclusions heading (optional), leave it empty if necessary {aaa}

% results heading (mandatory){aaa}

\maketitle

%-------------------------------------------------------------------------------------------------------
%NEW SECTION
%-------------------------------------------------------------------------------------------------------
\section{Introduction}
\label{sec:Introduction}

In the last several decades multi-wavelength galaxy surveys have played a key role in the quest to understand how galaxies form and evolve. In particular,   surveys at different redshifts have been instrumental to establish that the bulk of the stellar mass ($M_\star$) observed today was built up at $z\geq 2$. The most relevant observational results  are: 
{\em(1)} The star formation rate density in the Universe ($\rho_{\rm SFR}$) peaks at $\sim $ 3.5 Gyr  after the Big Bang, at $z\sim$ 2  \citep{lilly96, madau98, hopkinsbeacom06, fardal07, madau14}.  {\em(2)} The stellar mass density in the Universe ($\rho_{\star}$) evolved very little since  $z \sim 1$ \citep{perezgonzalez08, pozzetti10, ilbert13, muzzin13}. {\em(3)} Up to $z \geq$ 4  there exists a relation between SFR and  $M_\star$, known as the main sequence of star formation (MSSF), with a logarithmic slope that is below 1 \citep{brinchmann04, noeske07,daddi07, elbaz07, wuyts11, whitaker12, renzinipeng15,  tasca15, catalan15, canodiaz16}. 
{\em(4)}  The specific star formation rate (sSFR = SFR/$M_\star$), declines weakly with increasing galaxy mass \citep{salim07, schiminovich07}, decreasing rapidly at $z < 2$ \citep{rodighiero10, oliver10, karim11, elbaz11, speagle14}, and increasing  slowly at $z>2$ \citep{magdis10,stark13,  tasca15}.

These observational results cannot be easily explained by current models of galaxy formation,  which assume
that galaxies grow their mass by merging of dark matter halos, progressively assembling more massive systems to become a single massive galaxy \citep{naab16}. Moreover, galaxies can grow in a large extent by fresh gas supply from the cosmic web  \citep{keres05, dekelbirnboim09, lilly13}. In this context, the gas accretion and SFR  are expected to be associated with the specific accretion rate of  dark matter \citep{neistein06,birnboim07, neistein08b, dutton10}.  
However, the specific cosmological accretion rate of baryons declines with time as $\dot{M}_{acc}/M_{acc} \propto (1+z)^{2.5}$ \citep{neistein08b, dekel09}, while the observed sSFR $\propto (1+z)^3$ for galaxies at $z \leq 2$ \citep{oliver10, elbaz11}, and is constant at $z > 2$ \citep{magdis10,stark13}. Some studies have shown that this particular tension can be relaxed
assuming  different mechanisms, such as enhanced feedback from super-winds in starbursts  \citep{lehnert15}, or by  non-trivial modifications in the semi-analytic models,  that involve a suppressed SFR at $z >4$ (by enhanced feedback or reduced SFR efficiency) following an initial active phase at $z > 7$ \citep{weinmann11}. 

Despite the importance of the high $z$ work, still much of our knowledge about galaxy evolution comes from the study of nearby galaxies.
Surveys at low $z$ have been very useful to characterize the global properties of galaxies \citep[e.g.][]{blanton09,  kauffmann03, baldry10}, to retrieve  SFR and sSFR   as a function of galaxy mass \citep{brinchmann04, salim07, schiminovich07}, and the current value of $\rho_{\rm SFR}$ \citep{gallego95, brinchmann04, heavens04}. 
Moreover, alternatively to surveys at different redshifts, the SFH of the Universe can be inferred by analyzing the fossil record of the current stellar populations of nearby galaxies \citep{panter03, heavens04, cidfernandes05, ocvirk06, asari07, panter08, tojeiro11, koleva11, mcdermid15, citro16}. This method, that originally started by fitting optical colors of galaxies to study how their SFHs vary along the Hubble sequence \citep{tinsley68, tinsley72, searle73, gallagher84, sandage86}, is currently applied to fit UV-optical integrated spectra of galaxies.
%\citep{cidfernandes05, asari07, tojeiro11}.

More recently, integral field spectroscopy  (IFS) surveys, such as CALIFA \citep{sanchez12, husemann13, garciabenito15, sanchez16}, ATLAS3D \citep{cappellari11}, SAMI \citep{bryant15}, and MaNGA \citep{bundy15, law15},  allow the study of the spatially-resolved SFH of galaxies  from spatially-resolved spectroscopy \citep{perez13, cidfernandes14, sanchez-blazquez14PING, gonzalezdelgado14a, sanchez-blazquez14, gonzalezdelgado15, ibarra16, gonzalezdelgado16,  goddard16, gonzalezdelgado17, amorim17, garciabenito17, cortijoferrero17mergers, sanchez17, catalantorrecilla17}. 
In particular, with CALIFA data and using the non-parametric code \starlight\ \citep{cidfernandes05, lopezfernandez16},
we have obtained the radial distribution of the current SFR and sSFR  as a function of  Hubble type \citep{gonzalezdelgado16}. We have found that there is a correlation between the stellar mass surface density ($\mu_\star$) and the intensity of the star formation ($\Sigma_{\rm SFR}$, defined as the SFR per unit area), 
defining a local MSSF relation with a slope similar to that of the global relation between total SFR and $M_\star$ \citep{gonzalezdelgado16, canodiaz16}. This suggests that local processes are important in determining the star formation in disks. Furthermore, the radial profiles of $\Sigma_{\rm SFR}$ are very similar for all spirals. The radial profiles of current sSFR increase outwards, indicating that the recent shut down of the star formation in spirals is progressing inside-out.

We have also studied the spatially-resolved evolution of the SFR and $\Sigma_{\rm SFR}$ as a function of galaxy mass and Hubble type  \citep{gonzalezdelgado17}. We have found that:
{\em (1)}  Galaxies form very fast independently of their current stellar mass, with the SFR peaking at $z \geq 2$. 
{\em (2)} At any epoch, $\Sigma_{\rm SFR}$ scales with Hubble type.
$\Sigma_{\rm SFR}$ reaches the highest values ($>  10^3 \,M_\odot\,$Gyr$^{-1}\,$pc$^{-2}$) in the central regions of current ETGs, similar to those measured in high-redshift star-forming galaxies.  SFR increases sub-linearly with  $M_\star$,  such that 
the most massive galaxies have the highest absolute but lowest specific SFRs.
{\em (3)} Evidence of an early and fast quenching is found only in the most massive ($M_\star > 2 \times 10^{11} M_\odot$) E galaxies of the sample, but not in spirals of similar mass, suggesting that  halo mass is not the main mechanism for the shut down of star formation. Less massive E and disk galaxies show more extended SFHs and a slower quenching than the massive E. 

In this paper we explore a parametric approach to ``fossil cosmology'', whereby colors and spectral indices predicted by analytical descriptions of the SFH are fitted to the data and used to obtain the cosmic evolution of $\rho_{\rm SFR}$, sSFR, and $\rho_\star$.
We develop a methodology that combines the stellar spectral indexes from CALIFA data with photometry from GALEX and SDSS. In this way,  we can harness the power of full spectral fitting (via the more relevant stellar features in the spectra of galaxies) with the spectral energy distribution, by fitting the large baseline of stellar continuum provided by the UV and optical bands. This method makes the results more independent on the errors and uncertainties associated to the calibration of the galaxy spectra.

After verifying that our results are comparable to those inferred from the snapshots of galaxy evolution obtained by studies at different redshifts, we take advantage of the spatially resolved information in our data to investigate how different radial regions contribute to evolution of $\rho_{\rm SFR}$, sSFR, and $\rho_\star$. This is an issue more easily tackled with fossil record studies of nearby galaxies than with redshift surveys, where spatial resolution is observationally challenging. Furthermore, since the morphology of the galaxies in our sample are known, we can also study the role of the currently early and late type galaxies in the evolution of $\rho_{\rm SFR}$, sSFR, and $\rho_\star$.

This work  involves a number of assumptions and limitations with respect to studies based on snapshots at different redshifts that need to be mentioned here. Besides limitations related with the fossil record method itself, one should bear in mind that: {\em (a)} The Hubble sequence has evolved with time \citep{delgado-serrano10, cappellari16}, so that the progenitors of, say,  E and S0 in the local Universe are not necessarily also E and S0 in the past. {\em (b)} Due to stellar migrations \citep{roskar08, minchev14}, stars currently located at inner and outer regions in a galaxy may not be at these locations in the past. {\em (c)} The sample used in this study does not contain galaxies with strong features of interactions or mergers, but such events could well have happened in their past histories. Our analysis cannot say if the mass in a galaxy has grown through mergers or smooth accretion. 
Mergers do play a role in galaxy evolution \citep{lotz11}. It is known, for instance, that the merger rate increases rapidly with redshift \citep{bundy09,bluck09,bluck12,lopez-sanjuan12,lopez-sanjuan15,kaviraj15,rodriguez-gomez15}, possibly flattening at $z  > 1$
 \citep{williams11,man12,casteels14}. The role played by minor or major mergers in setting the current Hubble sequence is still unknown \citep{lofthouse17, martin17}.

Notwithstanding these caveats, this study offers relevant insight on how different regions in a galaxy contribute to $\rho_{\rm SFR}(t)$, sSFR$(t)$, and $\rho_\star(t)$, as well as how galaxy morphology relates to this cosmic evolution. 

This paper is organized as follows. Section \ref{sec:Data} describes the observations and properties of the galaxies analyzed here. In Sections \ref{sec:Method} and \ref{sec:Results} we explain our method and present the results of the SFH analysis. In Section \ref{sec:Comparison} we use these results to estimate the cosmic evolution of  $\rho_{\rm SFR}$, sSFR, and $\rho_\star$ and to compare them with results obtained from surveys at different redshifts.
Section \ref{sec:Discussion} then examines how present day morphology, an how inner and outer galaxy regions contribute to the  $\rho_{\rm SFR}(t)$, sSFR$(t)$, and $\rho_\star(t)$ budget. The evolution of the MSSF is also discussed.
Section \ref{sec:Summary} reviews our main findings. 
We adopt a flat cosmology for the relation between lookback time, $t(z)$, and redshift: $\Omega_M= 0.3$, $\Omega_\Lambda = 0.7$, and H$_0 = 70$ km s$^{-1}$ Mpc$^{-1}$. The units of $M_\star$ are $M_\odot$ throughout; we do not specify them for the sake of clarity.

%-------------------------------------------------------------------------------------------------------
%NEW SECTION
%-------------------------------------------------------------------------------------------------------
\section{Sample and data}
\label{sec:Data}

\subsection{Sample}
\label{sec:Sample}

The main characteristics of the  CALIFA mother sample are: {\em (a)} angular isophotal diameter between 
$45{\tt''}$  and $79{\tt''}$;  {\em (b)} redshift range $0.005 \leq z \leq 0.03 $;  {\em (c)}  color ($u-r < 5$) and magnitude ($-24 < M_r < -17$)  covering the whole color-magnitude diagram. A full description and characterization of the mother sample is given by \citet{walcher14}. Although the sample is not limited in volume, it can be "volume-corrected", allowing us to provide estimates of the stellar mass function \citep{walcher14} and other cosmological observables such as $\rho_{\rm SFR}$ \citep{gonzalezdelgado16}. The volume-corrected distribution functions of luminosities,  masses,  and sizes  are statistically compatible with estimates from the full SDSS when accounting for large-scale structure (see figures 8, 9, and 14 in \citealt{walcher14}).
%for the nearby Universe 

\begin{figure}
\includegraphics[width=0.5\textwidth]{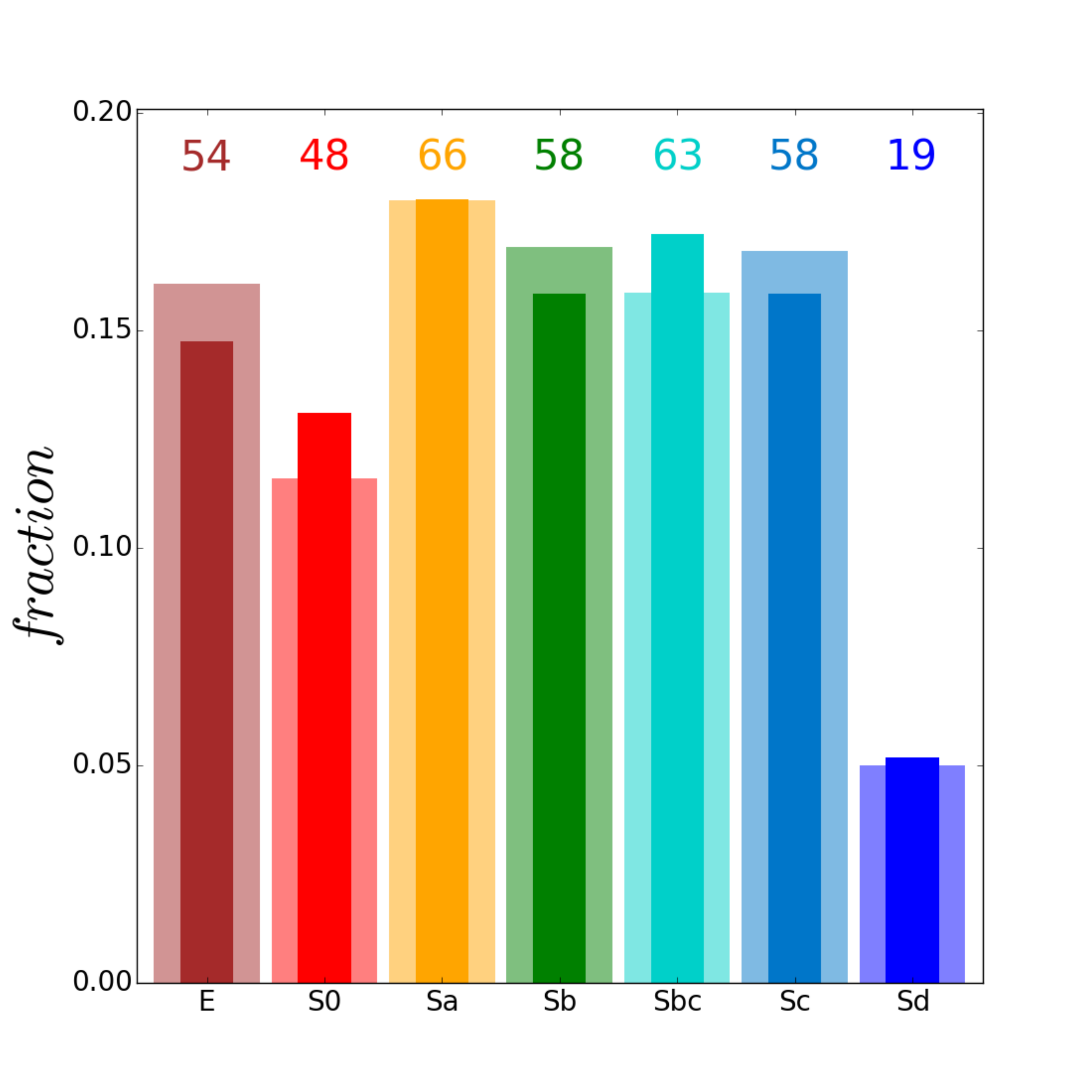}
\caption{ Comparison of Hubble type distributions in the CALIFA mother sample (939 galaxies, broad bars) and the 366 galaxies analyzed here (narrow, darker bars). Both histograms are normalized to unit sum. The number of galaxies in each morphology bin is labeled in color. 
 }
\label{fig:hist-type}
\end{figure}

The sample used in this study comprises all the galaxies from the CALIFA survey \citep{sanchez12, sanchez16} with COMBO cubes (the combination of the observations with the V1200 and V500 setups) in the data release 2 \citep{garciabenito15} for which there are GALEX images available \citep{martin05}.  This sub-sample comprises 366 galaxies, and is representative of the full CALIFA mother sample, covering seven bins  in morphology: E (54), S0 (48), Sa (66), Sb (58), Sbc (63), Sc (58) and Sd (19). Fig.\ \ref{fig:hist-type} shows that these proportions are very similar to the ones in the the mother sample. In terms of stellar masses, our sub-sample runs from  $\log M_\star = 8.6$ to 11.9 (for a Chabrier IMF).

%***FIG***FIG***FIG***FIG***FIG***FIG***FIG***FIG***FIG***FIG***
%***FIG***FIG***FIG***FIG***FIG***FIG***FIG***FIG***FIG***FIG***

\subsection{Observations and data reduction}
\label{sec:Observations}

The observations were carried out 
at the Calar Alto observatory with the 3.5m telescope and the Potsdam Multi-Aperture Spectrometer PMAS \citep{Roth05} in the PPaK mode \citep{verheijen04}. PPaK is an integral field spectrograph with a field of view of $74{\tt''} \times 64{\tt''}$  and 382 fibers of $2.7{\tt''}$  diameter each \citep{kelz06}. We use data calibrated with the version 1.5 of the reduction pipeline \cite{garciabenito15}. The spectra cover the 3700--7300 \AA\ range with the same resolution as the V500 grating ($\sim 6$ \AA\  of FWHM) and a spatial sampling of 1 arcsec/spaxel.

GALEX  images at FUV ($\lambda_{\rm eff} \sim $ 1542 \AA) and NUV ($\lambda_{\rm eff} \sim $ 2274 \AA) bands were retrieved from the GALEX website\footnote{https://asd.gsfc.nasa.gov/archive/galex}. We use the calibrated data products by \citet{morrissey07}, but we did the sky subtraction by averaging the intensity in several regions of the images. 
Galactic extinction was corrected for with $A_\lambda / E(B-V) = $ 8.24 and 8.20, for FUV and NUV respectively \citep{wyder07}.

Calibrated u, g, r, i, and z images from the 10$^{\rm th}$ SDSS data release  \citep{ahn14} were retrieved from the SDSS website\footnote{skyserver.sdss.org}. To correct for Galactic extinction we use the conversion from $E(B - V )$ to total extinction $A_\lambda$ tabulated by \citet{stoughton02}, namely: $A_\lambda/E(B-V) = $ 5.155, 3.793, 2.751, 2.086, 1.479 for filters u, g, r, i, and z, respectively.
These corrections, and those done on the FUV and NUV bands, are compatible with the Galactic extinction corrections applied to the CALIFA data cubes \citep{garciabenito15}.

The MONTAGE software\footnote{http://montage.ipac.caltech.edu} was used to perform a resampling of the SDSS and GALEX images to the same spatial scale as CALIFA, and also to align and cut the GALEX images using WCS to obtain processed FUV and NUV images with the same FoV as our CALIFA datacubes.

Because of the inferior spatial resolution of the UV images with respect to the optical images and spectra, the spatial analysis is reduced by analyzing only three radial regions: {\it (i)} between $R = 0$ and 2 HLR (the whole galaxy), {\it (ii)} inside $R < 0.5$ HLR (the central region, dominated by the galaxy bulge), and {\it (iii)} between $R = 1$ and 2 HLR, where HLR denotes the galaxy's Half Light Radius, measured as in \citet{cidfernandes13}. 

The spectral indices described in \ref{sec:ContraintsAndModels} were measured from CALIFA spectra extracted within these three apertures. Fluxes in the NUV, FUV, and ugriz images were measured in the same regions, after masking foreground stars and low signal-to-noise regions masked in the CALIFA data.

%-------------------------------------------------------------------------------------------------------
%NEW SECTION
%-------------------------------------------------------------------------------------------------------
\section{Star formation history analysis}
\label{sec:Method}

The method to derive the star formation history for each spectrum consists in finding out the best parameters of an analytical model for the SFH that fits simultaneously a combination of UV+optical photometry from GALEX and SDSS and spectral indices from CALIFA data.

\subsection{Observational constrains and stellar population models}
\label{sec:ContraintsAndModels}

In addition to the GALEX FUV and NUV photometry and SDSS ugriz photometry, we measure the H$\beta$ index, the 4000 \AA\ break D$_n$4000, and  [MgFe]$'$ in the CALIFA spectra. We use the Lick index definition \citep{worthey94} for  H$\beta$, and  \citet{balogh99} for  D$_n$4000. These are both good indicators of the stellar population age \citep{bruzualcharlot03, kauffmann03, gonzalezdelgado05}.  [MgFe]$'$\footnote{A combination of the Lick indices Mgb, Fe5270, and Fe5335  \citep{thomas03}} is a good tracer of the stellar metallicity in a galaxy  ($Z_\star$). The advantage of this index with respect to other  $Z_\star$-indicators (such as Mgb) is that it is almost independent of the $\alpha$/Fe ratio \citep{thomas03}. 

While  [MgFe]$'$ and D$_n$4000 are measured directly in the CALIFA spectra, the  H$\beta$ index needs to be corrected for nebular emission. In practice, we use the synthetic value of the index obtained from a full spectral fit with \starlight\ \citep{cidfernandes05}.

To model these 10 observables (7 photometric points plus three spectral indices) we use spectral models for simple stellar populations (SSP) from an updated version of the \citet{bruzualcharlot03} models (Charlot \& Bruzual 2007, private communication\footnote{http://www.bruzual.org/$\sim$gbruzual/cb07}), where the STELIB spectral library \citep{leborgne03} is replaced by a combination of the MILES \citep{sanchez-blazquez06,falcon-barroso11} and {\sc granada} \citep{martins05} libraries. This set of models is one denoted as base {\it CBe} in our previous works \citep{gonzalezdelgado15,gonzalezdelgado16}.  The evolutionary tracks are those collectively known as Padova 1994 \citep{alongi93,bressan93,fagotto94a,fagotto94b,girardi96}. The IMF is that of \citet{chabrier03}. The metallicity covers $\log Z_\star/Z_\odot = -2.3$, $-1.7$, $-0.7$, $-0.4$, 0, and $+0.4$, while ages run from 0 to 14 Gyr. 
The spectrum produced by one such SSP of initial mass $1 M_\odot$ is denoted by SSP$_\lambda(t,Z)$.

These models are combined with a parametric prescription for the SFH and a simple recipe for the effects of dust to predict the observables listed above. This process is similar to that followed by \citet{gallazzi06} to retrieve the stellar population properties of galaxies in the SDSS survey, and, more recently, by \citet{zibetti17} for CALIFA data.

\subsection{Method}

\subsubsection{General aspects}

We assume a generic SFH $=$ SFH$(t; \Theta)$, where $t$ is the lookback time and $\Theta$ is a vector of parameters including the stellar metallicity ($Z_\star$), a dust attenuation parameter ($\tau_V$), and parameters controlling the temporal behavior of the SFR $\psi(t)$.
For instance, in a conventional `tau-model'  $\psi(t)$  decays exponentially from an initial value $\psi_0$ at time $t_0$, with $\psi(t) = \psi_{0}e^{-(t_{0} - t)/\tau}$, where $\tau$ is SFR e-folding time. This particular model (one of the many we experimented with; see Appendix
\ref{sec:OtherParametricModels}) is characterized by five parameters: $\Theta = (\psi_0, t_0, \tau, Z_\star, \tau_V)$. 

The synthetic spectrum for a given parameterization and choice of $\Theta$ is computed with

\begin{equation}
\label{eq:ModelSpectrum}
L_\lambda(\Theta) =  e^{-q_{\lambda} \tau_V} \int \mathrm{SSP}_\lambda(t,Z) \, \psi(t; \Theta) \,  dt,
\end{equation}

\noindent where $q_\lambda \equiv \tau_\lambda / \tau_V$ denotes the attenuation law which, in our case, is the one by \citet{calzetti00}.

The goal is to explore the parameter space and constrain the parameters $\Theta$ that fit our data (indices + photometry). 
We use a Markov chain Monte Carlo (MCMC) method to sample the parameter space. For each sampled $\Theta$  we: 

\begin{itemize}
\item Compute the predicted model spectrum (equation \ref{eq:ModelSpectrum}) for a SFH forming a total of $1 M_\odot$ in stars.
\item Evaluate the corresponding observables.
\item Determine the corresponding total stellar mass formed by maximizing the likelihood of the scale dependent observables (the photometric fluxes, in our case).
\item Compare the observed and model observables, obtaining the likelihood of the data given $\Theta$.
\end{itemize}

In what follows we explain how this general method is implemented in practice.

\subsubsection{Bayesian inference}

Given a set of observations $O$ (7 broad band magnitudes and 3 spectral indices in our case), the general goal of an MCMC algorithm is to draw a set of samples $\{\Theta_{i}\}$ in the parameter space from the posterior probability density

\begin{equation}
p(\Theta|O) = \frac{ p(\Theta) p(O|\Theta) }{ p(O) },
\end{equation}

\noindent where $p(\Theta)$ is the prior distribution on $\Theta$ and $p(O|\Theta)$ is the likelihood function. The normalization $p(O)$ is independent of $\Theta$ once we have chosen the form of the generative model, and therefore it can be omitted from the analysis.

An advantage of a Bayesian analysis is that we can marginalize over uninteresting parameters. In our case we are more interested on parameters controlling the temporal behavior of the SFH than on $Z_\star$ or $\tau_V$.
Moreover, once the MCMC sample of $p(\Theta|O)$ is available we can also obtain the expected value of any function of $\Theta$. For example, the expected value of the luminosity-weighted mean log stellar age is computed from 

\begin{equation}
\label{eq:BayesianAverage}
\langle \log t \rangle_L = \sum_{\Theta_{i}}^{} p(\Theta_{i}|O) \langle \log t \rangle_L(\Theta_{i})
\end{equation}

\noindent  where $\langle \log t \rangle_L(\Theta_{i})$ is the mean age for a particular $\Theta_{i}$ and the sum runs over all $\Theta_i$'s sampled by the MCMC.

The formalism is also useful to evaluate typical uncertainties on the model parameters. For example, the uncertainties in $t_0$ and $\tau$ are calculated as:

\begin{equation}
\langle t_{0} \rangle = \sum_{t_{0}}^{} p(t_{0}|O)\, t_{0}; \quad \sigma^{2}_{\langle t_{0} \rangle}= \sum_{t_{0}}^{} p(t_{0}|O)\left[\langle t_{0} - \langle t_{0} \rangle\right]^{2}
\end{equation}

\begin{equation}
\langle \tau \rangle = \sum_{\tau}^{} p(\tau|O)\, \tau; \quad \sigma^{2}_{\langle \tau \rangle}= \sum_{\tau}^{} p(\tau|O)\left[\langle \tau - \langle \tau \rangle\right]^{2}
\end{equation}

These estimates can then be propagated to  estimate the associated uncertainties in properties such as  $\rho_{\rm SFR}$ and $\rho_\star$, as done in Sections \ref{sec:rho-SFR} and Sec.\ \ref{sec:rho-mass}.

\subsubsection{Likelihood function}

Assuming gaussian errors,  $p(O|\Theta) \propto e^{ -\chi^2(\Theta) / 2}$, so the computation of the likelihood $p(O|\Theta)$ reduces to the evaluation of the corresponding $\chi^2(\Theta)$, which in our case splits into two parts: One $\chi^2$ related to the $N_{\rm mag} = 7$ photometric magnitudes and another one related to the $N_{\rm Lick} = 3$ Lick indices. The latter reads

\begin{equation}
\chi^2_{\rm Lick} = \sum_{j=1}^{N_{\rm Lick}}\left( \frac{ I^{\rm obs}_j  - I^{\rm mod}_j(\Theta) }{ \epsilon(I^{\rm obs}_j) } \right)^2,
\end{equation}

\noindent where $I^{\rm obs}_j$  and $\epsilon(I^{\rm obs}_j)$ denote the observed index and its uncertainty, and $I^{\rm mod}_j(\Theta)$ is the predicted value for the SFH parameters encoded in $\Theta$. 

The photometric component of $\chi^2(\Theta)$ is 

\begin{equation}
\chi^2_{\rm mag} = \sum_{j=1}^{N_{\rm mag}}\left(\frac{O_{j} - M_{j} (\Theta ; 1M{\odot}) - A(\Theta) }{ \epsilon(O_j )}\right)^{2}
\end{equation}

\noindent where $O_j$ and $\epsilon(O_j )$ are the observed magnitude and its error, $M_{j}(1M_{\odot})$ is the magnitude expected for a galaxy forming $1 M_\odot$ in stars, and $A = - 2.5 \log M_\star^\prime$ defines the 
optimal\footnote{optimal in the sense of $\partial \chi^2_{\rm mag} / \partial M_\star^\prime = 0$ (see  \citealt{walcher11})
}
mass formed in stars ($M_\star^\prime$)  for parameters $\Theta$.

We combine $\chi^2_{\rm Lick}$ and $\chi^2_{\rm mag}$ as follows:

\begin{equation}
\label{eq:chi2_tot}
\chi^2_{\rm tot} = \chi^2_{mag} + \frac{N_{\rm mag}}{N_{\rm Lick}} \chi^2_{\rm Lick}.
\end{equation}

\noindent where the factor $N_{\rm mag} / N_{\rm Lick} = 7/3$ scales the two $\chi^2$'s to give the same weight to magnitudes and indices. Typically, the assumed errors in the magnitudes are 0.08, 0.06, 0.07, 0.05, 0.05, 0.05 and 0.06 mag, for FUV, NUV, u, g, r, i, z respectively. For the indices, the average errors are: $0.14$ \AA, $0.09$, and $0.03$ \AA, for H$\beta$, D$_n$4000, and  [MgFe]$'$, respectively.

%-------------------------------------------------------------------------------------------------------
%NEW SECTION
%-------------------------------------------------------------------------------------------------------

\section{Star formation histories as a function of Hubble type and radial region}
\label{sec:Results}

This section presents our results for the SFH as a function of the Hubble type and galaxy mass. We discuss the absolute and specific SFR, and mass fraction as a function of lookback time. The spatial analysis is simplified to three radial regions: $R= 0$--2 HLR,  $R< 0.5$ HLR, and $R =1$--2 HLR, corresponding to the whole-galaxy, the central region (dominated by the galaxy bulge), and outer regions (dominated by the disk). 
First we present our reference SFH law, the quality of the fits, and the stellar population properties obtained from the integrated spectra.

%-----------------------------------------------------------subsection-----------------------------------------------------------------------------------------------------------

\subsection{The delayed-$\tau$ model}

We have explored a total of nine different parameterizations for $\psi(t;\Theta)$. Appendix \ref{sec:OtherParametricModels} describes and compares these alternative models. In what follows we restrict our discussion to the results obtained with a delayed-$\tau$ model:

\begin{equation}
\label{eq:law}
\psi(t) = k \frac{ ( t_0 - t) }{ \tau } e^{ - ( t_0 - t) / \tau }
\end{equation}

\noindent where $t$ is the lookback-time, $ t_0$ is the lookback-time onset of the star formation, $\tau$ is the SFR e-folding time and $k$
is a normalization factor trivially related to the total mass formed in stars, $M_\star^\prime = \int \psi(t) dt$.  
Although not as popular as a simple exponential decay, this recipe has been used in several works (e.g., 
\citealt{bruzualkron80,chiosi80,gavazzi02,chiosi17,lee10,simha14}).

This model involves a total of four parameters, $\Theta = ( t_0, \tau, Z, \tau_V)$, plus the stellar mass $M_\star$ which is optimized as explained in the previous section. To quantify how well the model fits the data we may use the difference between the synthetic and observed quantity in units of its error: $\Delta \equiv ({\rm model} - {\rm data}) / {\rm error}$.  
The mean and one sigma values of $\Delta$ for the three spectral indices are $0.56 \pm 0.60$, $-0.38\pm1.06$, $0.66 \pm 0.94$ for D$_n$4000, [MgFe]$'$, and H$\beta$, respectively, indicating satisfactory fits. The same is true for the SDSS magnitudes, with $\Delta = 0.19 \pm1.15$, $-0.08\pm 0.71$, $-0.46\pm 0.29$, $-0.30\pm 0.28$ and $0.10\pm 0.55$ for u, g, r, i, and z, respectively. The GALEX bands are the least well fitted observables, with  $\Delta_{\rm FUV} = -1.38\pm 0.81$ and $\Delta_{\rm NUV} =1.84 \pm 0.98$.  
As shown in the Appendix, these differences between the observed and predicted UV fluxes are reduced only when a combination of two exponential functions is used for $\psi(t)$.

The quality of the fits obtained with the delayed-$\tau$ model is similar to those obtained with exponential or \citet{sandage86} models for the SFH; in terms of $\chi^2$, the best is a combination of two SFH models (see Fig.\ \ref{fig:qualityfits}). Nevertheless, we favor the delayed-$\tau$ model because it results in a better match of the cosmic evolution of published $\rho_{\rm SFR}$ and sSFR from galaxy surveys. 

For each galaxy, this analysis provides estimates of $t_0$, $\tau$, $Z_\star$, $\tau_V$, and $M_\star$, as well as the associated SFR$(t)$ and sSFR$(t)$ functions. Other properties, such as the mean stellar ages and metallicities, can be derived from these. In what follows we explore the results of these SFH fits. The emphasis is on statistical results obtained when grouping galaxies by their Hubble type.

%subsection-----------------------------------------------------------------------------------------------------------
 
\subsection{Global properties as a function of Hubble type}

%***FIG***FIG***FIG***FIG***FIG***FIG***FIG***FIG***FIG***FIG***
%\begin{figure*}[!ht]
\begin{figure}
\includegraphics[width=0.5\textwidth]{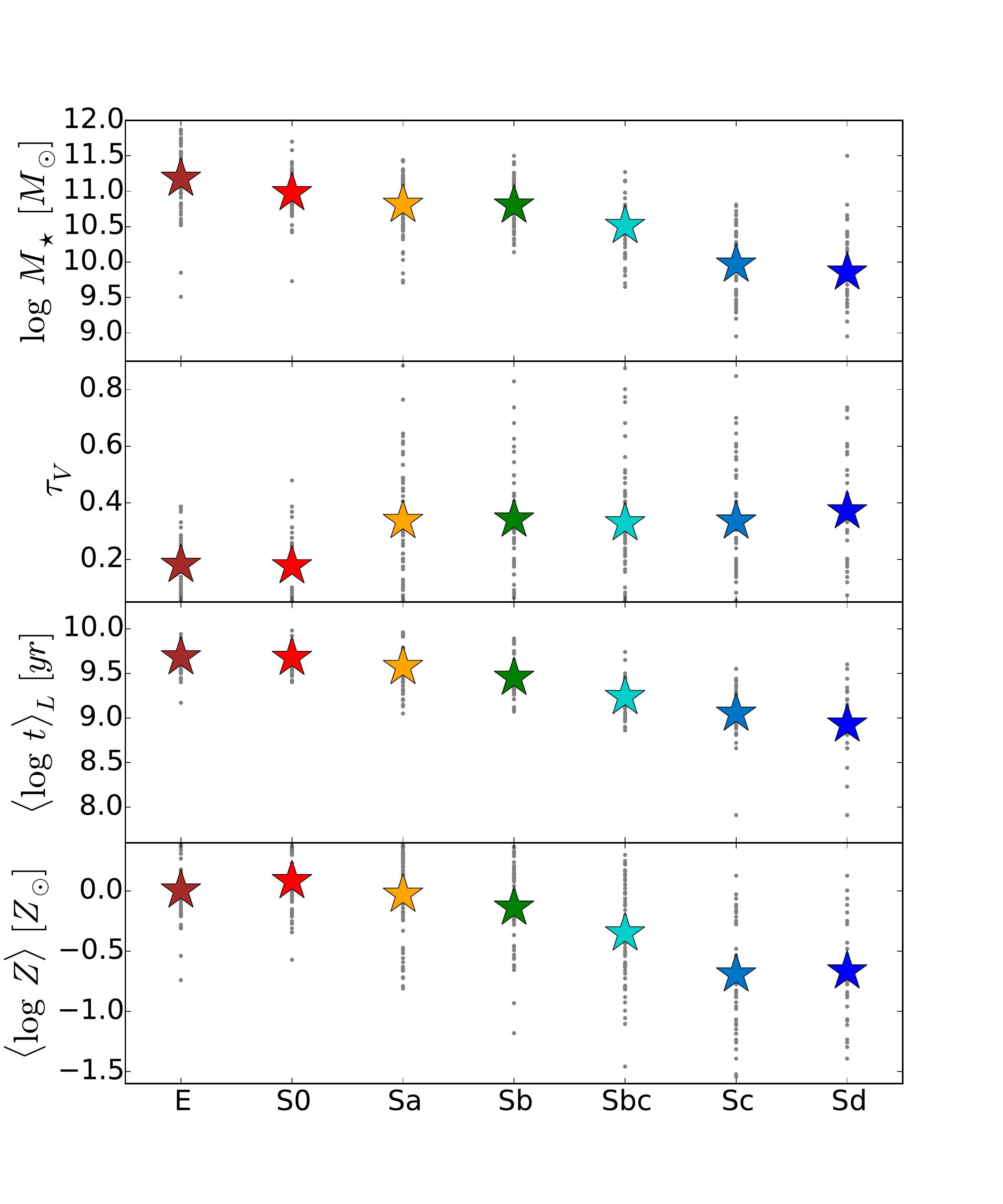}
\caption{Global stellar population properties as a function of the Hubble type obtained with  a delayed-$\tau$ model 
and Chabrier IMF. From top to bottom are the galaxy stellar mass ($M_\star$), attenuation ($\tau_V$), light-weighted averaged age ($\langle \log \  t \rangle_L$), and stellar metallicity ($\langle \log \  Z \rangle$). The  individual galaxy values are plotted as grey dots, and the average values for each Hubble type  as color coded stars. }
\label{fig:SPproperties}
\end{figure}
%***FIG***FIG***FIG***FIG***FIG***FIG***FIG***FIG***FIG***FIG***

In Fig.\ \ref{fig:SPproperties} we present the results for the stellar mass ($M_\star$), attenuation ($\tau_V$), luminosity-weighted age ($\langle \log \  t \rangle_L$), and  the  stellar metallicity ($\langle \log \  Z \rangle$)\footnote{For this model, the metallicity is calculated as:
\begin{equation}
\langle \log Z \rangle = \sum_{\Theta_{i}}^{} p(\Theta_{i}|O)  \log Z_\star (\Theta_{i})
\end{equation}}, 
 as a function of Hubble type. 
Dots represent the expected values (cf.\ equation \ref{eq:BayesianAverage}) for individual galaxies, while and coloured stars mark the average for each morphology bin.

As noticed in our previous works, $M_\star$ correlates with Hubble type, with average values of $\log M_\star$ = 11.18, 10.98, 10.81, 10.80, 10.51, 9.97, and 9.85 for E, S0, Sa, Sbc, Sc, and Sd, respectively. Extinction is not correlated with morphology, but spirals have higher  $\tau_V$ than  E and S0.  $\langle \log \ t \rangle_L$ and $\langle \log \ Z \rangle$ scale with the Hubble type, with early type galaxies (ETG) more metal rich and older than late type spirals.

These trends with the Hubble type are in good agreement with those derived by \citet{gonzalezdelgado14a, gonzalezdelgado15} using \starlight\ for the non-parametric spectral fits instead of the parametric models used in this paper.  The agreement is also quantitative when the SSP models used for the full spectral fits are the same ones used for the parametric analysis. The differences between non-parametric and parametric properties are (mean $\pm$ standard deviation) $0.16 \pm 0.17$ in $\log M_\star$ , $-0.12 \pm0.12$ in $\tau_V$, $0.12 \pm 0.20$ in $\langle \log \ t \rangle_L$, and  $-0.07 \pm 0.37$  in $\langle \log \ Z \rangle$. 

We note that in a few cases our parametric fits produce unrealistically small values of $t_0$. For 4 of our 54 E galaxies, for instance, the fits suggest $t_0 \sim 4$ Gyr and small $\tau$ values. These objects are better represented by a combination of an old population with some modest recent star formation, a mixture that is not well described by a delayed-$\tau$ model. This difficulty is analogous to the one described by  \citet{Trager2000b}, who find that single burst models for early type galaxies sometimes find too young ages, and that a more reasonable scenario is one where a "frosting" of young stars is added to a old "base" population (see also \citet{kaviraj07a}).  Overall, these cases are not numerous enough to affect our main results.

%---------------------------------------------------------subsection-------------------

\subsection{The evolution of the star formation rate}
\label{sec:SFR}

%***FIG***FIG***FIG***FIG***FIG***FIG***FIG***FIG***FIG***FIG***
%\begin{figure*}[!ht]
\begin{figure*}
\includegraphics[width=\textwidth]{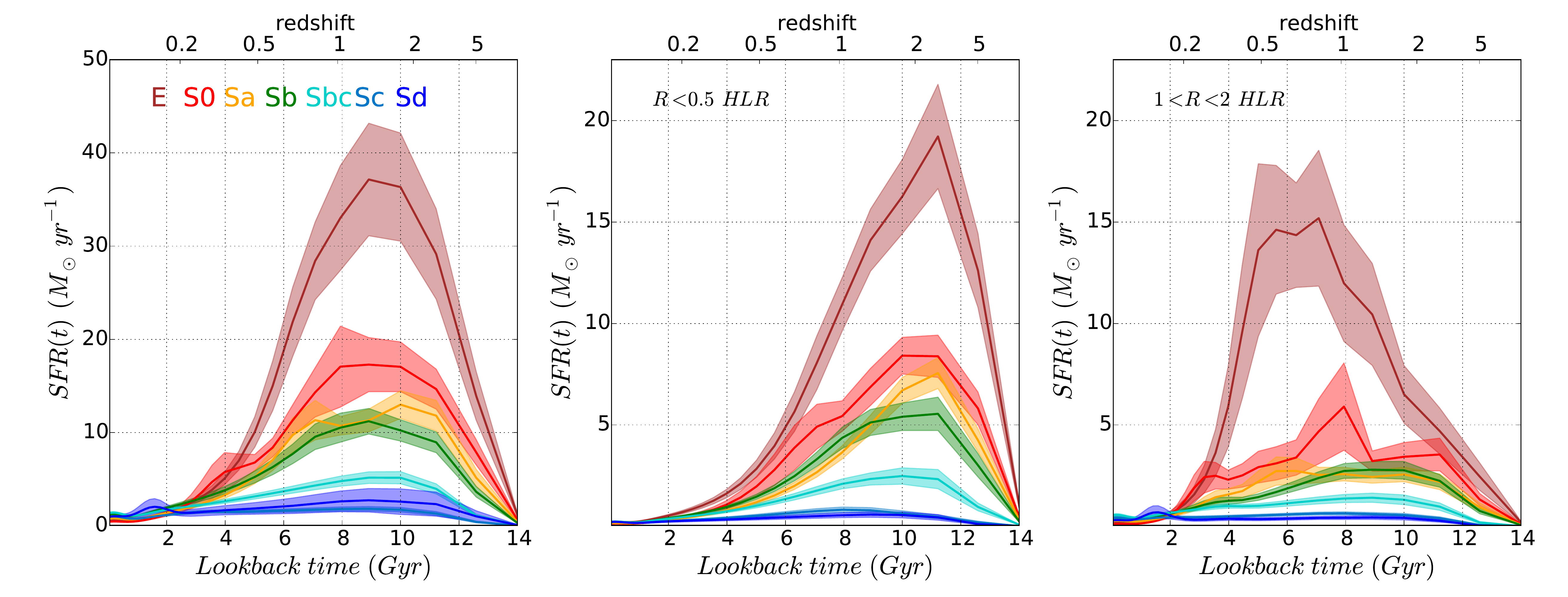}
\caption{The evolution of the  star formation rate (SFR) of CALIFA galaxies as a function of  Hubble type; for the whole galaxy (left), the inner (middle), and outer regions (right). Solid lines are the average sSFR($t$)  of the galaxies in each morphology bin removing only the 5$\%$ of galaxies with smallest $t_0$.  Shaded bands around the average curves represent $\pm$ the error in the mean. 
}
\label{fig:SFR}
\end{figure*}
%***FIG***FIG***FIG***FIG***FIG***FIG***FIG***FIG***FIG***FIG***

Figure \ref{fig:SFR} shows the time evolution of the SFR for seven morphology bins obtained for the global ($R \leq 2$ HLR, left panel), central ($R \leq 0.5$ HLR, middle panel), and outer regions ($1 \leq R \leq 2$ HLR, right panel) of the galaxy. In order to make a consistent comparison with results obtained with non-parametric methods (such as those obtained with \starlight, see figure 6 in \citealt{gonzalezdelgado17}), we  compute the mean star formation history for each Hubble type by averaging the SFR$(t)$ functions for individual galaxies in the same morphology group. The error in the mean is computed as the r.m.s.\ dispersion of the corresponding SFR$(t)$ values divided by the square root of the number of galaxies in each bin. 

These statistics are computed after excluding the 5\% of objects with smaller $t_0$ in each group. This minor correction is done only for cosmetic purposes. Since these outliers also have small $\tau$ values, their SFHs are concentrated in time, leading to artificial peaks in the mean SFR$(t)$ curves, which are meant to be typical of a given morphology group.

Figure \ref{fig:SFR} shows that the SFRs at any epoch scale with the Hubble type, as expected due to the  relations between SFR, $M_\star$, and morphology. E galaxies have the highest SFRs, reaching  $\sim 40 \  M_\odot\,$yr$^{-1}$ about 9 Gyr ago ($z\sim1.5$), while the lowest SFRs at the same epoch occur in Sd galaxies,  with only $\sim 3 \ M_\odot\,$yr$^{-1}$.
Note that although the shape of SFR$(t)$ for each individual galaxy follows a delayed-$\tau$ function, the mean SFR$(t)$ curve does not. This happens because of the dispersion on the fitted $ t_0$ and $\tau$ among galaxies in a same Hubble type bin. The end result is that the mean SFR$(t)$ curves shows a relatively broad plateau from $z \sim 2$ to 1, as well as some secondary peaks. This is evident not only for E, but also for S0 and late type spirals. At recent epochs, the SFRs have declined by a factor of $\sim 40$ for ETG, and $\sim 2$ for Sc with respect to their values at $z \sim 1$--2. 

The SFR in the inner regions (central panel) increases faster at earlier times and declines quickly after the peak, in particular for E's. At the peak of the star formation epoch (lookback time 10-12 Gyr) the inner region contributes significantly ($\sim55\%$) to the total SFR. In contrast, for the Sd galaxies, the $R <  0.5$ HLR region accounts for only $16\%$ of the total SFR.

The SFR in the outer regions (right panel) of spirals  shows a behavior similar to that in the inner regions, but with a rising phase that extends to more recent times, and  a declining phase that is also more extended in time. In ETG's, and in particular E's, the peak of star formation occurs between redshift 1 and 0.5. This result confirms our previous suggestion that E galaxies are actively forming or accreting stars in the outer regions 
between  $\sim 5$ and 8 Gyr ago \citep{gonzalezdelgado17}.

%-----------------------------------------------------------subsection-----------------------------------------------------------------------------------------------------------

\subsection{The star formation rate parameters}

%***FIG***FIG***FIG***FIG***FIG***FIG***FIG***FIG***FIG***FIG***
%\begin{figure*}[!ht]
\begin{figure}
\includegraphics[width=0.5\textwidth]{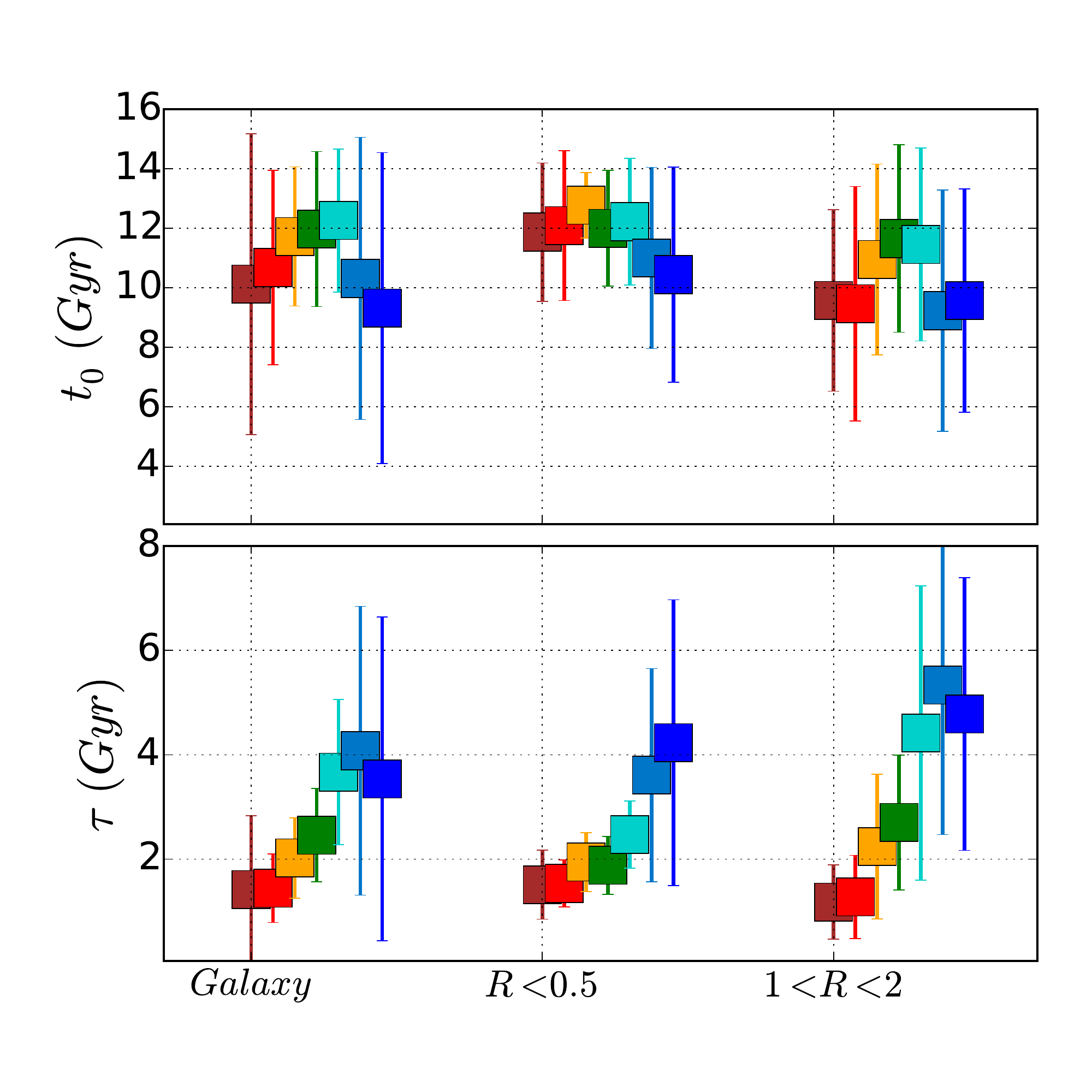}
\caption{Average and dispersion of the parameters $ t_0$ and $\tau$ of a delayed-$\tau$ model for the SFH, grouped into seven morphology bins. The three columns refer to results obtained for the whole galaxy (left), only the central $R < 0.5$ HLR regions (middle), and $1 < R < 2$ HLR (right).
}	
\label{fig:tau}
\end{figure}
%***FIG***FIG***FIG***FIG***FIG***FIG***FIG***FIG***FIG***FIG***

\begin{table}
\begin{tabular}{c c c c} % centered columns (4 columns)
\multicolumn{2}{c}{M1 model}& $t_{0}$ [Gyr] & $\tau$ [Gyr]\\ %
\hline
E  &  
\begin{tabular}{c}
$R<0.5$\\
galaxy\\
$1<R<2$
\end{tabular}
& \begin{tabular}{c}  $11.9\pm 2.3$ \\ $10.1\pm 5.1$\\ $9.6\pm 3.1$ \end{tabular}
& \begin{tabular}{c}  $1.5\pm 0.7$\\ $1.4\pm 1.4$\\ $1.2\pm 0.7$ \end{tabular}\\
\hline
S0  &  
\begin{tabular}{c}
$R<0.5$\\
galaxy\\
$1<R<2$
\end{tabular}
& \begin{tabular}{c}  $12.1\pm 2.5$ \\ $10.7\pm 3.3$\\ $9.5\pm 3.9$ \end{tabular}
& \begin{tabular}{c}  $1.5\pm 0.5$ \\ $1.4\pm 0.7$\\ $1.3\pm 0.8$ \end{tabular}\\
\hline
Sa  &  
\begin{tabular}{c}
$R<0.5$\\
galaxy\\
$1<R<2$
\end{tabular}
& \begin{tabular}{c}   $12.8\pm 1.1$ \\ $11.7\pm 2.3$\\ $10.9\pm 3.2$ \end{tabular}
& \begin{tabular}{c}   $1.9\pm 0.6$ \\ $2.0\pm 0.8$\\ $2.2\pm 1.4$ \end{tabular}\\
\hline
Sb  &  
\begin{tabular}{c}
$R<0.5$\\
galaxy\\
$1<R<2$
\end{tabular}
& \begin{tabular}{c}  $12.0\pm 2.0$ \\ $12.0\pm 2.6$\\ $11.6\pm 3.1$ \end{tabular}
& \begin{tabular}{c}   $1.9\pm 0.5$ \\ $2.5\pm 0.9$\\ $2.7\pm 1.9$ \end{tabular}\\
\hline
Sbc  &  
\begin{tabular}{c}
$R<0.5$\\
galaxy\\
$1<R<2$
\end{tabular}
& \begin{tabular}{c}   $12.2\pm 2.1$ \\ $12.3\pm 2.4$\\ $11.4\pm 3.2$ \end{tabular}
& \begin{tabular}{c}   $2.5\pm 0.6$ \\ $3.7\pm 1.4$\\ $4.4\pm 2.8$ \end{tabular}\\
\hline
Sc  &  
\begin{tabular}{c}
$R<0.5$\\
galaxy\\
$1<R<2$
\end{tabular}
& \begin{tabular}{c}   $11.0\pm 3.0$ \\ $10.3\pm 4.7$\\ $9.2\pm 4.1$ \end{tabular}
& \begin{tabular}{c}   $3.6\pm 2.0$ \\ $4.1\pm 2.8$\\ $5.3\pm 2.9$ \end{tabular}\\
\hline
Sd  &  
\begin{tabular}{c}
$R<0.5$\\
galaxy\\
$1<R<2$
\end{tabular}
& \begin{tabular}{c}   $10.4\pm 3.6$ \\ $9.3\pm 5.2$\\ $9.6\pm 3.7$ \end{tabular}
& \begin{tabular}{c}  $4.2\pm 2.7$ \\ $3.5\pm 3.1$\\ $4.8\pm 2.6$ \end{tabular}\\
\hline %inserts single line
\end{tabular}
\caption{Average and standard deviation of $ t_0$ and $\tau$, the  SFH parameters for a delayed-$\tau$ model  (eq.\ \ref{eq:law}) for seven bins in Hubble type. For each type we list the results obtained for the integrated spectra, and for spectra that include regions located at $R < 0.5$ HLR and $ 1 < R < 2$ HLR.
}
\label{tab:tautable}
\end{table}

Table \ref{tab:tautable} and Fig.\ \ref{fig:tau} shows the statistics (mean and standard deviation) for all values $ t_0$ and $\tau$ as a function of  Hubble type and radial region. Because the  5$\%$ of  outliers (excluded from Fig.\ \ref{fig:SFR}) are included in the statistics, the dispersion in  $ t_0$ is significant, in particular for E galaxies. 
The average values indicate that the  onset of  star formation occurs very early on: In Sd,  $t_0\sim9$ Gyr; in Sc, $t_0\sim10$ Gyr;  in Sa-Sb $t_0\sim12$ Gyr, and in E and S0 $t_0\sim10.1-10.7$ Gyr. The onset of  star formation in the inner regions is earlier  than in the outer regions, i.e., $t_0(R<0.5) > t_0(1<R<2)$.

The time scale $\tau$ increases  from 1.4  Gyr in E to  4.1 Gyr in Sc, indicating that the period of star formation is more extended in late  spirals than in ETG. The outer regions of spirals have a larger e-folding time than the inner regions, $\tau(R<0.5) < \tau(1<R<2)$; for instance, Sc galaxies typically have $\tau = 5.3$ Gyr at $1<R<2$ HLR, but 3.6 Gyr at $R < 0.5$ HLR. In ETGs the inner and outer values are quite similar, with $\tau \sim 1.2$ and 1.5 Gyr, respectively.

These results confirm our earlier finding that galaxies  form inside-out \citep{perez13, gonzalezdelgado14a, gonzalezdelgado15,garciabenito17}, because the peak of star formation, at $t = t_0 - \tau$, always occurs earlier in the inner  than in the outer regions, and this inner-outer delay is very similar in all Hubble types, with $(t_0-\tau)_{inner} - (t_0-\tau)_{outer} =2.2$ Gyr on average.

%---subsection-----------------------------------------------------------------------------------------------------------

\subsection{The evolution of the specific star formation rate}
\label{sec:sSFR}

%***FIG***FIG***FIG***FIG***FIG***FIG***FIG***FIG***FIG***FIG***
%\begin{figure*}[!ht]
\begin{figure*}
\includegraphics[width=\textwidth]{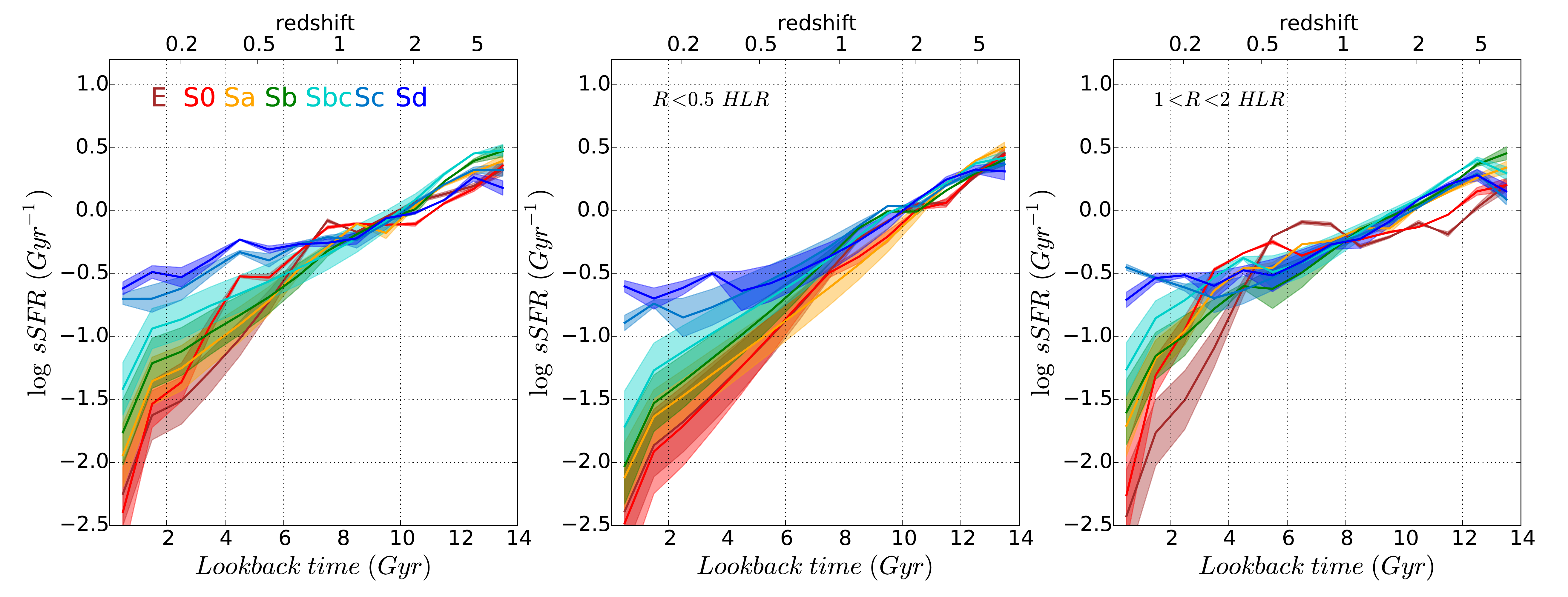}
\caption{As Fig.\ \ref{fig:SFR}, but showing the evolution of the specific star formation rate.}
\label{fig:sSFR}
\end{figure*}
%***FIG***FIG***FIG***FIG***FIG***FIG***FIG***FIG***FIG***FIG***

The SFH of a galaxy can  be expressed also through the time evolution of its specific star formation rate (sSFR), defined as the ratio between the current SFR and the stellar mass, sSFR $=$ SFR$(t = 0)/M_\star(t=0)$. This quantity provides information of the relative rate at which stars are  forming now with respect to the past. The sSFR$(t = 0)$ declines slowly with increasing $M_\star$ because the MSSF is sublinear (SFR $\sim M_\star^\alpha$, with $\alpha = 0.7$--0.9, \citealt{brinchmann04, salim04, renzinipeng15, catalantorrecilla15, gonzalezdelgado16}). 

In previous works, using the fossil record analysis based on the full spectral fitting of CALIFA data with \starlight, we were able to: {\em (a)} map the radial structure of the sSFR$(t=0)$, and  {\em (b)} study temporal variations of the sSFR. Our analysis has shown that sSFR decreases as the Universe evolves \citep{gonzalezdelgado17}, in line with redshift surveys \citep{speagle14}.

One important advantage of the methodology proposed here based on parametric SFH is that for each galaxy we have an analytical expression for SFR$(t)$ (eq. \ref{eq:law}), and the values of the sSFR$(t)$ are more easily derived. The sSFR$(t)$ for each galaxy is calculated as SFR$(t)/M_\star(t)$, where the SFR at each epoch is divided by $M_\star(t)$ (the  stellar mass of the galaxy at lookback time $t$), estimated as:

\begin{equation}
M_\star(t) = \int_{t_0}^{t} (1-R(t)) \ {\rm SFR}(t) \ dt
\end{equation}

\noindent
where the original mass formed in stars,  $ \int_{t_0}^{t}  {\rm SFR}(t) \ dt$, is corrected by the mass loss term $R(t)$. Rather than assuming a global correction,  we use the prediction given by the SSP models \citep{bruzualcharlot03} for each population of each galaxy, based on their ages and metallicities. Then, the individual sSFR$(t)$ are stacked as a function of  Hubble type. 

Figure \ref{fig:sSFR} shows our results, where the full lines are average curves and shaded bands are $\pm$ the error in the mean, i.e., the dispersion divided by the square root of the number of galaxies in each class. For all morphological types the sSFR$(t)$ curves decrease as the Universe evolves, but the slope is different for each Hubble type. Sd galaxies have the flattest slope,  while Sa, S0 and E show steeper slopes. 

The middle panel in Fig.\ \ref{fig:sSFR} shows the sSFR$(t)$ of regions located in the central 0.5 HLR. Although the behavior is similar to  that of the whole galaxy, the slope is steeper, producing the shut-down of the star-formation at earlier epochs. At $z > 2$, all the galaxies seem to have a common sSFR$(t)$ of $\sim 2$ Gyr$^{-1}$.
In contrast, for regions currently located in the disk of spirals, sSFR$(t)$ declines more slowly (right panel in Fig.\ \ref{fig:sSFR}),  indicating a longer period of star formation. The outer regions of E galaxies show a remarkable behavior: over the period   $0.5\leq z \leq 1$, sSFR$(t)$  run above those of spirals, while at $z > 1$ they are below. In previous works we suggested this period  $0.5 < z < 1$ as an epoch of growth of the envelope of E and S0 through mergers.

%-------------------------------------------------------------------------------------------------------
%NEW SECTION
%-------------------------------------------------------------------------------------------------------
\section{Comparison with galaxy redshift surveys}
\label{sec:Comparison}

We now put the SFHs presented in the last section in a cosmic context by applying volume corrections to obtain $\rho_{\rm SFR}(t)$, sSFR$(t)$, and  $\rho_\star(t)$. Our main goals in this section are: 1) To compare our results for the cosmic evolution of the SFR, sSFR and stellar mass obtained from the analysis of CALIFA data corresponding to the whole galaxy with the results from redshift surveys in the literature. These comparisons allow us to establish the advantages and also the limitations of the fossil record method to trace the evolution of these properties with respect to snapshots surveys of galaxy evolution obtained by studies at different redshifts. 2)  To discuss the capability of a delayed-$\tau$ model, with respect to other parameterizations of the SFH,  to trace the evolution of the star formation rate of the Universe.

\subsection{The cosmic star formation rate density}
\label{sec:rho-SFR}

%***FIG***FIG***FIG***FIG***FIG***FIG***FIG***FIG***FIG***FIG***
%\begin{figure*}[!ht]
\begin{figure}
\includegraphics[width=0.5\textwidth]{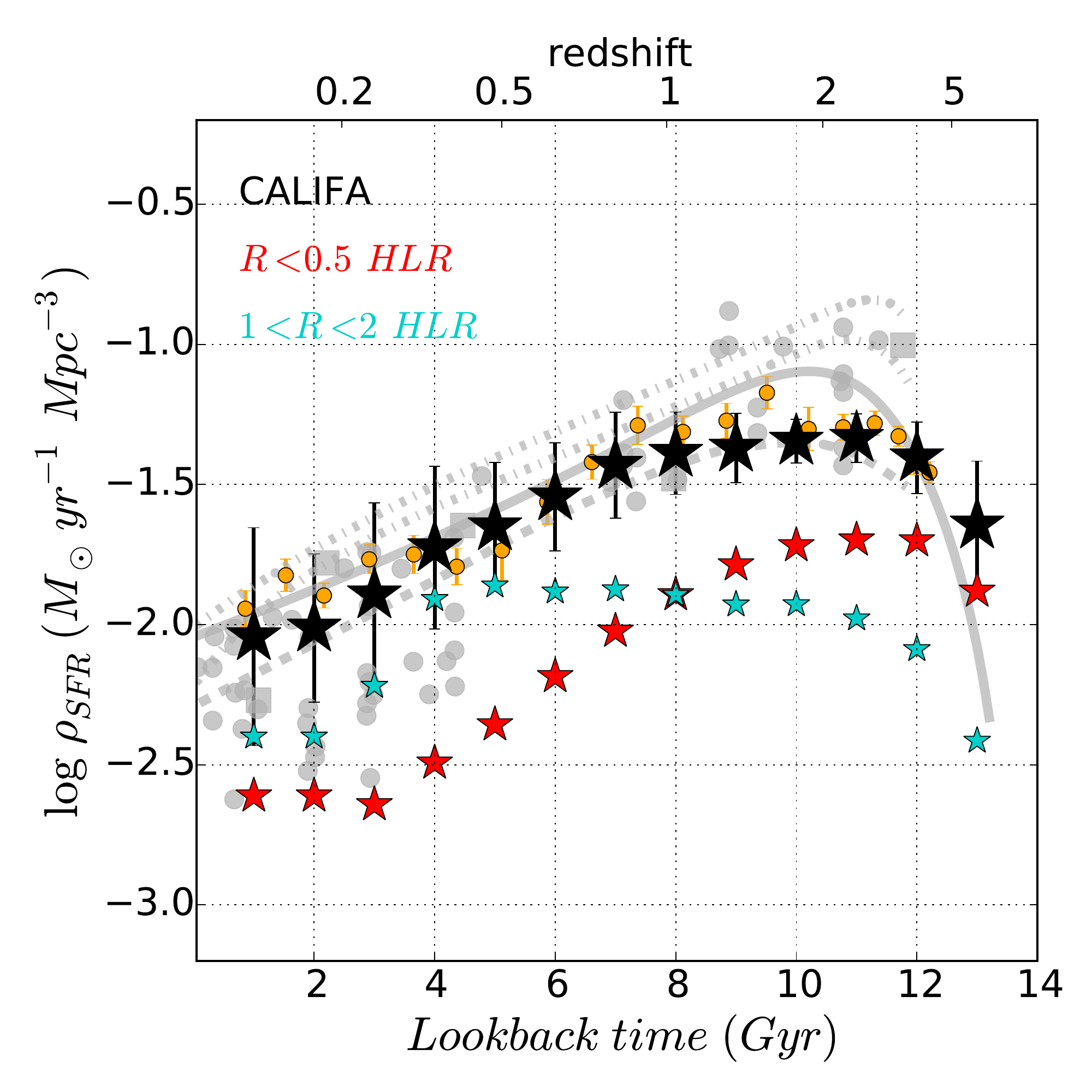}
\caption{  
The cosmic evolution of the star formation rate density, $\rho_{\rm SFR}$,  in the present study (black stars).  
Blue and red stars represent  the contribution to $\rho_{\rm SFR}$ of the regions between 1 and 2 HLR and within the inner 0.5 HLR, respectively. Other results are from recent determinations by \citet{gunawardhana13, gunawardhana15} and their compilation (gray points), and the redshift evolution of $\rho_{\rm SFR}$ from \citet{hopkinsbeacom06} (the top two gray dotted lines are $\pm 1 \sigma$ of their relation);  \citet{madau14} (middle gray full  line); \citet{fardal07} (bottom gray dashed line);  from the fossil record method applied to SDSS data by \citet{panter03} (gray squares); and from the work of \citet{driver17} for GAMA/G10-COSMOS/3D-HST data (orange dots). When needed, literature values have been scaled to a Chabrier IMF. 
}
\label{fig:cosmicSFR}
\end{figure}

\begin{table}
\resizebox{0.5\textwidth}{!} {\begin{tabular}{c c c c c c} % centered columns (4 columns)
\multicolumn{2}{c}{$\log\, \rho_{\rm SFR}$ [$M_{\odot}yr^{-1}Mpc^{-3}$]}& $z=0$ & $z=1$ & $z=2$ & $z=5$ \\
\hline %inserts double horizontal lines
\textbf{M1}&
\begin{tabular}{c}
$R<0.5$\\
Galaxy\\
$1< R < 2$\
\end{tabular}&  
\begin{tabular}{r}
$-2.61 \pm 0.13$\\
$-2.04 \pm 0.38$\\
$-2.40 \pm 0.11$\\
\end{tabular}& 
\begin{tabular}{r}
$-1.89 \pm 0.11$\\
$-1.39 \pm 0.14$\\
$-1.89 \pm 0.31$\\
\end{tabular}& 
\begin{tabular}{r}
$-1.72 \pm 0.09$\\
$-1.34 \pm 0.07$\\
$-1.93 \pm 0.14$\\
\end{tabular}& 
\begin{tabular}{r}
$-1.70 \pm 0.08$\\
$-1.41 \pm 0.12$\\
$-2.09 \pm 0.25$\\
\end{tabular}\\

\hline
Driver et al.                   & & $-1.94$ & $-1.31$ & $-1.30$ & $-1.46$\\
Madau $\&$ Dickinson  &  & $-2.05$ & $-1.27$ & $-1.10$ & $-1.41$\\
Fardal et al.                  &  & $-2.28$ & $-1.42$ & $-1.35$ & \\
\hline\\[0.25cm] % inserts single horizontal line

\multicolumn{2}{c}{$\log\, sSFR$ [$Gyr^{-1}$]}& &  & &  \\
\hline %inserts double horizontal lines
\textbf{M1}&
\begin{tabular}{c}
$R<0.5$\\
Galaxy\\
$1< R < 2$\
\end{tabular}&  
\begin{tabular}{r}
 $-1.59 \pm 0.04$\\
$-1.27 \pm 0.13$\\
$-1.16 \pm 0.05$\\
\end{tabular}& 
\begin{tabular}{r}
$-0.34 \pm 0.02$\\
$-0.27 \pm 0.04$\\
$-0.36 \pm 0.03$\\
\end{tabular}& 
\begin{tabular}{r}
$-0.06 \pm 0.03$\\
$-0.08 \pm 0.12$\\
$-0.13 \pm 0.02$\\
\end{tabular}& 
\begin{tabular}{r}
$0.13 \pm 0.02$\\
$0.18 \pm 0.02$\\
$0.08 \pm 0.02$\\
\end{tabular}\\
\hline

Elbaz et al. &  & $-1.12$ & $-0.21$ & $0.18$ & \\
\hline\\[0.25cm] % inserts single horizontal line

\multicolumn{2}{c}{$\log\, \rho_\star$ [$M_{\odot}Mpc^{-3}$]} &  &  &  &  \\
\hline %inserts double horizontal lines
\textbf{M1}&
\begin{tabular}{c}
$R<0.5$\\
Galaxy\\
$1< R < 2$\
\end{tabular}&  
\begin{tabular}{r}
$7.73 \pm 0.13$\\
$8.18 \pm 0.25$\\
$7.70 \pm 0.26$\\
\end{tabular}& 
\begin{tabular}{r}
$7.64 \pm 0.16$\\
$7.99 \pm 0.19$\\
$7.37 \pm 0.13$\\
\end{tabular}& 
\begin{tabular}{r}
$7.46 \pm 0.19$\\
$7.78 \pm 0.24$\\
$7.10 \pm 0.18$\\
\end{tabular}& 
\begin{tabular}{r}
$7.01 \pm 0.43$\\
$7.24 \pm 0.46$\\
$6.47 \pm 0.34$\\
\end{tabular}\\
\hline
Driver et al.                   &  & $8.30$ & $8.08$ & $7.90$ & $7.70$\\
Madau $\&$ Dickinson &  & $8.56$ & $8.32$ & $8.01$ & $5.92$\\
\hline\\[0.25cm] % inserts single horizontal line

\multicolumn{2}{c}{$\log\, \rho_\star^\prime$ [$M_{\odot}Mpc^{-3}$]} &  &  &  &  \\
\hline %inserts double horizontal lines
\textbf{M1}&
\begin{tabular}{c}
$R<0.5$\\
Galaxy\\
$1< R < 2$\
\end{tabular}&  
\begin{tabular}{r}
$8.02 \pm 0.13$\\
$8.47 \pm 0.24$\\
$7.99 \pm 0.25$\\
\end{tabular}& 
\begin{tabular}{r}
$7.87 \pm 0.16$\\
$8.21 \pm 0.18$\\
$7.58 \pm 0.14$\\
\end{tabular}& 
\begin{tabular}{r}
$7.66 \pm 0.19$\\
$7.97 \pm 0.25$\\
$7.29 \pm 0.18$\\
\end{tabular}& 
\begin{tabular}{r}
$7.14 \pm 0.42$\\
$7.35 \pm 0.44$\\
$6.60 \pm 0.34$\\
\end{tabular}\\
\hline

\hline
\end{tabular}}
\caption{$\rho_{\rm SFR}$,  sSFR, and $\rho_\star$ for redshifts $z = 0, 1, 2, 5$ obtained in this work. Results from \citet{driver17}, \citet{madau14}, \citet{fardal07},  and \citet{elbaz11} are included. The result from  \citet{driver17} included in column $z=5$ in this table was measured, in fact, at a lookback time of 12.2 Gyr.
}
\label{tab:tablecosmic} 
\end{table}

One of the fundamental results obtained over the last two decades of observations from multiwavelength galaxy surveys is that the star formation rate density of the Universe peaked approximately 3.5 Gyr after the Big Bang, at $z \sim 2$, and declined thereafter \citep{lilly96, madau98, hopkinsbeacom06, fardal07, madau14}.
Here, we extrapolate the SFR computed at each cosmic epoch from the CALIFA galaxies to derive the cosmic evolution of the SFR density ($\rho_{\rm SFR}$)  and how it breaks up into contributions from current inner and outer galaxy regions, and into the different Hubble types.

CALIFA, as many other samples, is not volume-limited, but it can be volume-corrected using the $V_{max}$ method of \citet{schmidt68}. The volume available per galaxy, $V_{max}$,  was calculated for the CALIFA mother sample assuming that the ratio between apparent and linear isophotal size of a galaxy depends only on its angular diameter distance \citep{walcher14}.
In \citet{gonzalezdelgado16}, we used this method to derive a SFR density in the local Universe  $\rho_{\rm SFR} = (0.0105 \pm 0.0008) \ M_\odot\,$yr$^{-1}\,$Mpc$^{-3}$  (for a Salpeter IMF), in very good agreement with independent estimates. Furthermore, we showed that most of the current star formation  occurs in the disk of spirals. Now we extend this study by calculating  $\rho_{\rm SFR}$ at different cosmic epochs using the 366 galaxies that belong to the CALIFA mother sample. 
We transform our SFR$(t)$ estimates into $\rho_{\rm SFR}(t)$  by adding SFR$(t) / V_{max}$ at each epoch and correcting the result by the fraction $937 / 366$ to emulate what would be obtained for the full CALIFA mother sample of 937 galaxies.

Figure \ref{fig:cosmicSFR} places our values (black stars) in the $\rho_{\rm SFR}$ vs. lookbacktime  (or redshift) diagram. The calculations are done at 1 Gyr  time step,  and the error in each epoch is obtained by propagating the dispersion of $t_0$ and $\tau$ of each galaxy.  $\rho_{\rm SFR}$ shows a clear increase from $z = 0$ to  1, a plateau between $z=1$ and 3, and a decrease at higher redshift. 

Figure \ref{fig:cosmicSFR} includes (grey lines) the evolution of $\rho_{\rm SFR}$ from \citet{madau14}, \citet{hopkinsbeacom06}, and \citet{fardal07}. It also includes $\rho_{\rm SFR}$ from the compilation of \citet{gunawardhana15,gunawardhana13}, and the results obtained by \citet{panter03} from the fossil record method applied to  SDSS data (grey dots and squares). When necessary, the literature results are scaled to a Chabrier IMF.  Our estimations are similar to the values from \citet{fardal07}, but higher by 0.24 dex at $z \sim$ 0 (see also Table \ \ref{tab:tablecosmic}).
 In contrast, our $\rho_{\rm SFR}$ at $z=0$ is in agreement with \citet{madau14} when their values are scaled by the change of IMF (Salpeter in \citealt{madau14} and Chabrier in this work); but it is below by 0.24 dex at $z=2$.

Our results are also in excellent agreement with those obtained  by \citet{driver17} using GAMA/G10-COSMOS/3D-HST data (orange points in Fig.\ \ref{fig:cosmicSFR}).  Their analysis (which takes in account the effects of dust, AGN, and cosmic variance) results in a plateau in $\rho_{\rm SFR}$ between $z = 3$ and 1, similar to that found in our analysis. 
At $z\sim2$, our results and those from  \citet{driver17}  are  a factor 1.7 below the results from \citet{bourne17}. This latter study reports that obscured star formation dominates the total $\rho_{\rm SFR}$ in massive galaxies at all redshifts,  exceeding unobscured star formation by a factor $>$ 10. We can think of three reasons for the discrepancy at $z\sim2$: (i) The sample in \citet{bourne17} is dominated by dust obscured massive galaxies, and these may not be the progenitors of the CALIFA galaxies, which are representative of the current Hubble sequence. (ii) We cannot account for an extra dust obscuration at high redshift because our analysis does not assume a dust content evolution in the stellar populations. (iii) The \citet{bourne17} study is contaminated by AGN, which boost $\rho_{\rm SFR}$ at high redshift (as suggested by \citealt{driver17}). 

On the whole, we conclude that our fossil-record estimates of $\rho_{\rm SFR}(t)$ are in good general agreement with the cosmological studies.

%-------------------------------------------------------------------------------------------------------

\subsection{The cosmic specific star formation rate}
\label{sec:EvolutionsSFR}

%***FIG***FIG***FIG***FIG***FIG***FIG***FIG***FIG***FIG***FIG***
%\begin{figure*}[!ht]
\begin{figure}
\includegraphics[width=0.5\textwidth]{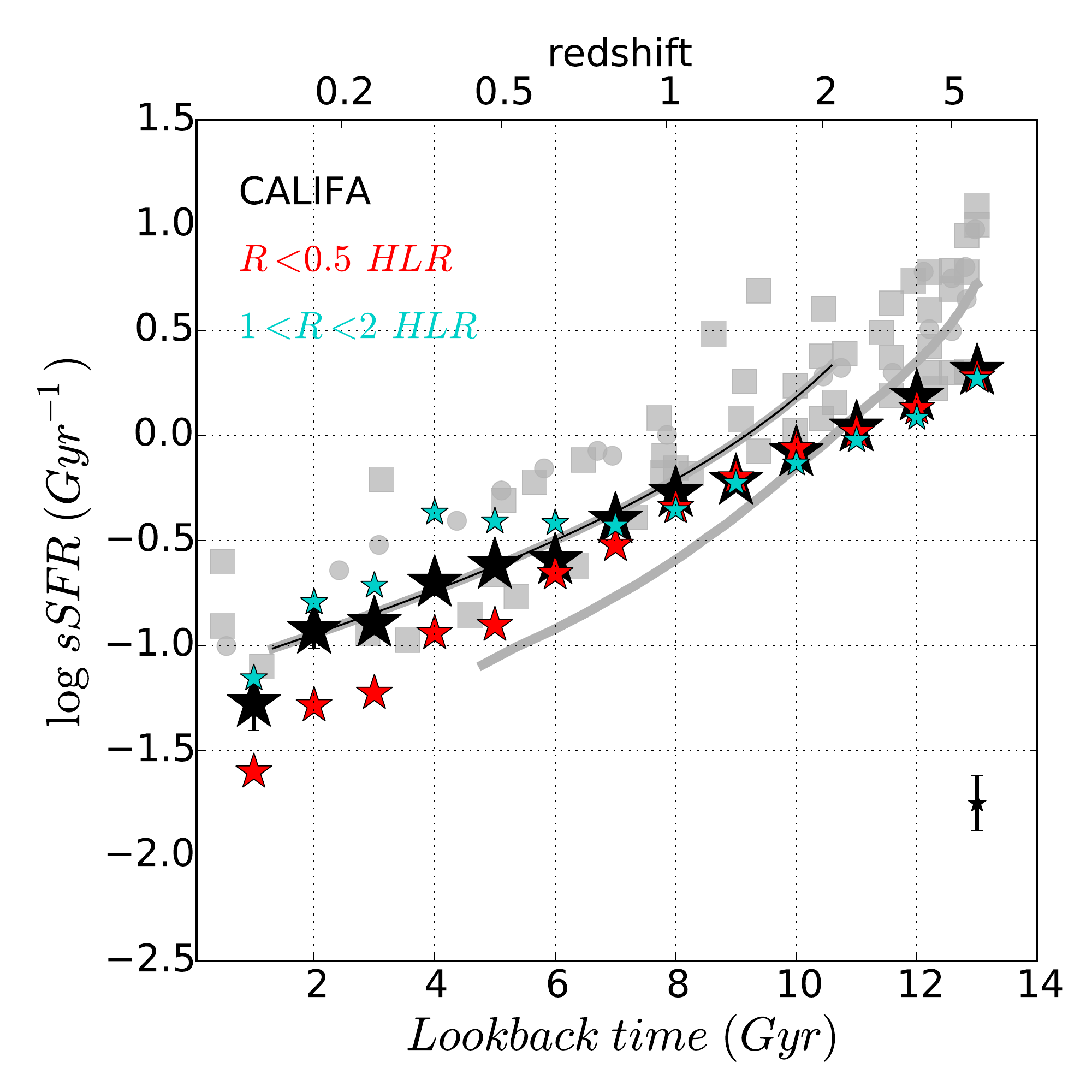}
\caption{  
The cosmic evolution of the sSFR($t$) in the present study (black stars). The grey line is from \citet{madau14}, the dark-grey line is the $(1+z)^3$ relation for $z<2$ from  \citet{elbaz07}, and grey symbols are from a compilation by  \citet{lehnert15}.  Red and blue stars show the sSFR($t$) corresponding to regions at the present epoch located at $R < $ 0.5 HLR and $1 < R< 2$  HLR. The  bars are the error in the mean, computed as the r.m.s.  of the sSFR($t$) values at each epoch divided by the square root of the number of galaxies in each bin
}
\label{fig:sSFR_all}
\end{figure}
%***FIG***FIG***FIG***FIG***FIG***FIG***FIG***FIG***FIG***FIG***

In Sec.\ \ref{sec:sSFR} and Fig.\ \ref{fig:sSFR} we have presented our results on the evolution of the sSFR($t$) as a function of  Hubble type.  Here, we discuss the average sSFR at each cosmic epoch,  obtained weighting each galaxy by $w_i = V_{max,i}^{-1} / \sum_j V_{max,j}^{-1}$,  where the sum runs over all galaxies. This volume weighted cosmic $\langle {\rm sSFR}(t) \rangle$ is shown as  black stars in 
Fig.\ \ref{fig:sSFR_all}.  As for $\rho_{\rm SFR}$, the calculation is done in  1 Gyr intervals. In Table \ref{tab:tablecosmic} we present the results for $z =  0$, 1, 2, and 5.

As expected,  $\langle {\rm sSFR}(t) \rangle$ decreases with cosmic time.  
 This result is in agreement with galaxy surveys at different redshifts. To illustrate this, Fig.\ \ref{fig:sSFR_all} includes the evolution of sSFR($t$) obtained by \citet{madau14} (grey line), after scaling to a Chabrier IMF. It also includes results (grey dots and squares) from the compilation done by \citet{lehnert15} in their figure 2, based on measurements by \citet{elbaz07, daddi07, dunne09, rodighiero10, oliver10, elbaz11} in galaxy surveys at $z \leq2$, and by \citet{feulner05, stark09, magdis10, stark13, ilbert13} at $z \geq2$. Fig.\ \ref{fig:sSFR_all} also shows (dark-grey line) the best-fit relation, sSFR$(t) = 26 \ t^{2.2}$, from \citet{elbaz11} over the  range $0\leq z\leq 2$,  equivalent to sSFR$(z) = (1+z)^3/t_{H_0}$, where $t_{H_0}$ is the Hubble time at $z= 0$. At $z > 2$, the cosmological galaxy surveys show that there is a plateau at 2 Gyr$^{-1}$ that is in tension with the current galaxy-formation models \citep{weinmann11}. Several solutions have been proposed to bring into agreement models and observations. For example, to explain the sSFR plateau at $z > 2$,  \citet{lehnert14} have argued that at these high redshifts the star formation must be self-regulated by high pressures, generated by the intense star formation itself, and it is the increase of the angular momentum with cosmic time what causes a decrease in the surface density of the accreted gas, and a decrease of sSFR as the Universe evolves.

Our results are in  good agreement with  cosmological surveys, and they follow well the time evolution curve for $z <1$  proposed by \citet{elbaz11}. Although our value at $z= 0$ is $\sim 0.26$ dex below the \citet{elbaz11} curve,  it is compatible with the lower envelope of individual values from the \citet{lehnert14} compilation, and above the \citet{madau14} curve.
At $z=1$, we derive $\langle {\rm sSFR}(t) \rangle =  0.55$ Gyr$^{-1}$,  just 10\% smaller than in \citet{elbaz11}. 
 At $1<z <2$ our values are below those from the fit estimated by \citet{elbaz11}. But at $z >2$ our $\langle {\rm sSFR}(t) \rangle$ is compatible with the lower envelope of the observed galaxy surveys at $z > 2$; e.g. $\langle {\rm sSFR}(t) \rangle = 0.74$, and 1.44, for $z = 2$ and 5, respectively.

\subsection{The evolution of the cosmic stellar mass density}
\label{sec:rho-mass}

%***FIG***FIG***FIG***FIG***FIG***FIG***FIG***FIG***FIG***FIG***
%\begin{figure*}[!ht]
\begin{figure}
\includegraphics[width=0.5\textwidth]{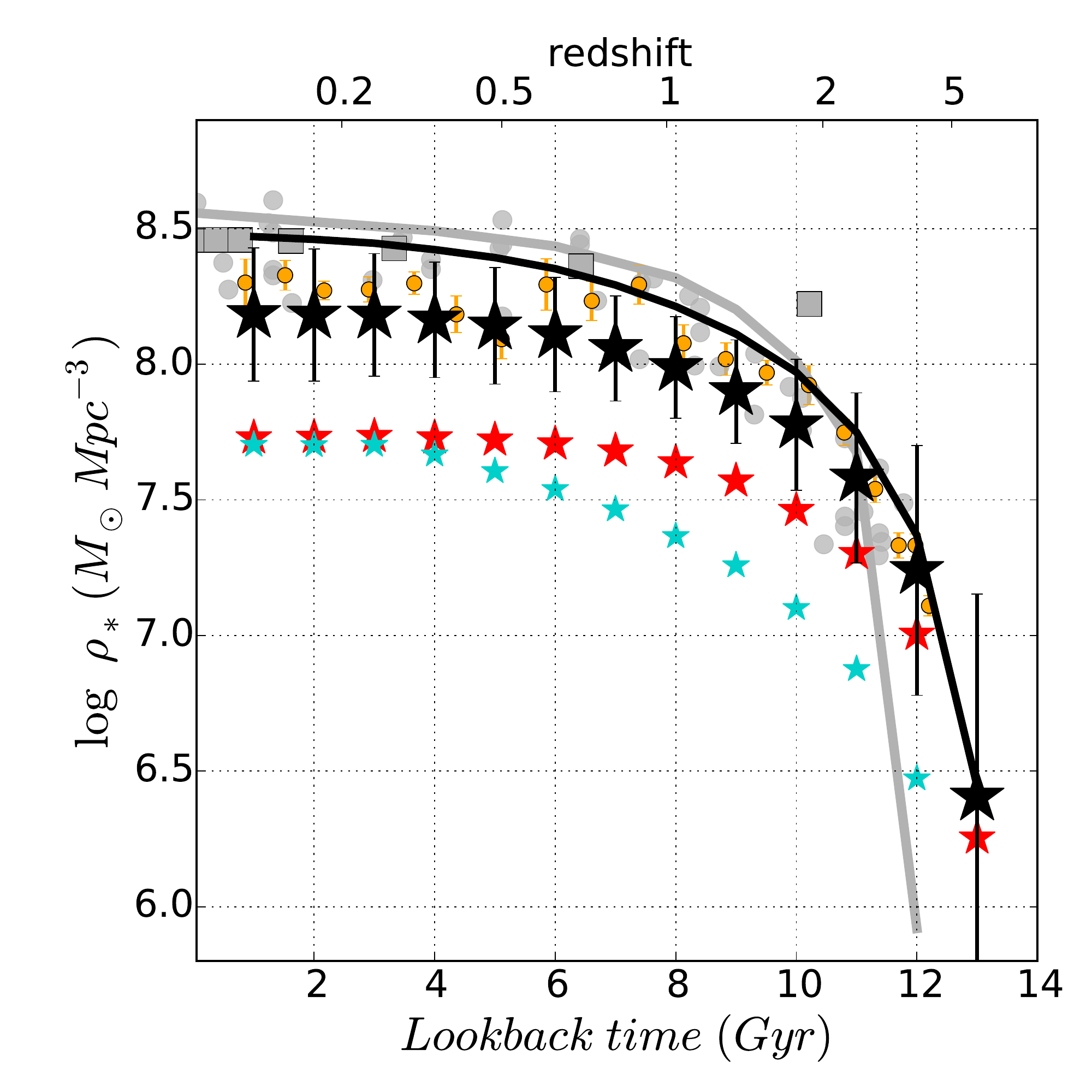}

\caption{  
 The cosmic evolution of the stellar mass volume density, $\rho_\star$ (black stars), obtained with $M_\star$(t) (corrected by mass-loss); $\rho_\star^\prime$ (black line) obtained with $M_\star^\prime$(t) (mass formed). Orange points are from the GAMA survey by \citet{driver17}. The grey line is from \citet{madau14}, grey squares are from \citet{panter07}, and grey dots are from different galaxy surveys as explained in the text.  Blue and red stars represent  the contribution to $\rho_\star$ from the galaxy outer regions, $1\leq R\leq2$ HLR, and inner regions, $R\leq 0.5$ HLR, respectively.
}
\label{fig:Massdensitycosmic}
\end{figure}
%***FIG***FIG***FIG***FIG***FIG***FIG***FIG***FIG***FIG***FIG***

The stellar mass density, $\rho_\star$, is also a relevant observational indicator of the physical processes that regulate  the mass assembly in galaxies across  cosmic time. Surveys, and in particular those done at near-infrared wavelengths,  have provided important information about the stellar mass function at redshifts $z > 2$ (e.g. \citealt{dickinson03, fontana03, rudnick03}). Others that combine observations at different wavelengths were able to study the time evolution of the mass function at high and intermediate redshifts (e.g. \citealt{muzzin13, ilbert13, moustakas03}). One important conclusion from these studies was that the mass function has evolved very little since $z = 1$ (e.g. \citealt{pozzetti10}). 

At low-redshift, the SDSS survey  also provided relevant information about the stellar mass function (e.g. \citealt{baldry08, baldry12}). It has been very useful to construct a distribution of the stellar mass as a function of  age  \citep{gallazzi08} and to derive $\rho_\star$ at $z = 0$, as well as to derive the cosmic evolution of $\rho_\star (t)$ by using the fossil record of the stellar population  \citep{panter07}.

In our analysis, as explained in Section \ref{sec:sSFR}, the mass locked in stars up to a lookback time $t$ is easily obtained from the SFR$(t)$ fits. The volume density $\rho_\star$ is then obtained from these $M_\star(t)$ functions just as 
 SFR$(t)$ was used to derive $\rho_{\rm SFR}$ in Section \ref{sec:EvolutionsSFR}. 
Fig.\ \ref{fig:Massdensitycosmic} places our results in the $\rho_\star$ vs. lookback time  (or redshift) diagram, where the black stars and black line are the cosmic stellar mass density with and without correction for mass loss, respectively. 
Errors are obtained by propagating the dispersion of $t_0$ and $\tau$ of each galaxy.
In agreement with galaxy redshift surveys, $\rho_\star$ shows little evolution since $z=1$, but a fast increase at higher $z$. As a reference, the $50\%$ point is reached at $\sim9$ Gyr lookback time.

Figure \ref{fig:Massdensitycosmic} also includes (grey line) the evolution of $\rho_\star$ from \citet{madau14}, after scaling to a Chabrier IMF. It also includes (grey dots) $\rho_\star$ from the compilation of \citet{madau14} (their figure 11 and Table 2) based on measurements  at different redshifts by \citet{gallazzi08, li09, moustakas13} for $z < 1$, \citet{arnouts07, perezgonzalez08, kajisawa09, marchesini09, pozzetti10, reddy09, ilbert13, muzzin13} for  $0.1 <  z < 4$, \citet{caputi11, yabe09, gonzalez11, lee12, labbe13} for $3 \leq z \leq 5$. The results obtained by \citet{panter03} from the fossil record method applied to  SDSS data are included as grey squares. 
Orange points are the recent results from GAMA/G10-COSMOS/3D-HST by  \citet{driver17}.
When necessary, the literature results are scaled to a Chabrier IMF. We note that $\rho_\star$ by \citet{madau14} is on average $\sim$ 0.2 dex higher than most of the data reported in the literature for $0 < z < 3$.

Comparing our results to galaxy redshift  surveys,  we found: a) at the highest redshifts (13 and 12 Gyr ago), $\rho_\star$ is higher than in  \citet{madau14}; b) but at $z \leq 1$,  our $\rho_\star$ is below the \citet{madau14} curve and in agreement with the lower envelope of the grey dots. Our $\rho_\star$ is in very good agreement with the recent results from GAMA/G10-COSMOS/3D-HST multi-wavelength study  \citep{driver17}. In particular, for $z > 2$, our results follow better the  \citet{driver17} points than the \citet{madau14} results, because they do not show the sudden rapid increase of the \citet{madau14} curve. Furthermore, our values for lookback time at 12 and 13 Gyr (1.7$\times$10$^7$ and 2.5$\times$10$^6$ $M_\odot$ Mpc$^{-3}$, respectively) are also in good agreement with the results  reported by  \citet{duncan14}, \citet{grazian15}, and  \citet{song16}, which range from 2$\times$10$^7$ to 3$\times$10$^6$ $M_\odot$ Mpc$^{-3}$.

The volume density $\rho_\star^\prime$ from $M_\star^\prime(t)$ (mass formed in stars) has a  
 similar behavior to $\rho_\star$. The difference between $\rho_\star$  and $\rho_\star^\prime$ measures the stellar mass formed in galaxies that is returned to the interstellar medium due to stellar evolution. At  expected,  $\rho_\star^\prime > \rho_\star$ by 0.22 dex ($z= 1$) to 0.29 dex ($z=0$); this is because  for most galaxies in the sample $t_0 > 10$ Gyr,  and according with SSP models by  \citet{bruzualcharlot03}, the stellar population needs only $\sim$4 Gyr  to lose $\sim$0.45 of its original mass, for a Chabrier IMF. 
 Reassuringly, \citet{driver17} compute $\rho_\star^\prime$  by integrating $\rho_{\rm SFR}$, and comparing with $\rho_\star$ they obtain a returned mass fraction of $0.50 \pm 0.07$.

%-------------------------------------------------------------------------------------------------------------------------------------------

\subsection{SFH of galaxies: fossil cosmology vs redshift galaxy surveys}

So far we have shown the capability of the delayed-$\tau$ model and nearby galaxies to trace the cosmic evolution of SFR and sSFR of the Universe. This SFH has been proposed to be an accurate representation of SFH of galaxies on the main sequence \citep{speagle14}. \citet{madau14} have proposed a consistent picture in which the star formation rate density peaks at 3.5 Gyr after the Big Bang and then  declines exponentially with an e-folding time of 3.9 Gyr.  Similarly to \citet{speagle14}, \citet{madau14}  derived their model from the fit to the evolution of $\rho_{\rm SFR}$ that was obtained from compilations of SFR using different SFR indicators and galaxy surveys, from the nearby Universe to high redshift ($z \sim 5$). 

In our case, we have derived SFR$(t)$ by assuming that the SFH of nearby galaxies is well represented by a delayed-$\tau$ SFH. This model is able to provide estimations of cosmic star formation rate density and  evolution of the sSFR that are compatible with  $\rho_{\rm SFR}$ and sSFR derived from galaxy surveys.  However, other parametric and non-parametric SFHs are able to fit equally well  the observational constrains (see Appendix) and provide, to a first approximation, good estimates of $\rho_{\rm SFR}$ and sSFR \citep[e.g][]{gonzalezdelgado16, gonzalezdelgado17}. Here, we discuss the similarities and differences between the different models by comparing the mass fractions, $m(t)$,  obtained with different parametric and non-parametric SFH applied to the CALIFA sample and with the mass fraction derived by \citet{madau14}.

For each galaxy, $m(t)$ is derived by dividing the stellar mass formed in each epoch by the total mass formed up to that time. Previously in \citet{gonzalezdelgado17}, using a non-parametric method to derive the SFHs of CALIFA galaxies, we obtained that the highest mass fractions invariably occur at the earliest times. Subsequent star formation varies systematically with $M_\star$, with the low $M_\star$ galaxies forming stars over extended periods of time, and high $M_\star$ galaxies exhibiting the fastest decline in $m(t)$. The behavior with morphology mimics the behavior with $M_\star$; for all morphologies,  $m(t)$ peaks at the earliest epoch and subsequent star formation increases systematically. Here,  with the parametric delayed-$\tau$ SFH, we confirm that independently of  morphology, $m(t)$ has a maximum at $z\geq 2$, $\sim10$ Gyr ago, with a small shift of $\lesssim -0.1$ dex for Sc and Sd galaxies. 

\begin{figure}
\includegraphics[width=0.5\textwidth]{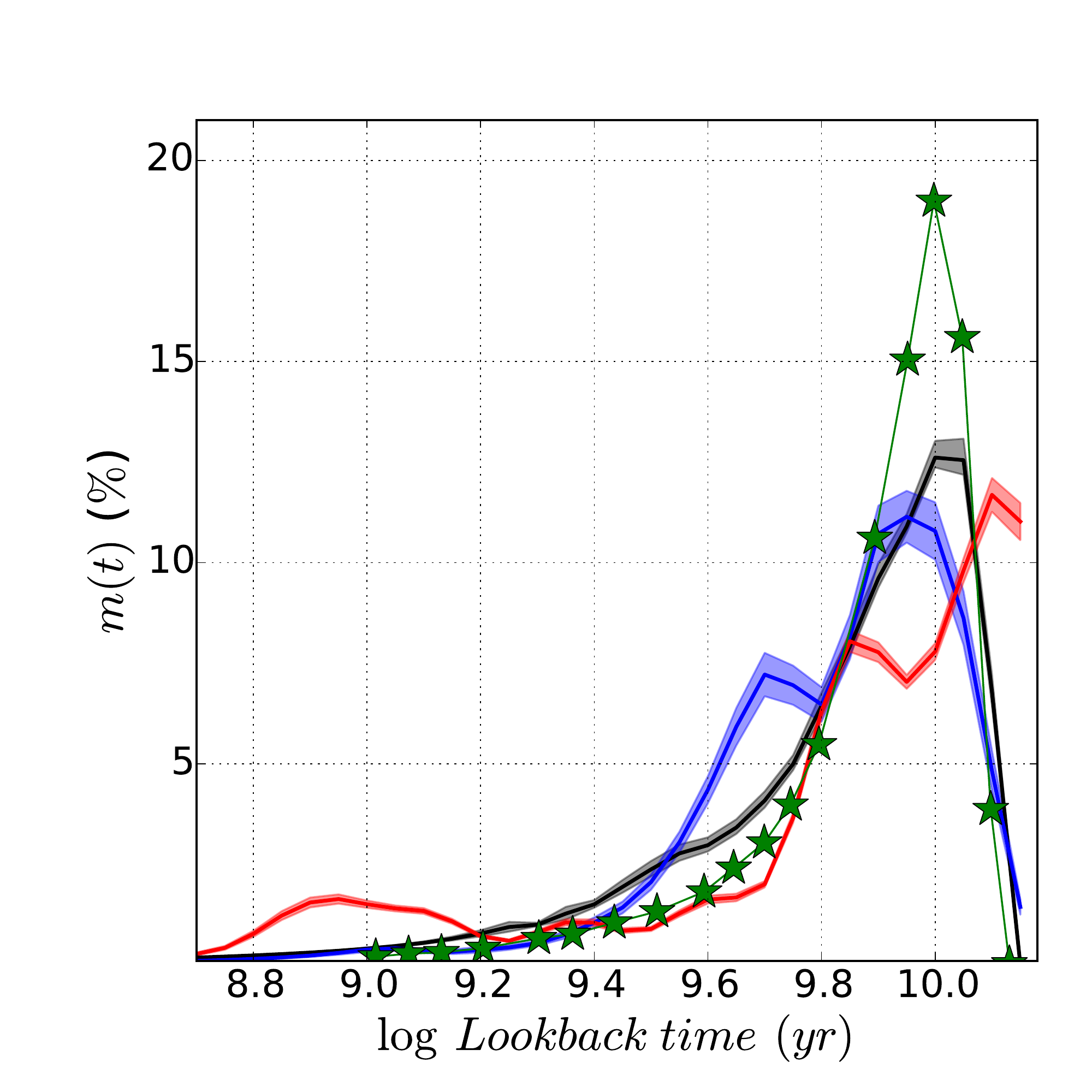}
\caption{ Evolution of the mass fraction, $m(t)$, obtained with parametric SFHs: delayed-$\tau$ (black), a combination of two exponential SFR (blue), and \starlight\ non-parametric (red), are compared with $m(t)$ from \citet{madau14} (green stars). The shaded bands around the mean curves represent $\pm$ the error in the mean.
}
\label{fig:massfraction_mean}
\end{figure}

In order to obtain a global representation of the SFH of all the galaxies and compare with the results of \citet{madau14}, the average $m(t)$ is obtained by weighting the mass fraction from each galaxy with $w_i = V_{max,i}^{-1} / \sum_j V_{max,j}^{-1}$. Fig.\ \ref{fig:massfraction_mean} shows a comparison of our results to those of \citet{madau14}  (the histogram on their figure 11, and the exponential law that they proposed). From our CALIFA sample, $m(t)$ is obtained for: a) delayed-$\tau$ model; (b) a combination of two exponential SFR;  and (c) using the non-parametric SFH derived from the \starlight\ code. In addition to the description of the method used to obtain (b) and (c) models, a comparison of other properties related to the mass assembly in galaxies is also given in the Appendix.

The result from the delayed-$\tau$ model is quite similar to the exponential model of \citet{madau14}, with the highest fraction peaking $\sim 3.5$ Gyr after the Big Bang, followed by a nearly exponential decline.  Notice that there are slightly larger fractions at lookback times between 2 and 7 Gyr ago. This behavior is somewhat different to that obtained using two exponential SFR laws; while the maximum of $m(t)$ occurred early on, at $\sim 8$ Gyr ago ($z \sim 1$), there is a secondary epoch, $\sim 5$ Gyr ago, where the fraction of mass assembled is relevant before decaying to the actual values, this is not seen by \citet{madau14}.  For the \starlight\ non-parametric results, $m(t)$ has two peaks 11 and 7 Gyr ago, which  cover the same period of time than the peak of the delayed-$\tau$ model; then, $m(t)$ declines in a similar way as the \citet{madau14} model. However, $m(t)$ derived from \starlight\ shows other peak at very recent times, $\sim 1$ Gyr,  connected to the rejuvenation epoch that late spirals experienced in the last 2 Gyr. 
 
In \citet{gonzalezdelgado17} we suggest that galaxies can grow in two different modes. For the early evolution of Sa-Sbc spirals and the entire evolution of E and S0, the logarithm of their mass fraction, $\log m(t)$, declines linearly with $\log t$, i.e.,
as an exponential mode in $m(t)$ vs. $\log t$. 
For the late types (Sc-Sd) and low-mass galaxies, $\log m(t)$ vs. $\log t$ is almost constant.
We suggested that the first mode could represent the transition between the formation of a thick and a thin disk. The thick disk is a self-regulated mode, where strong outflows and turbulence drive the high intensity of the star formation rate that occurs very early on in the evolution \citep{lehnert15}. The second mode can be associated to the formation of the thin disk that is regulated by secular processes; a phase driven by self-gravity and energy injection from the stellar population is not relevant for global regulation \citep{lehnert15}.

It is interesting to notice that these two modes are again reproduced here by  \starlight\ using different constrains and SSPs models. However,  while the exponential mode is well derived by the parametric SFH (delayed-$\tau$ model, or the two exponential SFRs) and the exponential mode by \citet{madau14}, the second mode for galaxy growth (not so relevant in terms of the total mass assembled for most of the galaxies) is not  reproduced by the delayed-$\tau$ model. This second mode of star formation is relevant in thin disks, the main contributors to $\rho_{\rm SFR}$ and sSFR at $z=0$ \citep{gonzalezdelgado16}. This could explain why our delayed-$\tau$ model underestimates  the  sSFR at lookbacktime $\lesssim$ 1 Gyr.

%***FIG***FIG***FIG***FIG***FIG***FIG***FIG***FIG***FIG***FIG***
%\begin{figure*}[!ht]
%***FIG***FIG***FIG***FIG***FIG***FIG***FIG***FIG***FIG***FIG***

%-----------------------------------------------------------------------------------------------------------------------------------------------------

\section{Discussion: The cosmic evolution of the spatially-resolved SFR and stellar mass}
\label{sec:Discussion}

Reassured by the satisfactory agreement between our estimates of $\rho_{\rm SFR}(t)$, sSFR$(t)$ and $\rho_\star(t)$ and those obtained in galaxy surveys, we now go a step further and explore the spatially resolved and morphological information for our data set.  The main goals of this section are to discuss: 1) The evolution of $\rho_{\rm SFR}$, sSFR and $\rho_\star$ separately for the inner and outer regions of galaxies. 2) The role of Hubble types in the budget of $\rho_{\rm SFR}$, sSFR, and $\rho_\star$. 3) The evolution of the main sequence of star formation, and the comparison of the global relation with that obtained for  the inner regions of galaxies.

\subsection{The spatially-resolved evolution of $\rho_{\rm SFR}$, sSFR, and $\rho_\star$}

Our data allow us to discuss the evolution of $\rho_{\rm SFR}$, sSFR, and $\rho_\star$ separately for the currently inner and outer regions of galaxies. This information is interesting because spatially-resolved spectroscopic observations of high redshift galaxies are difficult to obtain, and this analysis is an attempt to infer the contributions of different galaxy sub-components, bulge and disk, to the total budget of $\rho_{\rm SFR}$, sSFR, and $\rho_\star$. 

To evaluate the contribution of the different galaxy regions to  $\rho_{\rm SFR}$, we compute the contribution of  zones  located at $R <  0.5$ HLR (``inner regions'', mainly dominated by the current bulge component), and $1 < R <  2$ HLR (``outer regions'',  dominated by the disks).  These contributions are plotted in Figs.\ \ref{fig:cosmicSFR},   \ref{fig:sSFR_all}, and  \ref{fig:Massdensitycosmic} as red and blue points, respectively, and they are listed in Table \ \ref{tab:tablecosmic} for $z=$ 0, 1, 2, and 5. The main results are:

\begin{itemize}

\item $\rho_{\rm SFR}$ (Fig.\ \ref{fig:cosmicSFR}): 
In the local Universe ($z\sim 0$)  the ``disk" regions dominate $\rho_{\rm SFR}$ \citep{gonzalezdelgado16}, while at higher redshift the main contributors are the  inner regions. At $z=0$, $27\%$ of the $\rho_{\rm SFR}$  comes from the inner regions ($R <  0.5$ HLR), while outer ones  ($1 < R <  2$ HLR) contribute with $44\%$. With redshift, the central contribution increases its relevance to the total SFR density, competing with the disk dominated  regions at $z\sim1$, where $\rho_{\rm SFR} (R <  0.5$ HLR) = $\rho_{\rm SFR} (1 < R <  2$ HLR), and dominating at higher $z$.  These central  regions contribute  $51\%$ of $\rho_{\rm SFR}$ at $z\sim5$.

\item sSFR$(t)$ (Fig.\ \ref{fig:sSFR_all}):
At  $z < 1$, the outer galaxy regions have higher sSFR$(t)$ than inner regions. This can be simply understood as a consequence of regions located in the disks being the major contributors to $\rho_{\rm SFR}$ at these redshifts. At higher $z$, regions located today in the inner 0.5 HLR (presumably associated to the present day bulge) and outwards of 1 HLR (belonging today to galaxy disks) have similar sSFR$(t)$. These results suggests that all the regions are equally efficient in forming stars and growing their mass at high redshift.

\item $\rho_\star$ (Fig.\ \ref{fig:Massdensitycosmic}):
In the local Universe ($z < 0.2$), the outer and inner regions contribute to $\rho_\star$ in very similar amounts. This agrees with our previous finding that the ratio of half mass to half light radii is close to unity (HMR/HLR $\sim 0.8$ on average; \citet{gonzalezdelgado14a}). At higher redshifts the contribution from inner galaxy regions increases with respect  to that from the outer zones by a factor 2.3 and 3.5 at $z = 2$ and 5 (Table \ \ref{tab:tablecosmic}). Thus, the central regions of galaxies are the main place where  $\rho_\star$ was built.  
Comparing  the blue and red stars it is clear that  the central regions of galaxies have grown their mass more rapidly than the outer regions. Taking  as reference the 50\% point, for the central regions it was reached at lookback time 9 Gyr, and 6 Gyr ago for the outer regions. This conclusion is in agreement with our previous findings that galaxies  grow their mass inside out \citep{perez13, garciabenito17}.

\end{itemize}

\subsection{The role of morphology in the evolution of $\rho_{\rm SFR}$ and $\rho_\star$}

The Hubble sequence has evolved over time \citep[e.g]{delgado-serrano10}. In particular, early type galaxies (E, S0, and Sa) can be the end product of later type spirals transformed by mergers (e.g., \citealt{cappellari16} and references therein). Bearing in mind that our fossil record analysis cannot trace such morphological transformations, we now discuss the evolution of the SFR and stellar mass density as a function of the present day morphological type.

\begin{figure*}
\includegraphics[width=0.3\textwidth]{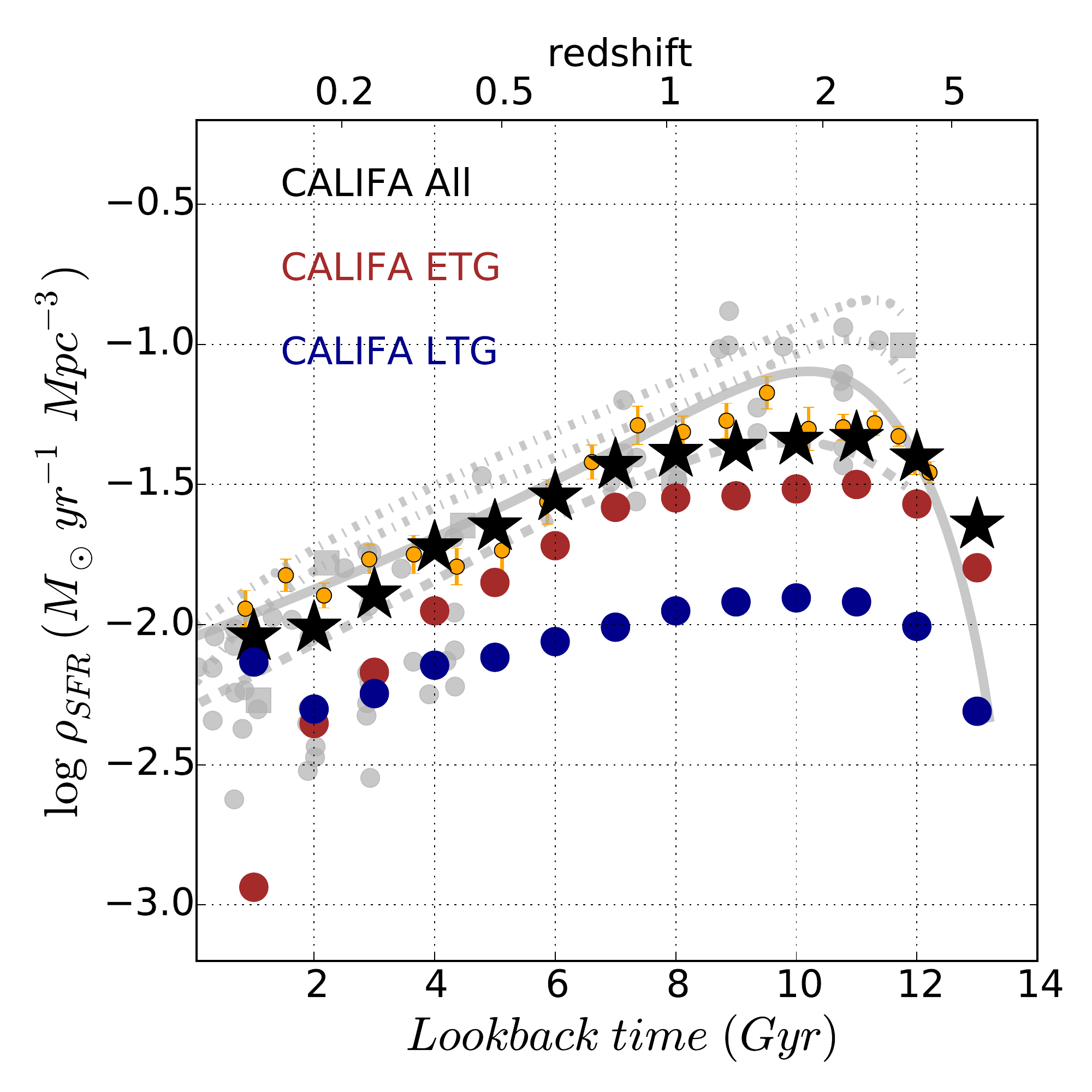}
\includegraphics[width=0.3\textwidth]{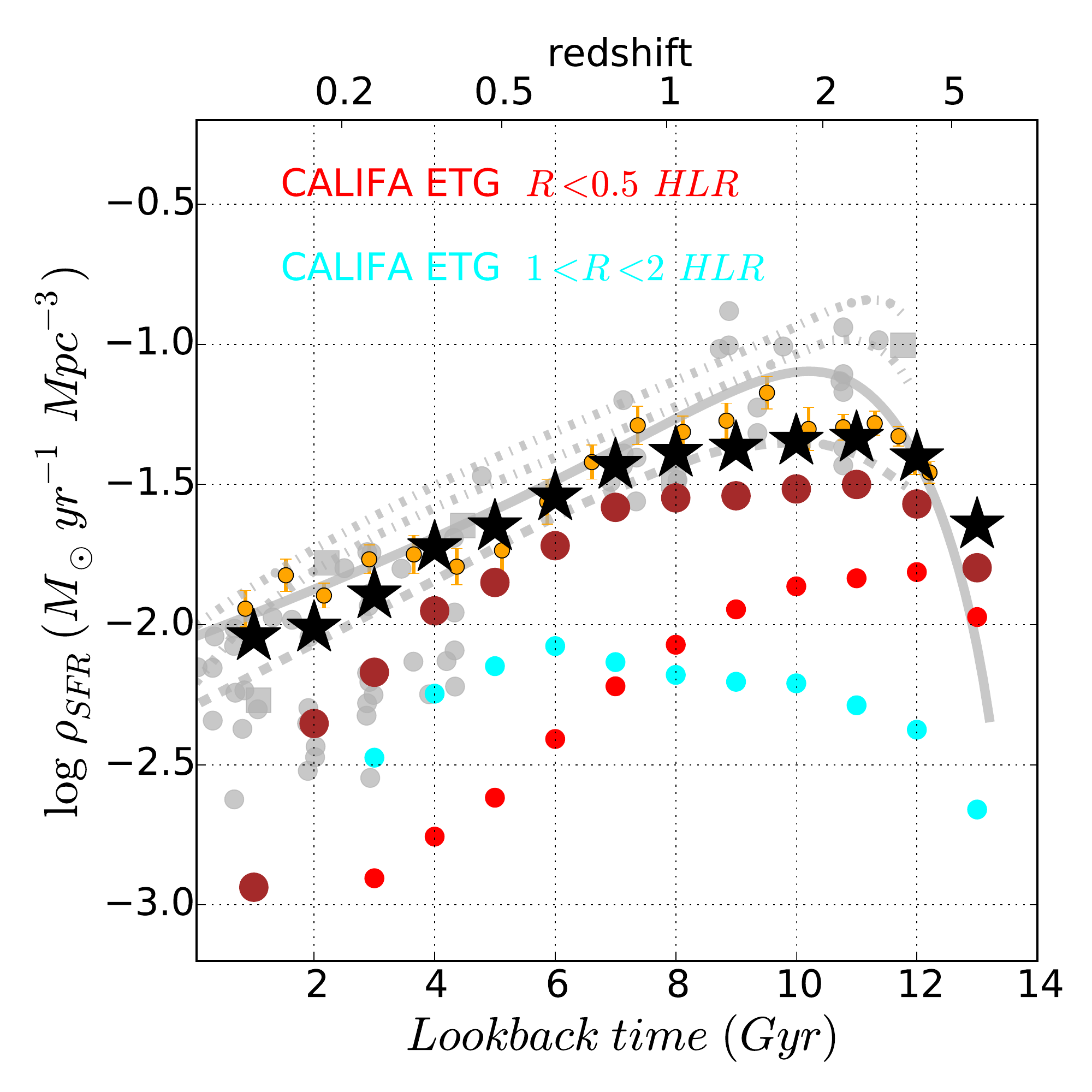}
\includegraphics[width=0.3\textwidth]{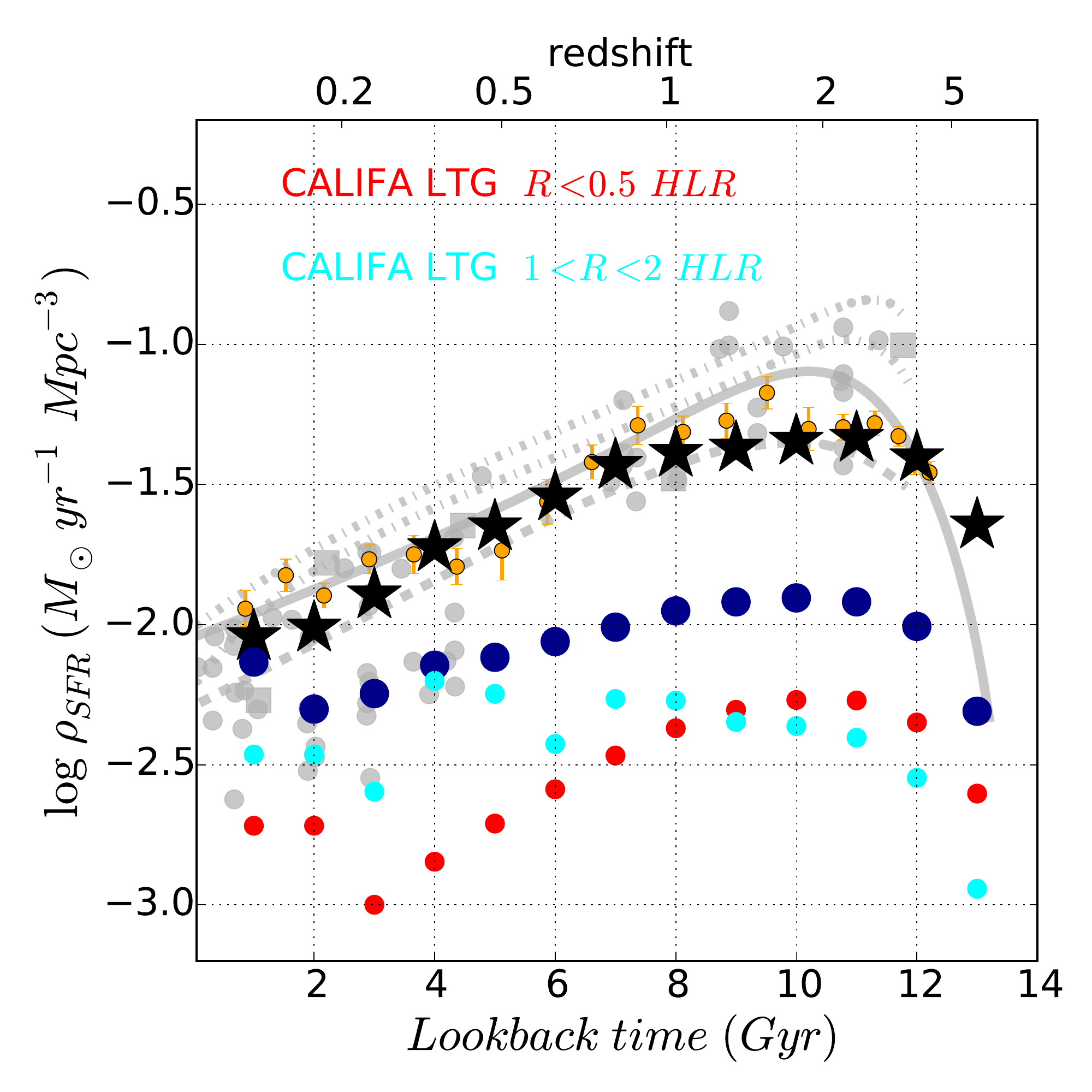} \\
\includegraphics[width=0.3\textwidth]{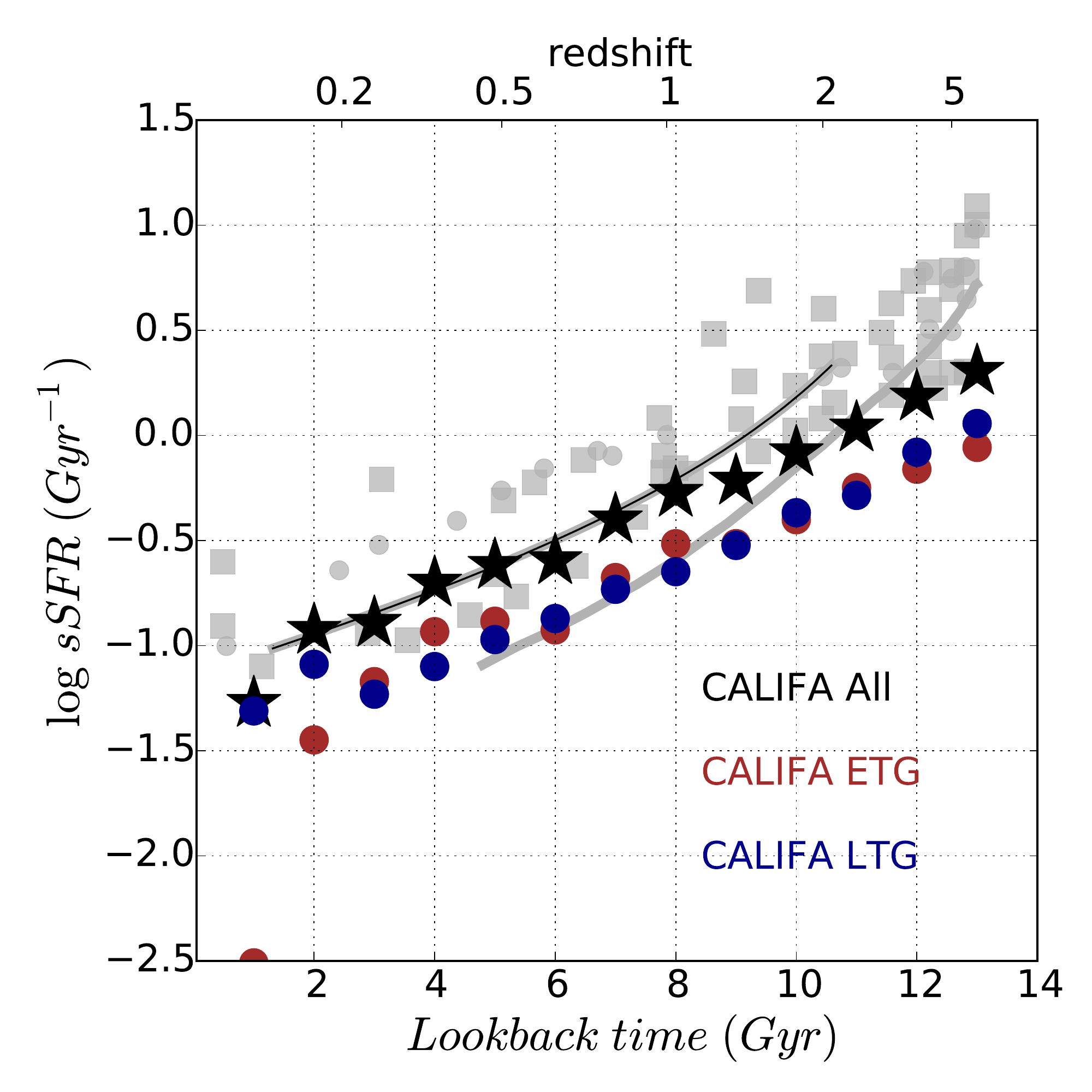}
\includegraphics[width=0.3\textwidth]{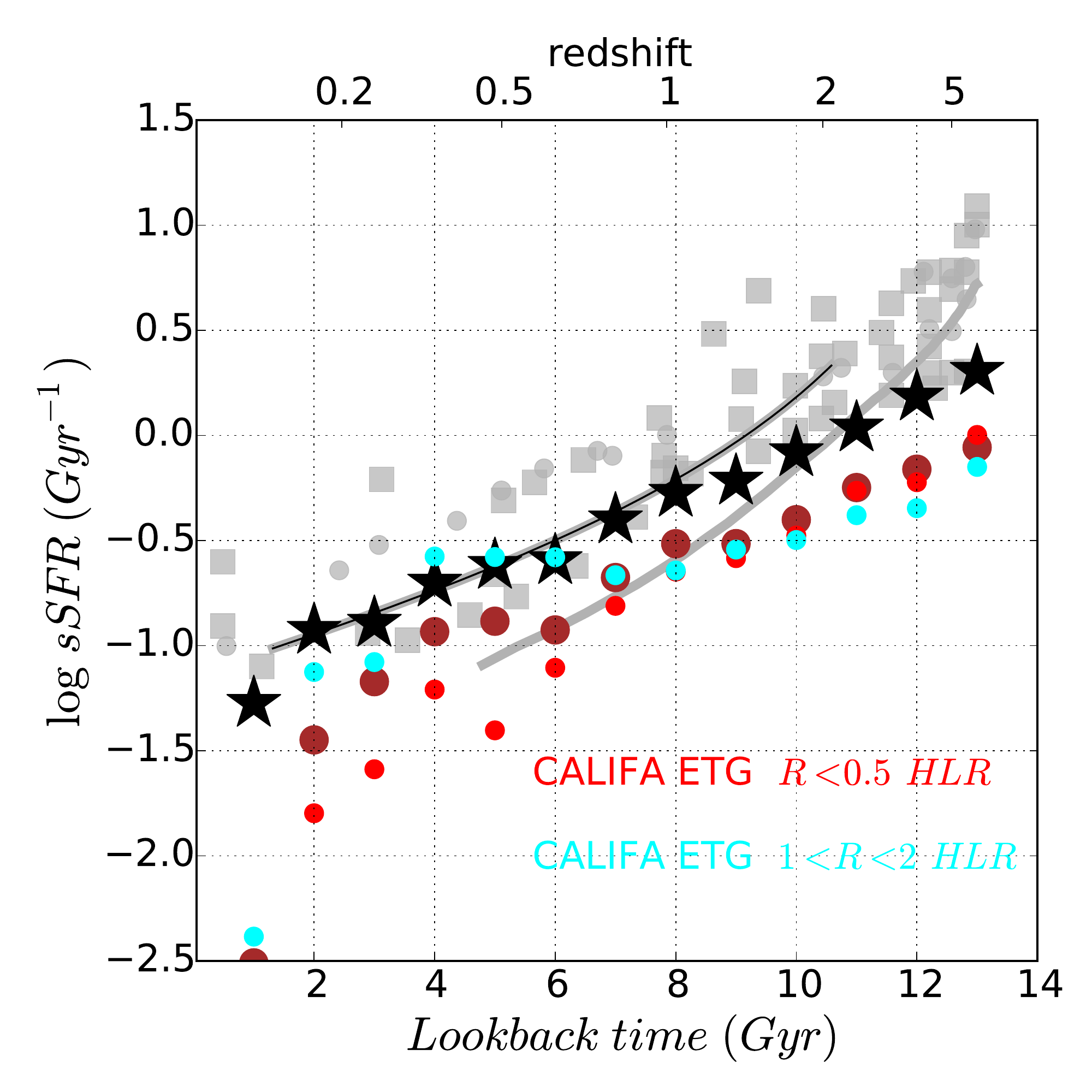}
\includegraphics[width=0.3\textwidth]{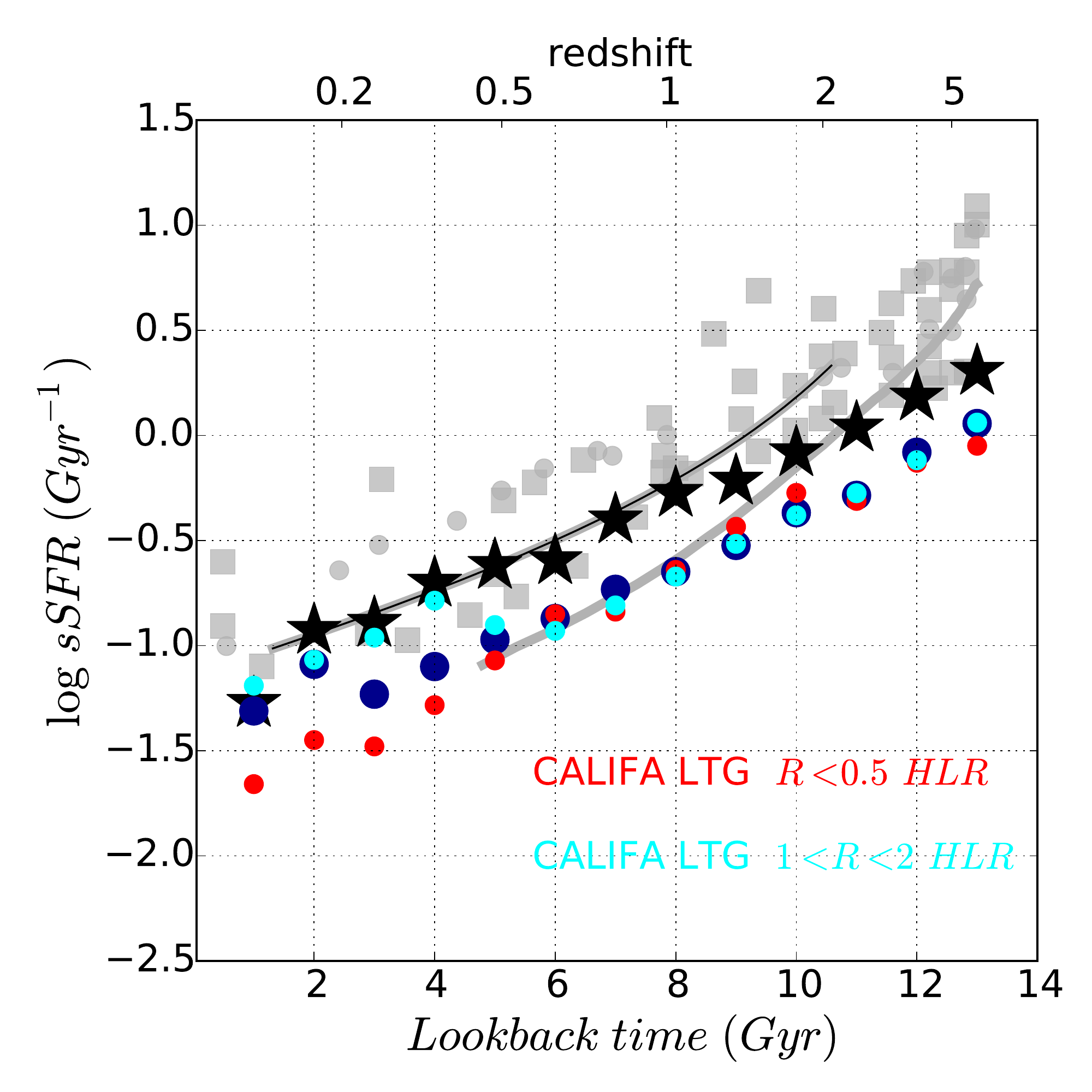} \\
\includegraphics[width=0.3\textwidth]{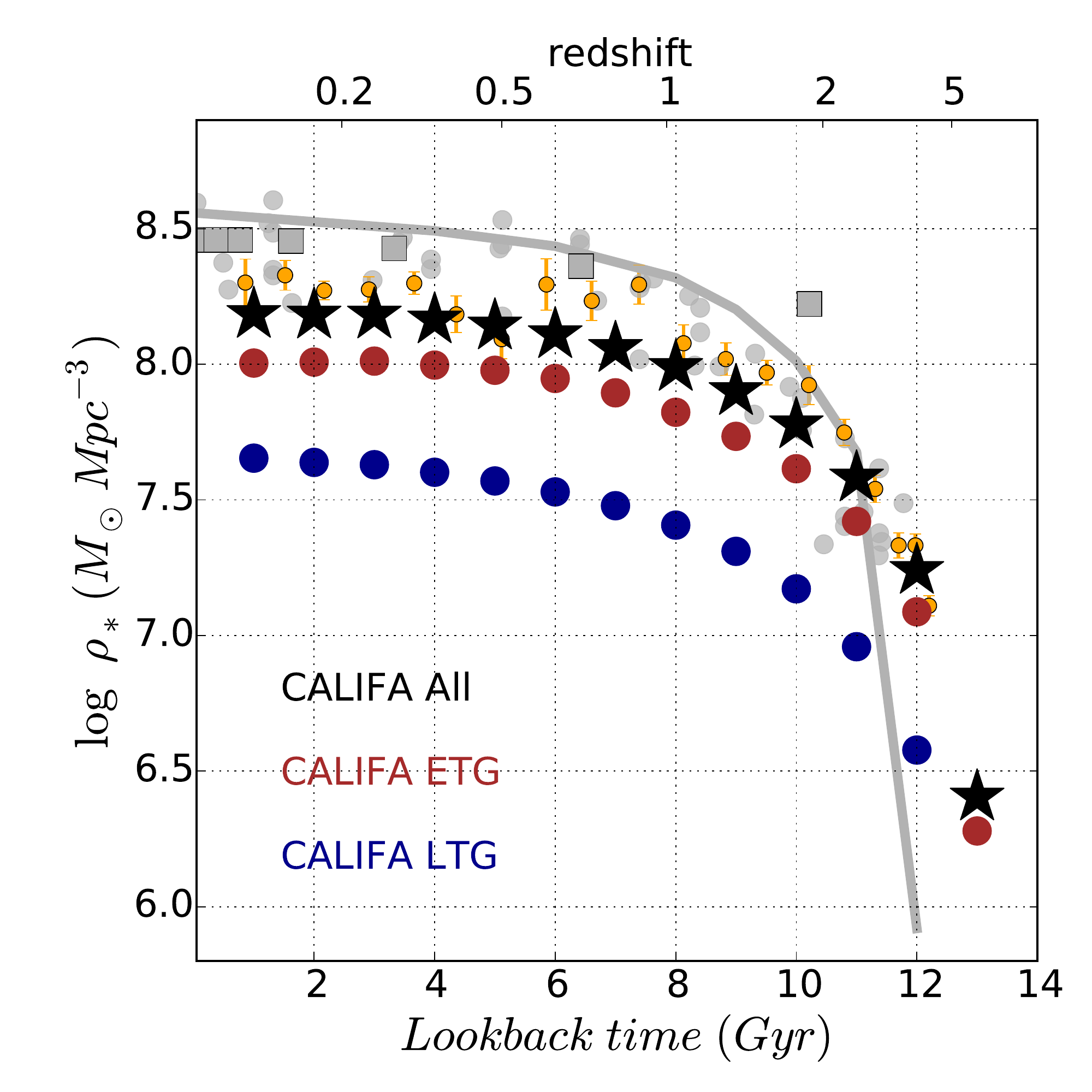}
\includegraphics[width=0.3\textwidth]{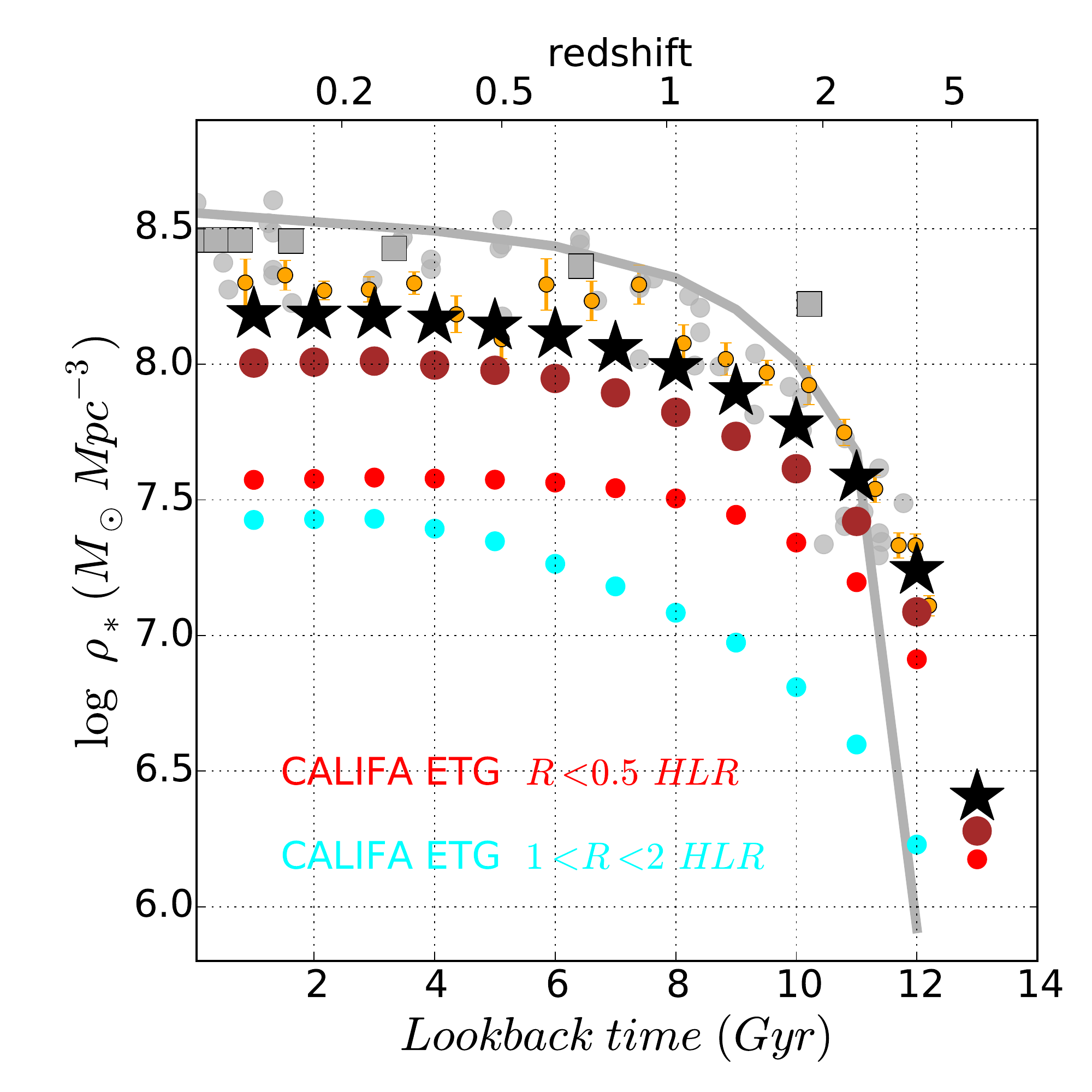}
\includegraphics[width=0.3\textwidth]{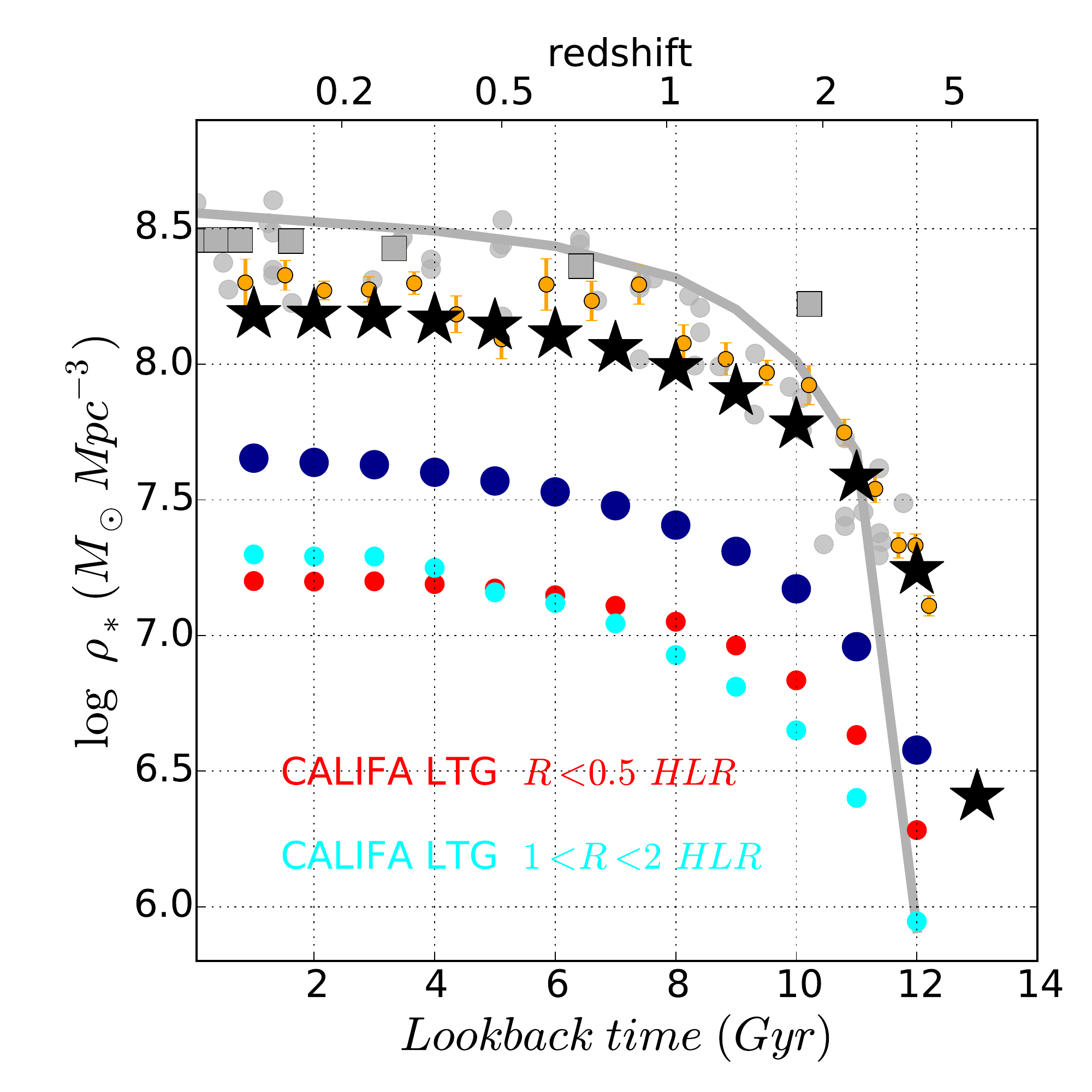}
\caption{ The cosmic evolution of the star formation rate density, $\rho_{\rm SFR}$ (upper panels), the sSFR(t) (middle panels), and the stellar mass volume density, $\rho_\star$ (lower panels), as a function of the morphology: early (E-S0-Sa) and late (Sb-Sbc-Sc-Sd) type galaxies. The contribution of inner ($R < 0.5$ HLR)  and outer ($1 < R < 2$ HLR)  regions of ETG and LTG are in the central and right panels, respectively.
}
\label{fig:rho_morphology}
\end{figure*}

Previously,  we have seen in \citet{gonzalezdelgado16} that the local Universe ($z=0$), Sbc, Sc, and Sd galaxies dominate the $\rho_{\rm SFR}$ budget. We found that galaxies of these morphologies together they contribute  $\sim 66\%$ of $\rho_{\rm SFR}$, while Sa and Sb galaxies contribute  $\sim 29\%$.
Here, we discuss the contribution of {\em currently} early (ETG; E, S0, and Sa), and late  (LTG, Sb, Sbc, Sc, and Sd) type galaxies to the evolution of $\rho_{\rm SFR}$ and $\rho_\star$. Table  \ref{tab:tableETG} and Table \ref{tab:tableLTG} list their contributions, and Fig.\ \ref{fig:rho_morphology} shows the results. A summary of the results are:

\begin{itemize}

\item $\rho_{\rm SFR}$: 
Present day  ETG  are  the main contributors to $\rho_{\rm SFR}$, except in the local universe (Fig.\ \ref{fig:rho_morphology} top-left panel). At high redshift, the progenitors of ETG dominate the SFR budget with $\sim 69\%$ of $\rho_{\rm SFR}$, while the progenitors of late type spirals contribute with $\sim 26\%$. However, at $z=0$, 
LTG contribute with 81$\%$ of $\rho_{\rm SFR}$, while current ETG contribute with less than 12$\%$ to the total $\rho_{\rm SFR}$.
The inner regions of the progenitors of present ETG contribute very little to the SFR density at the local universe ($\sim 3 \%$), but their contributions increase with redshift, with $21\%$ at $z=1$ to $40 \%$ at $z=4$. However, the inner regions of LTG contribute little to $\rho_{\rm SFR}$, being $\sim 11\%$ at any epoch, except at recent time, that contributes up to $21 \%$ of the total $\rho_{\rm SFR} (z=0)$. Furthermore, the outer regions of LTG are the major contributors to the SFR density at $z=0$, at any other epoch,  the outer regions of LTG or ETG contribute with $\leq 16\%$ of the total budget $\rho_{\rm SFR}$.

\item sSFR$(t)$: 
Except at low redshift, the sSFR evolves similarly for ETG and LTG. At $z=0$, LTG have larger sSFR$(t)$  than ETG. This is a consequence of their larger contribution to $\rho_{\rm SFR}$. At higher redshifts the inner regions of ETG and LTG evolve in similar ways, and the same happens for the outer regions of ETG and LTG.

\item $\rho_\star$: 
At any epoch, the progenitors of present day ETG are the dominant population in terms of stellar mass (Fig.\ \ref{fig:rho_morphology} bottom-left panel). They contribute to the total budget of $\rho_\star$ by 69$\%$ at $z=5$ to 66$\%$ at $z=0$; but LTG contribute with $< \sim 30\%$ of the stellar mass density of the universe. The inner regions of ETG and their progenitors are also the main contributor to $\rho_\star$,  going from $25\%$ at $z=0$ to $47\%$ at $z=5$. The inner regions of LTG, however, contribute very little to the stellar mass density of the universe, being $\sim 10\%$ of the total $\rho_\star$ at any epoch. The outer regions of ETG or LTG contribute with $\leq 17\%$ to the total budget of $\rho_\star$.

\end{itemize}

Thus, we can conclude that while in the local Universe the current SFR density is dominated by disks of LTG, at $z > 1$ the SFR density is dominated by the central components of present day early type galaxies. Moreover, the inner regions of ETG are the major contributors to the total stellar mass density.

%--------------------------------------------------------------------------------------------------------------------------------------------
%***FIG***FIG***FIG***FIG***FIG***FIG***FIG***FIG***FIG***FIG***
%\begin{figure*}[!ht]
%***FIG***FIG***FIG***FIG***FIG***FIG***FIG***FIG***FIG***FIG***

%-----------------------------------------------------------------------------------------------------------------------------------------------------------------
 %#####early type galaxies######
\begin{table}
\resizebox{0.5\textwidth}{!} {\begin{tabular}{c c c c c c} % centered columns (4 columns)
 \multicolumn{2}{c}{$\log\, \rho_{\rm SFR}$ [$M_{\odot}yr^{-1}Mpc^{-3}$]} & $z=0$ & $z=1$ & $z=2$ & $z=5$ \\
\hline %inserts double horizontal lines
\textbf{M1}&
 \begin{tabular}{c}
 $R<0.5$\\
 ETG\\
 $1< R < 2$\
 \end{tabular}&  
 \begin{tabular}{r}
$-3.51 \pm 0.09$\\
$-2.94 \pm 0.15$\\
$-3.46 \pm 0.25$\\
 \end{tabular}& 
 \begin{tabular}{r}
$-2.07 \pm 0.10$\\
$-1.55 \pm 0.16$\\
$-2.18 \pm 0.42$\\
 \end{tabular}& 
 \begin{tabular}{r}
$-1.86 \pm 0.09$\\
$-1.52 \pm 0.09$\\
$-2.21 \pm 0.19$\\
 \end{tabular}& 
 \begin{tabular}{r}
$-1.81 \pm 0.06$\\
$-1.57 \pm 0.15$\\
$-2.37 \pm 0.23$\\
\end{tabular}\\
 
 \hline

\multicolumn{2}{c}{$\log\, sSFR$ [$Gyr^{-1}$]}& &  & &  \\
\hline %inserts double horizontal lines
\textbf{M1}&
 \begin{tabular}{c}
 $R<0.5$\\
ETG\\
 $1< R < 2$\
 \end{tabular}&  
 \begin{tabular}{r}
$-2.77 \pm 0.01$\\
$-2.51 \pm 0.07$\\
$-2.38 \pm 0.02$\\
 \end{tabular}& 
 \begin{tabular}{r}
$-0.65 \pm 0.06$\\
$-0.55 \pm 0.12$\\
$-0.64 \pm 0.05$\\
 \end{tabular}& 
 \begin{tabular}{r}
$-0.48 \pm 0.04$\\
$-0.40 \pm 0.08$\\
$-0.49 \pm 0.06$\\
 \end{tabular}& 
 \begin{tabular}{r}
$-0.22 \pm 0.05$\\
$-0.16 \pm 0.07$\\
$-0.35 \pm 0.06$\\
 \end{tabular}\\
\hline

 \multicolumn{2}{c}{$\log\, \rho_\star$ [$M_{\odot}Mpc^{-3}$]} &  &  &  &  \\
\hline %inserts double horizontal lines
\textbf{M1}&
 \begin{tabular}{c}
 $R<0.5$\\
 ETG\\
 $1< R < 2$\
 \end{tabular}&  
 \begin{tabular}{r}
$7.57 \pm 0.11$\\
$8.00 \pm 0.26$\\
$7.42 \pm 0.42$\\
 \end{tabular}& 
 \begin{tabular}{r}
$7.51 \pm 0.13$\\
$7.82 \pm 0.22$\\
$7.08 \pm 0.25$\\
 \end{tabular}& 
 \begin{tabular}{r}
$7.34 \pm 0.15$\\
$7.61 \pm 0.28$\\
$6.81 \pm 0.34$\\
 \end{tabular}& 
 \begin{tabular}{r}
$6.91 \pm 0.34$\\
$7.08 \pm 0.51$\\
$6.23 \pm 0.67$\\
\end{tabular}\\
\hline

\hline
\end{tabular}}
\caption{$\rho_{\rm SFR}$,  sSFR and $\rho_\star$ for redshifts $z = 0, 1, 2, 5$ obtained for early (E, S0, Sa)  type galaxies.
}
\label{tab:tableETG} 
\end{table}

%#####late type galaxies######
\begin{table}
\resizebox{0.5\textwidth}{!} {\begin{tabular}{c c c c c c} % centered columns (4 columns)
 \multicolumn{2}{c}{$\log\, \rho_{\rm SFR}$ [$M_{\odot}yr^{-1}Mpc^{-3}$]}& $z=0$ & $z=1$ & $z=2$ & $z=5$ \\
\hline %inserts double horizontal lines
\textbf{M1}&
 \begin{tabular}{c}
 $R<0.5$\\
 LTG\\
 $1< R < 2$\
 \end{tabular}&  
 \begin{tabular}{r}
$-2.72 \pm 0.12$\\
$-2.13 \pm 0.44$\\
$-2.46 \pm 0.07$\\
 \end{tabular}& 
 \begin{tabular}{r}
$-2.37 \pm 0.19$\\
$-1.95 \pm 0.12$\\
$-2.27 \pm 0.19$\\
 \end{tabular}& 
 \begin{tabular}{r}
$-2.27 \pm 0.07$\\
$-1.91 \pm 0.07$\\
$-2.36 \pm 0.09$\\
 \end{tabular}& 
 \begin{tabular}{r}
$-2.35 \pm 0.14$\\
$-2.00 \pm 0.12$\\
$-2.55 \pm 0.23$\\
 \end{tabular}\\
 
 \hline

\multicolumn{2}{c}{$\log\, sSFR$ [$Gyr^{-1}$]}& &  & &  \\
\hline %inserts double horizontal lines
\textbf{M1}&
 \begin{tabular}{c}
 $R<0.5$\\
 LTG\\
 $1< R < 2$\
 \end{tabular}&  
 \begin{tabular}{r}
$-1.66 \pm 0.08$\\
$-1.31 \pm 0.20$\\
$-1.19 \pm 0.08$\\
 \end{tabular}& 
 \begin{tabular}{r}
$-0.64 \pm 0.04$\\
$-0.65 \pm 0.07$\\
$-0.67 \pm 0.03$\\
 \end{tabular}& 
 \begin{tabular}{r}
$-0.27 \pm 0.06$\\
$-0.37 \pm 0.06$\\
$-0.38 \pm 0.05$\\
 \end{tabular}& 
 \begin{tabular}{r}
$-0.13 \pm 0.06$\\
$-0.08 \pm 0.05$\\
$-0.12 \pm 0.05$\\
 \end{tabular}\\
\hline

 \multicolumn{2}{c}{$\log\, \rho_\star$ [$M_{\odot}Mpc^{-3}$]} &  &  &  &  \\
\hline %inserts double horizontal lines
\textbf{M1}&
 \begin{tabular}{c}
 $R<0.5$\\
LTG\\
 $1< R < 2$\
 \end{tabular}&  
 \begin{tabular}{r}
$7.20 \pm 0.11$\\
$7.65 \pm 0.21$\\
$7.29 \pm 0.42$\\
 \end{tabular}& 
 \begin{tabular}{r}
$7.05 \pm 0.16$\\
$7.41 \pm 0.15$\\
$6.93 \pm 0.15$\\
 \end{tabular}& 
 \begin{tabular}{r}
$6.83 \pm 0.22$\\
$7.17 \pm 0.21$\\
$6.65 \pm 0.24$\\
 \end{tabular}& 
 \begin{tabular}{r}
$6.28 \pm 0.41$\\
$6.58 \pm 0.54$\\
$5.95 \pm 0.61$\\
\end{tabular}\\
\hline

\hline
\end{tabular}}
\caption{$\rho_{\rm SFR}$,  sSFR and $\rho_\star$ for redshifts $z = 0, 1, 2, 5$ obtained for late (Sb, Sbc, Sc, Sd)  type galaxies.
}
\label{tab:tableLTG} 
\end{table}
%--------------------------------------------------------------------------------------------------------------------------------------------

%-------------------------------------------------------------------------------------------------------

\subsection{The evolution of the main sequence of star formation}

%***FIG***FIG***FIG***FIG***FIG***FIG***FIG***FIG***FIG***FIG***
%\begin{figure*}[!ht]
\begin{figure*}
\includegraphics[width=\textwidth]{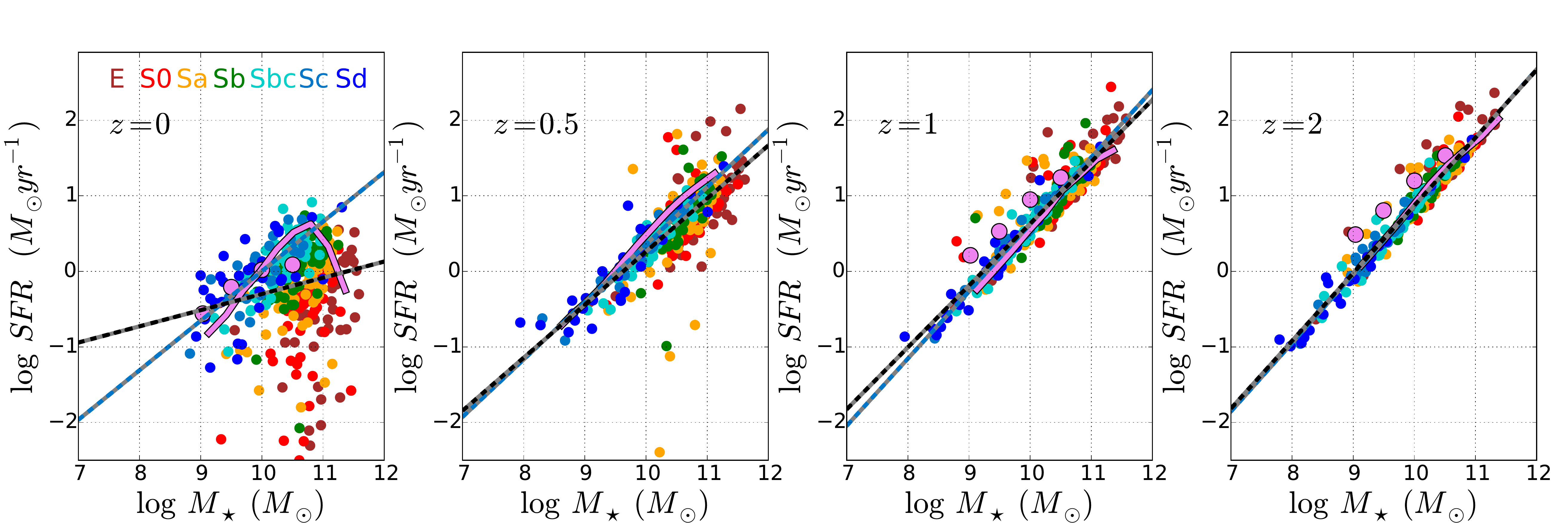}
\includegraphics[width=\textwidth]{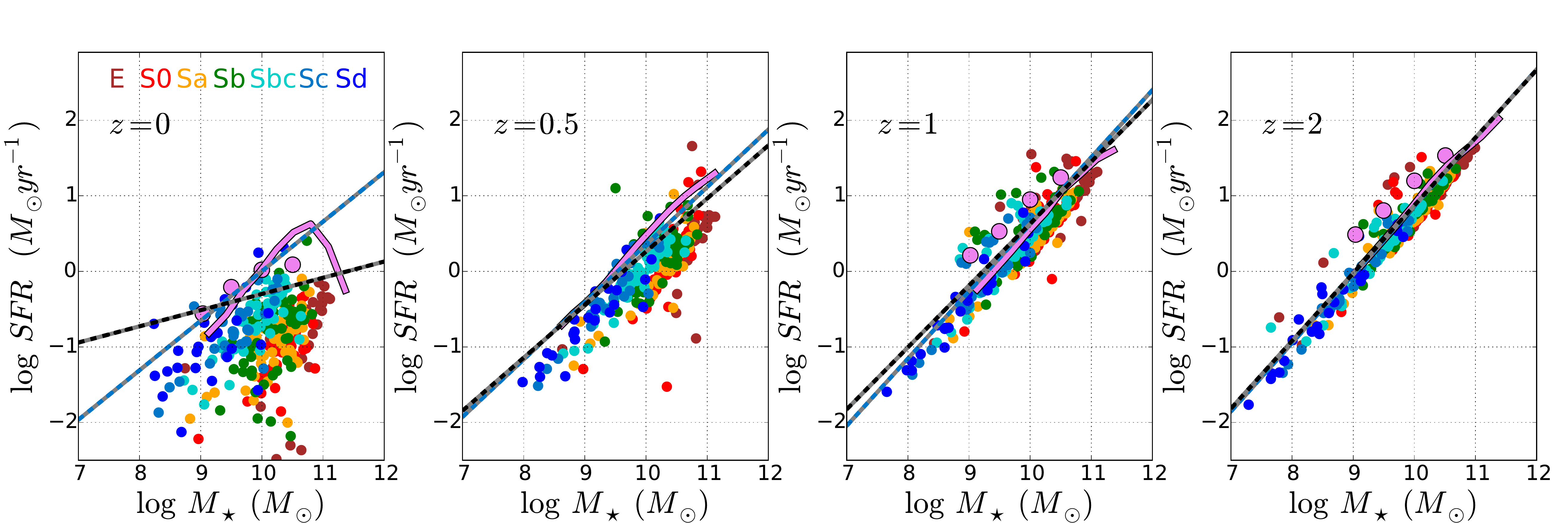}
\caption{  
Upper panels: The evolution of  MSSF from z= 2 (right panel) to z= 0 (left panel). Galaxies are color coded according with the morphology. The blue line is a linear fit to Sc galaxies. The black line is a fit to all  galaxies. 
Lower panels:  as in the upper panels but using only  the contributions  to  SFR and $M_\star$ of the galaxies current inner regions ($R\leq$ 0.5 HLR).
Blue-gray and black-gray lines (in upper and lower panels) are  fits to $\log$  SFR $-  \ \log \ M_\star$ for Sc and for all the galaxies in this study. Big purple dots are from the compilation by \citet{behroozi13}, and the purple line is the MSSF from the $Illustris$ simulation by \citet{sparre15} and \citet{tomczak16} . 
}
\label{fig:MSSF}
\end{figure*}
%***FIG***FIG***FIG***FIG***FIG***FIG***FIG***FIG***FIG***FIG***

Our analysis allows us to retrieve the correlation between  SFR and $M_\star$ at different redshifts, and to investigate the evolution of the  slope and normalization of the MSSF. Furthermore, the effect of spatially-sampling in the MSSF.
Studies have shown that the correlation found in the nearby Universe \citep{brinchmann04, salim07, renzinipeng15, catalantorrecilla15, gonzalezdelgado16}  persists up to $z \sim 7$ \citep{noeske07, daddi07, pannella09, peng10, wuyts11, whitaker12}. 

At $z = 0$, it is well proved that MSSF relation is  sub-linear, although its slope varies depending on the galaxy sample selection and on the indicator used to estimate the SFR \citep{speagle14}. Observationally,  there is also evidence that the slope and normalization of MSSF evolves with lookback time, although the role of the galaxy sample selection and the effect of extinction is not yet understood \citep{noeske07, elbaz07, salim07, schiminovich07, pannella09, whitaker12}. For example,  by studying a large sample of star forming galaxies out to $z = 2.5$ \citet{whitaker12}  found that the slope of the relation decreases at higher redshifts as a consequence of the sample being more biased to dusty star forming galaxies. Moreover since, on average, star forming galaxies  in the past had larger SFR than today \citep{madau14}, it is expected that the MSSF changes with time, at least in the zero point. 
More recently, \citet{speagle14} have used a compilation of data from 25 different surveys from the literature and they analyze, in a highly consistent framework, the evolution of the slope and zero point of the MSSF relation from $z$ = 0 to $z$ = 6. After accounting for the scatter between the different SFR indicators, they found that the slope and zero-point of the MSSF are likely time dependent.
Thus, the slope (and zero-point) of the MSSF increases (decreases) with redshift and lookback time. In contrast, \citet{lee15} have found that the star forming galaxies in the COSMOS field at $z\leq$1.3 observed at  far-infrared wavelengths follow a broken-law, SFR $\propto M_\star^{0.88\pm0.06}$ below the turnover mass of $10^{10}$, and SFR $\propto M_\star^{0.27\pm0.04}$ above $10^{10}$. 
Later, using data from the FourStar Galaxy Evolution Survey (ZFOURGE) at $0.5 < z < 4$  \citet{tomczak16} found that the turnover mass  ($M_0$) can range from $10^{9.5}$ to $10^{10.8}$, with evidence that $M_0$ increases with redshift.

With a completely different approach to redshift surveys, our analysis allows us to investigate the evolution of MSSF and to compare with  galaxy redshift surveys. Indeed, our results on SFR$(t)$ presented in Section \ref{sec:SFR}  and  \citet{gonzalezdelgado17} confirm that  SFR($t$)  declines since $z \lesssim 2$, and $M_\star(t)$ has grown little since $z \lesssim 1$ (\citealt{garciabenito17, perez13}). Thus, we should expect that the MSSF at earlier epochs is shifted up  with respect to the current MSSF. Furthermore, the decreasing scatter with increasing redshift can be understood because galaxies that today are out of the MSSF (because the star formation has been quenched) in the past could have been actively forming stars and be well placed on the MSSF. 

Figure \ref{fig:MSSF} (upper panel) presents our MSSF extracted at four redshifts: $z = 0$, 0.5, 1, and 2. It also shows the relations found by fitting only current star forming Sc galaxies, and all the galaxies of the sample. The first result to notice is that, while at $z = 0$ the relation fitted for Sc and all the galaxies are very different,  mainly due to ETGs that are off down  the MSSF, at higher redshift, the two fits are quite similar. 
Considering only Sc galaxies we obtain a slope, $a = 0.66\pm0.17$, and a zero-point, $b= -6.55\pm0.09$, at $z = 0$, and  $a = 0.91\pm0.04$ and $b= -8.20\pm0.03$ at $z = 2$. If all the galaxies are included, the slope (zero-point) also increases (decreases) significantly with redshift (see Table \ref{tab:mssf}).

There are several other interesting results: a) the scatter in the MSSF decreases significantly with redshift. The progenitors of all the galaxies in the sample were in the MSSF at $z \geq 2$; b) the progenitors of E and S0 were in or above the MSSF at $z=1$. 

Our results are in agreement with those of \citet{speagle14} and \citet{chiosi17}, in the sense that the MSSF slope is time dependent and flattens from high to low redshift. However, this slope does not decrease linearly with the age of the Universe, because our data show relatively little change from  $z=2$ to $z = 1$. The flattening from $z = 1$ to 0, though, is equal to their prediction.

Figure \ref{fig:MSSF} also shows (purple line) the results for the MSSF by the $Illustris$ cosmological hydrodynamical simulation of galaxy formation \citep{nelson15}, and observational results from a compilation by  \citet{behroozi13} that we took from figure 1 in \citet{sparre15}. The results for $z = 0.5$ are taken from the $Illustris$ simulation at $0.5 < z < 1$ as presented in figure 6 in \citet{tomczak16}.
Fig.\ \ref{fig:MSSF} shows that the $Illustris$ simulation at $z =$ 1 and 2 follow very well our results, as well as the trend of galaxies  going out from the main sequence at $z \leq 0.5$. At any redshift, the points from the compilation by  \citet{behroozi13} have a flatter slope than our results and the $Illustris$ simulation, and at $z=1, 2$, they are above the MSSF traced by $Illustris$ and our MSSF fit. However these points are located in a region populated by the progenitors of current ETG (E, S0, Sa) that were above the MSSF at $z= 1, 2$. 

Other high redshift galaxy surveys have also shown that passive  galaxies do exist up to at least $z\sim2$, in particular at high masses (e.g. \citealt{cimatti04}, \citealt{glazebrook04}),  that the red sequence of galaxies was present at $z \approx 1$ (e.g. \citealt{wuyts11}).  However, Fig.\ \ref{fig:MSSF} (upper panels) shows that current ETGs are not out of the MSSF until relatively recent, $z \leq$0.5. To check how this result depends on the spatial sampling, we obtain the SFR$^{in}$(t) and $M_\star^ {in}$(t) for the galaxy regions that today are located in the central 0.5 HLR (lower panels). The lines are the same as in the upper panels. We found that at $z \geq$1, the inner regions were still on the MSSF, and it is at lower redshift that these regions migrate off the MSSF. Thus, our results  suggest that the massive passive and dead galaxies detected at $z \geq 2$ by high redshift galaxy surveys cannot be the progenitors of current ETGs in the CALIFA survey.  Spectroscopically confirmed quiescent galaxies at $z>2$ have masses already in excess of $\log \ M_\star \sim 10.5 - 11$ (e.g. \citealt{toft12, belli17}); thus, if they further grow through mergers, their descendants may be more massive galaxies not sampled in the CALIFA volume.

Overall, our results point to the idea that the MSSF could be a natural consequence of the cold mode accretion of galaxies \citep{birnboimdekel03}, where the supply of gas feeding the star formation in galaxies follows the dark matter halo accretion rate, with  coupled baryons and dark matter halo.  However, because the slope is $< 1$ even at very high redshift,  feedback may also play a role in making the relation sub-linear.

\begin{table}
\resizebox{0.5\textwidth}{!} {\begin{tabular}{c c c c c c} % centered columns (4 columns)

& $ax +b$ & $z=0$ & $z=0.5$ & $z=1$ & $z=2$ \\
\hline %inserts double horizontal lines
\textbf{Sc}&
\begin{tabular}{c}
$a$\\
$b$\
\end{tabular}&
\begin{tabular}{r}
$0.66\pm 0.17$\\
$-6.55\pm 0.09$\\
\end{tabular}&
\begin{tabular}{r}
$0.76\pm 0.09$\\
$-7.25\pm 0.05$\\
\end{tabular}&
\begin{tabular}{r}
$0.89\pm 0.05$\\
$-8.27\pm 0.04$\\
\end{tabular}&
\begin{tabular}{r}
$0.91\pm 0.04$\\
$-8.20\pm 0.03$\\
\end{tabular}\\
\hline
\hline
& & &  & &  \\
\textbf{All galaxies}&
\begin{tabular}{c}
$a$\\
$b$\
\end{tabular}&
\begin{tabular}{r}
$0.21\pm 0.05$\\
$-2.44\pm 0.03$\\
\end{tabular}&
\begin{tabular}{r}
$0.70\pm 0.03$\\
$-6.76\pm 0.03$\\
\end{tabular}&
\begin{tabular}{r}
$0.82\pm 0.02$\\
$-7.56\pm 0.02$\\
\end{tabular}&
\begin{tabular}{r}
$0.90\pm 0.01$\\
$-8.09\pm 0.02$\\
\end{tabular}\\
\hline
\end{tabular}}
\caption{Parameters of $\log$  SFR  ($M_\odot$  yr$^{-1})$ = a $\log \ M_\star$ (M$_\odot)$ + b fits of the MSSF at different redshifts obtained for Sc  and  all the galaxies.
}
\label{tab:mssf}
\end{table}

%-------------------------------------------------------------------------------------------------------

%-------------------------------------------------------------------------------------------------------
%NEW SECTION
%-------------------------------------------------------------------------------------------------------

\section{Summary and conclusions}
\label{sec:Summary}

Using  the stellar populations fossil record method  for a sample of 366  CALIFA galaxies with GALEX images, we obtained the cosmic evolution of the absolute and specific star formation rate in galaxies, and galaxy mass.  These properties were estimated for galaxies with stellar mass in the range $\sim$ 10$^9$  to  10$^{12}$  (for a Chabrier IMF), stacking the results as a function of  Hubble type (E, S0, Sa, Sb, Sbc, Sc and Sd). A bayesian method based on parametric SFHs that fits simultaneously the FUV, NUV, and u,g,r,i,z-bands, and the D$_n$4000, H$\beta$, and [MgFe]$'$ indices from the data cubes is also presented. In the main body of the paper, we discuss the results obtained by using a parametric delayed-$\tau$ SFH. In the appendix, the results obtained with other parameterizations (formed by a single or by a combination of two laws) are compared with the delayed-$\tau$ model. Furthermore, we compare these results with those obtained by fitting the UV band and the full CALIFA spectra  using a recent version of the non-parametric code \starlight\ \citep{lopezfernandez16}. 
The fits  are processed to derive the time evolution of  SFR$(t)$, and sSFR$(t)$  for three different regions in each galaxy at the present epoch: a) 0$-$2 HLR, b) 0$-$0.5 HLR, c) 1$-$2 HLR, as representative of the "whole" (integrated) galaxy, the  innermost regions (dominated in most of the galaxies by the spheroidal  or bulge component), and outer regions (dominated by the disk in spirals). 

Our main results are:

\begin{itemize}

\item At any epoch, the SFR  scales with the current Hubble type, as expected from the dependence of  SFR with $M_\star$, and of galaxy mass with  morphology. The highest SFR ($\sim40$ $M_\odot$ yr$^{-1}$) occurred in  E galaxies at $z\sim 1-2$. The lowest SFR at similar epochs occurs in late spirals Sd ($\sim 3$ $M_\odot$ yr$^{-1}$). SFR in the inner regions of  E    peak  at $z \sim 2$, while it peaks at $z < 1$ in the outer regions. 
The SFR peak in early spirals (Sa, Sb, Sbc) occurred  earlier  than in E and S0. The SFR peak  in the inner regions occurred at a similar epoch in E, S0, and early spirals;  earlier  than for late type spirals. 

\item  
These results are a consequence of the values obtained for the parameters $t_0$ and $\tau$. For Sa-Sbc $t_0\sim 12$ Gyr, $t_0\sim10$ Gyr for E and S0, and $t_0\sim10-9 $ Gyr in Sc and Sd galaxies. For the present inner regions, $t_0-\tau$ is larger than that of the whole galaxy and of the present outer regions, indicating that the inner regions formed earlier than outer ones, and that galaxies formed inside-out. The e-folding time, $\tau$ increases with  morphological type, and  is higher in the outer than in the inner regions of spirals, indicating that star formation is more extended in time in late than in early type spirals.

\item The CALIFA sample is well suited to compute the evolution of $\rho_{\rm SFR}$ and $\rho_\star$ which is in agreement with results obtained from galaxy surveys, in particular with the recent estimations obtained from GAMA + G10-COSMOS/3D-HST by \citet{driver17}.  At $z\leq 0.5$ the majority of $\rho_{\rm SFR}$ takes place in the outer regions of galaxies, but at higher redshifts the present inner regions ($<0.5$ HLR) play a major role in building $\rho_{\rm SFR}$ and  $\rho_\star$ dominating at $z > 2$. In terms of morphology, while at $z=0$ late spirals dominate the $\rho_{\rm SFR}$ budget, at $z > 2$ the progenitors of the present  E and S0 are the major contributors to the SFR density and $\rho_\star$. Taken as reference the $50\%$ point of $\rho_\star$, the inner regions reached this value at 9 Gyr lookback time, and the outer regions at 6 Gyr. This delay between the inner and outer regions confirms that galaxies  grow inside-out.

\item The sSFR declines rapidly as the Universe evolves, although the slope depends on the morphology, steeper for early than for late type galaxies.
In the inner regions, sSFR declines with time more rapidly than in the outer regions in early spirals (Sa-Sbc) and E and S0, suggesting an earlier epoch for the shut down of the star formation. 

\item The MSSF is traced up to $z > 2$. We find that the slope evolves with time, in agreement with cosmological galaxy surveys. The slope flattens from 0.9$\pm 0.01$ at $z= 2$ to 0.66$\pm0.17$ at the present epoch (when only Sc galaxies are considered). Our estimations of the evolution of the MSSF are in good agreement with the predictions by the $Illustris$ simulation. They suggest that the MSSF is a natural consequence of a cold mode accretion, although feedback may also play some role to set the slope of the correlation below to 1.

\item  The comparison of nine different parametric SFHs (described in the Appendix) indicates that a delayed-$\tau$ SFR is the model that provides better match between our results and those from the snapshots of galaxy  evolution obtained by studies at different redshifts for $\rho_{\rm SFR}$, sSFR($t$), and $\rho_\star$.
The average SFH of galaxies, as represented by  $m(t)$ vs lookback time of CALIFA galaxies, confirms that globally galaxies grow their mass mainly in a mode that is well represented by a delayed-$\tau$, where the maximum peaks at high redshift ($z \sim 2$), and then declines exponentially with an e-folding time of $\sim$3.9 Gyr. This result is in agreement with the model proposed by \citet{madau14} and our previous results using non-parametric SFH \citep{gonzalezdelgado17}.

\end{itemize}

These results show again the uniqueness of the CALIFA survey to 
characterize the  the cosmic evolution of the spatially-resolved SFR and stellar mass of galaxies. 
The fossil record of the stellar populations of this sample of nearby galaxies has been very successful to derive $\rho_{\rm SFR}$, sSFR, and $\rho_\star$, that are in good agreement with the results from the snapshot galaxy surveys in a large range of  redshifts. Thanks to the spatially-resolved data, we have retrieved the contributions of different regions of early type and spirals to $\rho_{\rm SFR}$, and $\rho_\star$.

%---------------------------------------------------------------------------------------------
%NEW SECTION
%---------------------------------------------------------------------------------------------
\begin{acknowledgements} 
CALIFA is the first legacy survey carried out at Calar Alto. The CALIFA collaboration would like to thank the IAA-CSIC and MPIA-MPG as major partners of the observatory, and CAHA itself, for the unique access to telescope time and support in manpower and infrastructures.  We also thank the CAHA staff for the dedication to this project.
Support from the Spanish Ministerio de Econom\'\i a y Competitividad, through projects AYA2016-77846-P, AYA2014-57490-P, AYA2010-15081, and Junta de Andaluc\'\i a FQ1580. 
We also thank the Viabilidad, Dise\~no, Acceso y Mejora funding program, ICTS-2009-10, for funding the data acquisition of this project. RCF thanks the hospitality of the IAA and the support of CNPq. RGD acknowledges the support of CNPq (Brazil) through Programa Ci\^encia sem Fronteiras (401452/2012-3). RCF and NVA acknowledge the support from the CAPES CsF--PVE project 88881.068116/2014-01. SFS thanks the ConaCyt programs IA-180125 and DGAPA IA101217 for their support to this project.
We thank the support of the IAA Computing group, and to the referee for useful comments. 

\end{acknowledgements}

%\clearpage

%\clearpage

% Bibliography
\bibliographystyle{aa}
\bibliography{Califa8_SFH}

%---------------------------------------------------------------------------------------------
%NEW SECTION
%---------------------------------------------------------------------------------------------
\clearpage
\appendix

\section{Comparing results from parametric and non-parametric SFHs}

In this appendix  we compare the SFHs obtained with the delayed-$\tau$ model with those obtained with other models, parametric and non-parametric  (the \starlight\ code). We express the SFH by the mass fraction, $m(t)$, 
defined for each galaxy as the ratio of the mass formed in each epoch to the total  mass of the galaxy today. We also derive the mass growth curve (see e.g. \citealt{perez13, garciabenito17}), which provides useful information about how the mass is assembled in a galaxy as a whole and in different regions. From these curves, we compare the epoch at which galaxies assembled  $80\%$ of their current mass, $t_{80}$. Finally, we estimate $\rho_{\rm SFR}$, $\rho_\star$, and sSFR at different redshifts, and we compare the results from the different models with those  from   surveys at different redshifts. The goal of this Appendix is to show that delayed-$\tau$ model is the most simple, meaningful, and representative parameterization of the SFHs.

\subsection{delayed-$\tau$ vs non-parametric SFH with \starlight}

Our previous analysis obtained with \starlight\ for the CALIFA  sample presented in \citet{gonzalezdelgado17} was done  by fitting only the CALIFA spectra, using composite stellar population models built with SSPs by \citet{vazdekis10} and \citet{gonzalezdelgado05} assuming a Salpeter IMF. In terms of $m(t)$, we obtain that  the highest mass fractions invariably occur at the earliest times, and  subsequent star formation varies systematically with $M_\star$ and morphology, with the low $M_\star$ and also the later spiral galaxies forming stars over  more extended periods of time, while high $M_\star$ and early type galaxies exhibit the fastest decline in $m(t)$. 

In order to be consistent with the data, here we use  \starlight\ but now fitting  the GALEX (FUV, NUV) photometry and the full CALIFA spectra.\footnote{Instead of GALEX and SDSS photometry and the spectroscopic indexes, D$_n$4000, H$\beta$, and [FeMg]$^\prime$  used in M1.} The fits are done  with the new version of \starlight\ by \citet{lopezfernandez16} in combination with the same collection of SSPs by  Charlot  \& Bruzual (2007) used  with M1. We also assume  a Chabrier IMF.

\begin{figure}
\includegraphics[width=0.5\textwidth]{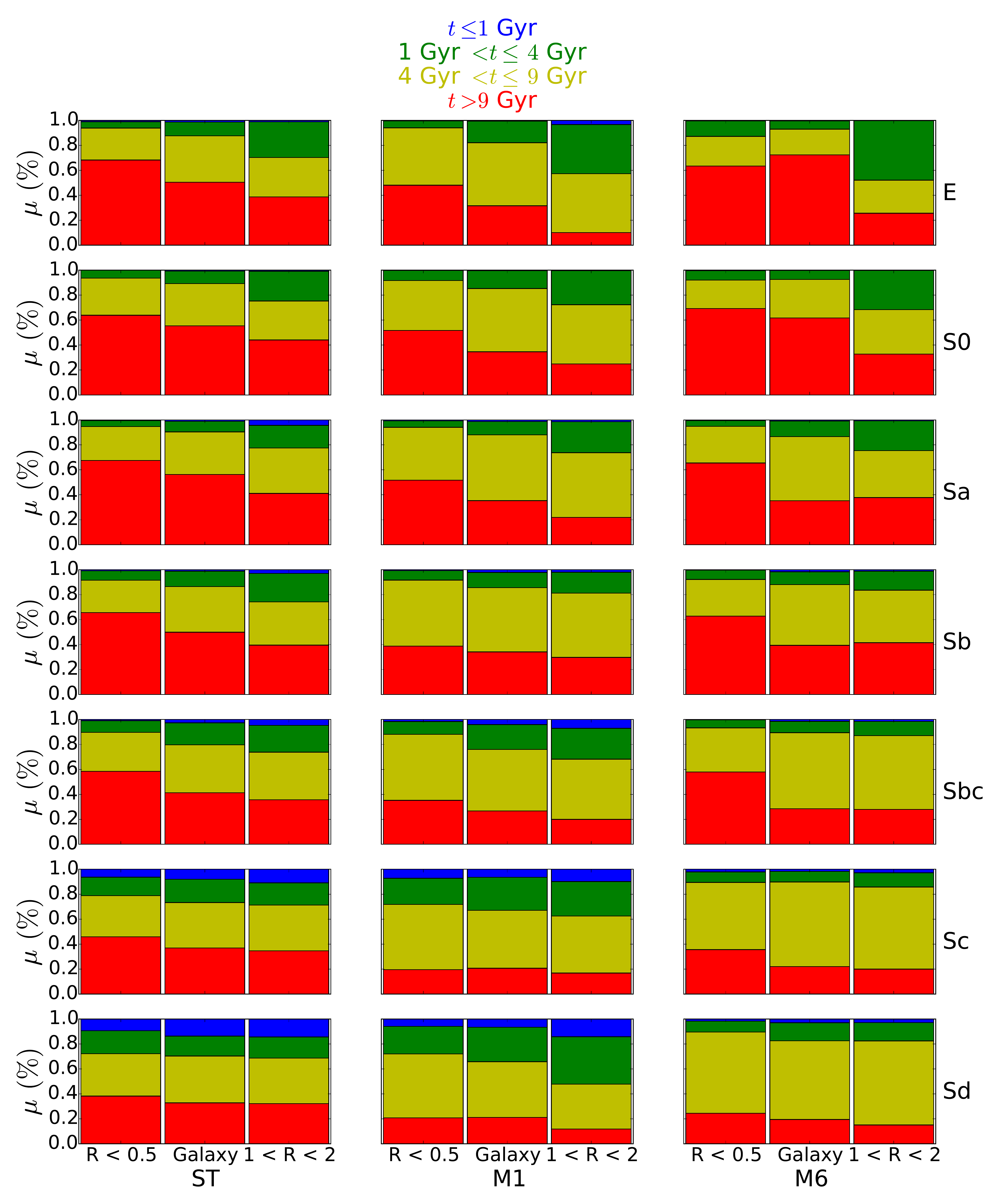}
\caption{ Results of mass fraction $m(t)$ grouped in four different age ranges (color coded as indicated on the top) from  \starlight\ (ST, left), the delayed-$\tau$ model (M1, middle), and a two-exponential model (M6, right).
}
\label{fig:massfraction_bars}
\end{figure}

To quantify the differences between the \starlight\ and the delayed-$\tau$ model we have discretized $m(t)$ in a few relevant age ranges. Fig.\ \ref{fig:massfraction_bars} presents the results for four age ranges: $t \leq 1$ Gyr (blue), $1 < t \leq 4 $ Gyr (green) , $4 < t \leq 9$ Gyr (yellow), and $t \geq 9$ Gyr (red). Each panel shows three bar charts  corresponding to the spatial regions: $R < 0.5$ HLR (left), the whole galaxy (middle), and the outer regions $1 < R < 2$ HLR. The results from the delayed-$\tau$ model (M1) and \starlight\ (ST) are presented in the middle and left columns.

%***FIG***FIG***FIG***FIG***FIG***FIG***FIG***FIG***FIG***FIG***
%\begin{figure*}[!ht]
%***FIG***FIG***FIG***FIG***FIG***FIG***FIG***FIG***FIG***FIG***

As with ST, most of the stellar mass in M1 is formed very early on and with very little mass in stars younger than 1 Gyr. Notice that the major differences occur in a similar way for all the morphological types. Thus, we can conclude that with ST non-parametric SFH  we derive a larger fraction of old stellar populations ($t > 9 $ Gyr and/or $4 < t < 9 $ Gyr) and less intermediate and young components ($1 < t < 4$ Gyr and/or $t < 1$ Gyr). To comment in more detail these differences, we discuss the results for the inner regions:

\begin{itemize}

{\item  $m(t \leq 1)$: this mass fraction increases from E to Sd, although the fraction is lower in the inner regions than in the outer ones. With ST, $m(t \leq 1)$ $\leq 1 \%$ in Sbc and earlier types,  increasing to $m(t \leq 1)$ $\sim 6\%$ and $9\%$ for Sc and Sd, respectively. Using M1,  $m(t \leq 1)$ $\leq 2 \%$ in Sbc and earlier types increasing to $m(t \leq 1)$ $\sim 7 \%$ and $6\%$ for Sc and Sd galaxies, respectively. }

{\item  $m(1< t \leq 4 )$: on average, we obtain lower mass fractions of this component in the inner than in the outer regions for all the Hubble types and all the models. Using ST and M1, the mass fraction of this component increases from early to late type galaxies. For E galaxies, $m(1< t \leq 4)$ $\sim 5 \%$ and $6 \%$, for ST and M1, respectively. On the other hand, for Sd galaxies, $m(1 < t \leq  4)$ $\sim 19 \%$ and  $22\%$, respectively.}

{\item  $m(4< t \leq 9)$: with ST and M1, we obtain similar mass fractions of this component for all the Hubble types 
With ST, $20 \% \leq$ $m(4< t \leq 9)$  $\leq 30 \%$. For M1, we obtain $3\% \leq$ $m(4< t \leq 9)$ $\leq 46 \%$ for E, S0, and Sa galaxies; $m(4< t \leq 9)$  $\sim 53 \%$ for Sb, Sbc, and Sc; and $m(4< t \leq 9)$ $\sim 52 \%$ for Sd galaxies.}

{\item  $m(t > 9)$: for all the models the old component decreases from E to Sd galaxies and  is larger in the inner region. For E galaxies, $m(t > 9)\sim 74 \%$ and  $48\%$, for ST and M1, respectively. For Sd galaxies, $m(t > 9)\sim 42 \%$ and $ 21\%$, respectively.}

\end{itemize}

Other interesting results come from the comparison of  the epoch at which  galaxies  assemble  $80 \%$ of their mass, $t_{80}$. From our previous analysis with \starlight, \citet{perez13} and \citet{garciabenito17} found that the most massive galaxies  assembled their mass earlier than the low mass galaxies, a signature of downsizing. We have also  obtained that $t_{80}$ for the inner regions is higher than for the outer regions, suggesting that galaxies  assemble their mass  inside-out.   

\begin{figure}
\includegraphics[width=0.5\textwidth]{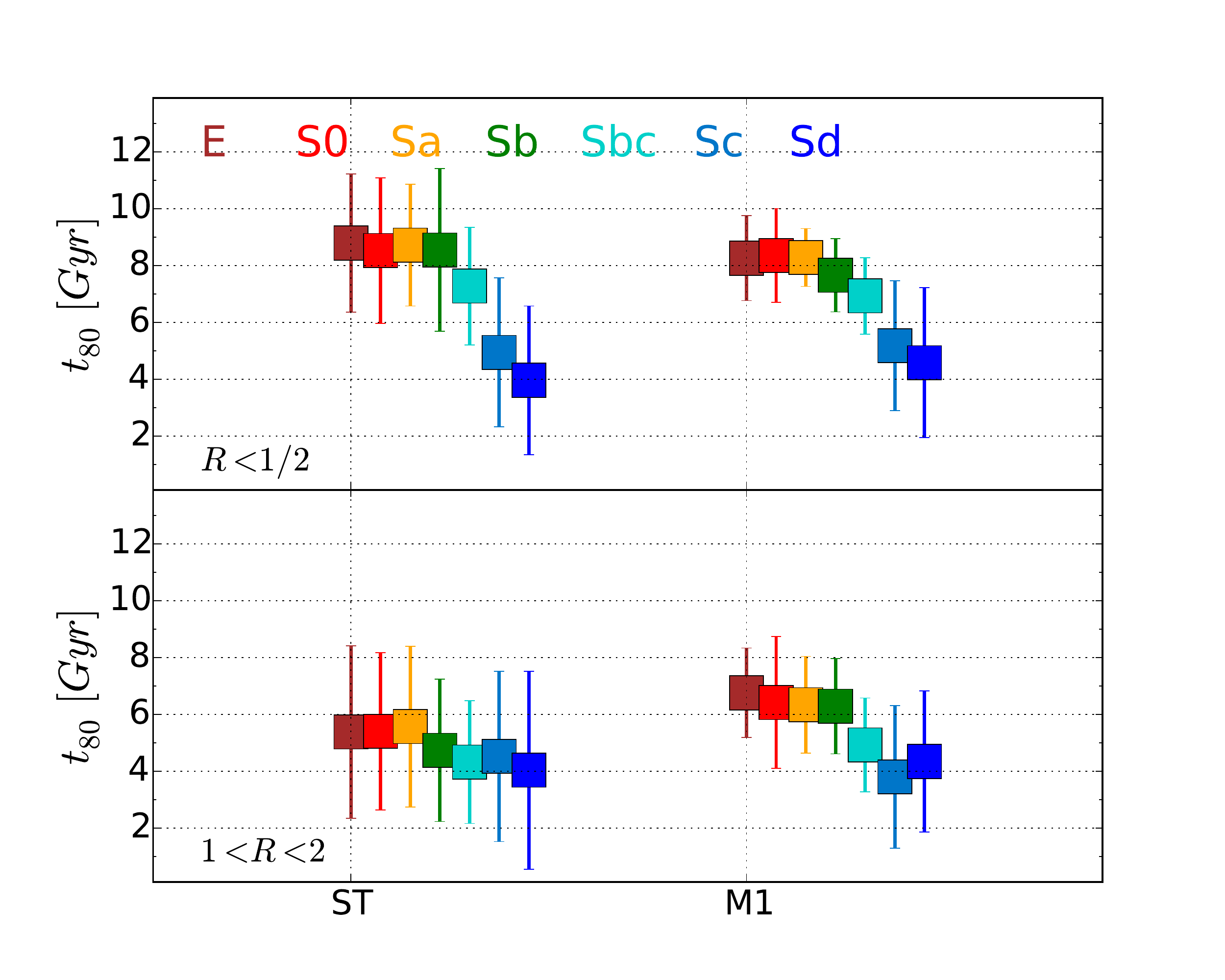}
\caption{ Lookback time at which galaxies  assemble  $80 \%$ of their mass, $t_{80}$, from \starlight\,  and from M1, corresponding to the inner $R < 0.5$ HLR and outer regions $1 < R < 2$ HLR.
}
\label{fig:t80_models}
\end{figure}

Figure \ref{fig:t80_models} compares the results from M1 and ST for $t_{80}$ in the inner and outer regions. They are:

\begin{itemize}

{\item Inner regions: 1)  M1, $t_{80}$ decreases with Hubble type, with E, S0, and Sa having similar values, $t_{80}\sim 9-8$ Gyr, while $t_{80}\sim5$ Gyr for Sc and Sd. 2) The results with ST are very similar,  $t_{80}$ decreasing from early  to late types. For Sb and earlier types, $t_{80}\sim 8-9$ Gyr,  $t_{80}\sim 5$ Gyr for Sc, and $t_{80}\sim 4$ Gyr for Sd, slightly lower than with M1.}

{\item Outer regions: 1)   M1, $t_{80}$ decreases with the morphology, lower for later types. The values for the outer regions are lower than for the inner ones for all types. For Sb and earlier types $t_{80}\sim6-7$  Gyr, and for Sd galaxies $t_{80}$ $\sim4$  Gyr. 2) With ST, $t_{80}$ decreases similarly with  Hubble type, from $t_{80}\sim5-6$ Gyr for Sa and later types to $t_{80}\sim 4$ Gyr for Sd. }

\end{itemize}

%***FIG***FIG***FIG***FIG***FIG***FIG***FIG***FIG***FIG***FIG***
%\begin{figure*}[!ht]
%***FIG***FIG***FIG***FIG***FIG***FIG***FIG***FIG***FIG***FIG***

\subsection{delayed-$\tau$ vs other parametric SFHs}
\label{sec:OtherParametricModels}
 
%***FIG***FIG***FIG***FIG***FIG***FIG***FIG***FIG***FIG***FIG***
%\begin{figure*}[!ht]
\begin{figure}
\includegraphics[width=0.5\textwidth]{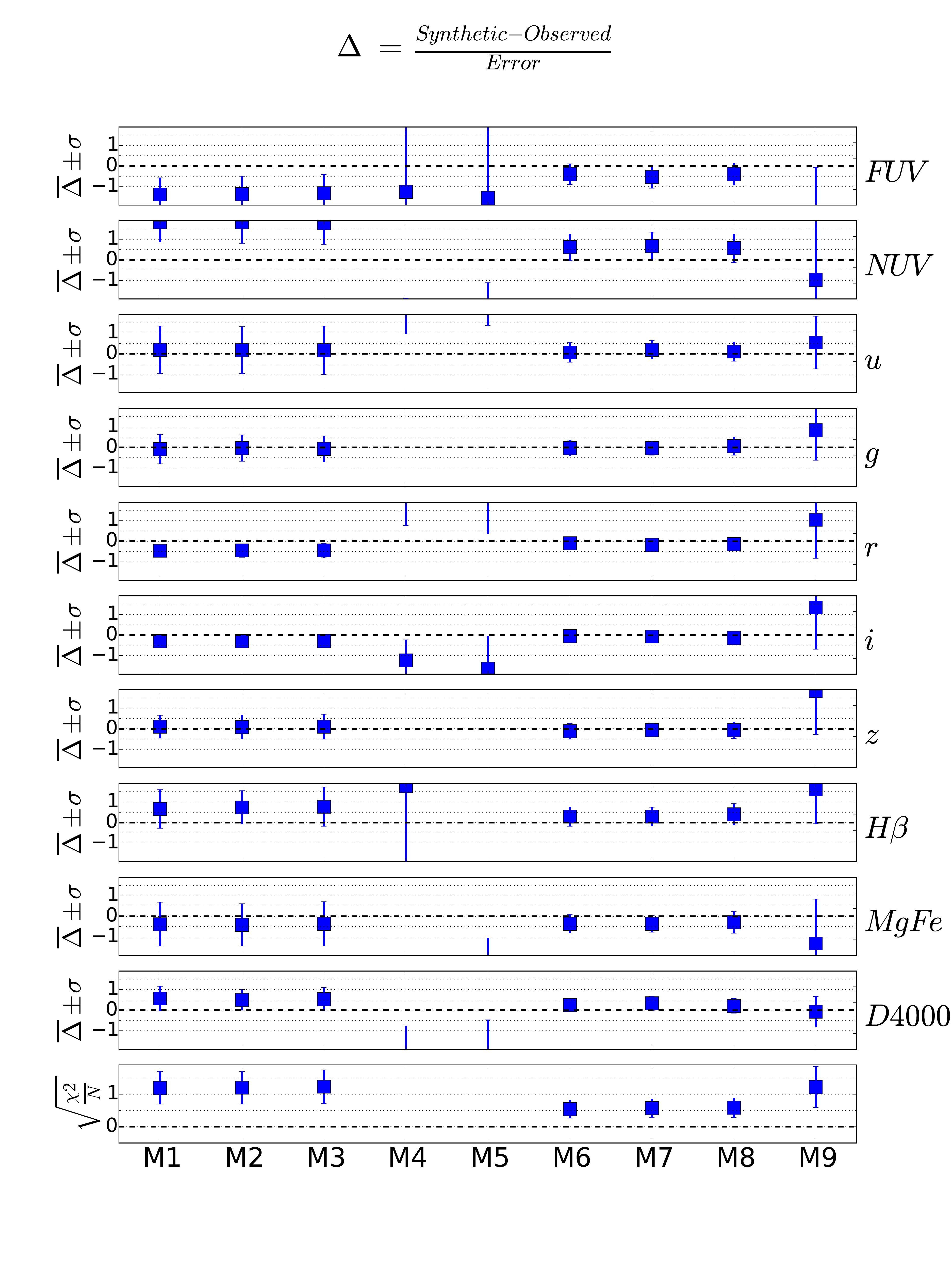}
\caption{ Quality of the fits using the different models. Each panel shows $\Delta \pm \sigma$ for each of the observables. $\Delta$ is the average of the difference between the synthetic and the observable quantity divided by the error. The bottom panel, however, shows the reduced $\chi^2$.
}
\label{fig:qualityfits}
\end{figure}
%***FIG***FIG***FIG***FIG***FIG***FIG***FIG***FIG***FIG***FIG***

Other parametric SFHs have been used to fit the observational constrains. They are of two types: 
 
a) One single function, as in the $\tau$ model:

\begin{itemize}
\item A $\tau$ model (M2): $$\psi(t) = \psi_{0}e^{-(t_{0} - t)/\tau}$$
\item Sandage model \citep{sandage86} (M3): $$\psi(t) = \frac{A}{\tau^{2}}(t_{0} - t)e^{-(t_{0}-t)^{2}/2\tau^{2}}$$
\item Linearly-rising (M4): $$\psi(t) =  \psi_{0} - \frac{d\psi}{dt}(t_{0}-t)$$
\item Power rising (M5): $$\psi(t) =  \psi_{0} (t_{0}-t)^{\alpha}$$
\end{itemize}

b) A combination of two  functions: 
\begin{itemize}
\item Two exponentials (M6): $$\psi_{old}(t) = \psi_{0}e^{-(t^{old}_{0} - t)/\tau^{old}}$$
$$\psi_{young}(t) = \psi_{0}e^{-(t^{young}_{0} - t)/\tau^{young}}$$
\item Two exponentials, with old component fixed  (M7): $$\psi_{old}(t) = \psi_{0}e^{-(14Gyr - t)/\tau^{old}}$$
$$\psi_{young}(t) = \psi_{0}e^{-(t^{young}_{0} - t)/\tau^{young}}$$
\item An exponential plus a Sandage component (M8): 
$$\psi_{old}(t) = \psi_{0}e^{-(t^{old}_{0} - t)/\tau^{old}}$$
$$\psi_{young}(t) = \frac{A}{[\tau^{young}]^{2}}(t^{young}_{0} - t)e^{-(t^{young}_{0}-t)^{2}/2[\tau^{young}]^{2}}$$
\item Constant SFR plus an exponential declining (M9): 
$$\psi_{old}(t) = \psi_{0}$$
$$\psi_{young}(t) = \psi_{0}e^{-(t_{0} - t)/\tau}$$
\end{itemize}

Figure \ref{fig:qualityfits} shows the quality of the fits; we conclude: a) M2 and M3 provide similar quality in the fits that M1; b) M4 and M5 give very poor fits; c) the quality of the fits with M6, M7, and M8 are somewhat better than with M1; d) M9 provides fits of similar quality to M1. Although M1 is not the model with the lowest $\chi^2$,  the results related with the star formation history of galaxies and stellar population properties are better than with the other models, as explained below; so we use M1 as the reference model.

Figure \ref{fig:massfraction_bars} shows a detailed comparison between M6, M1, and \starlight. Table A.1 lists the results for all the models. %:

\begin{figure}
\includegraphics[width=0.5\textwidth]{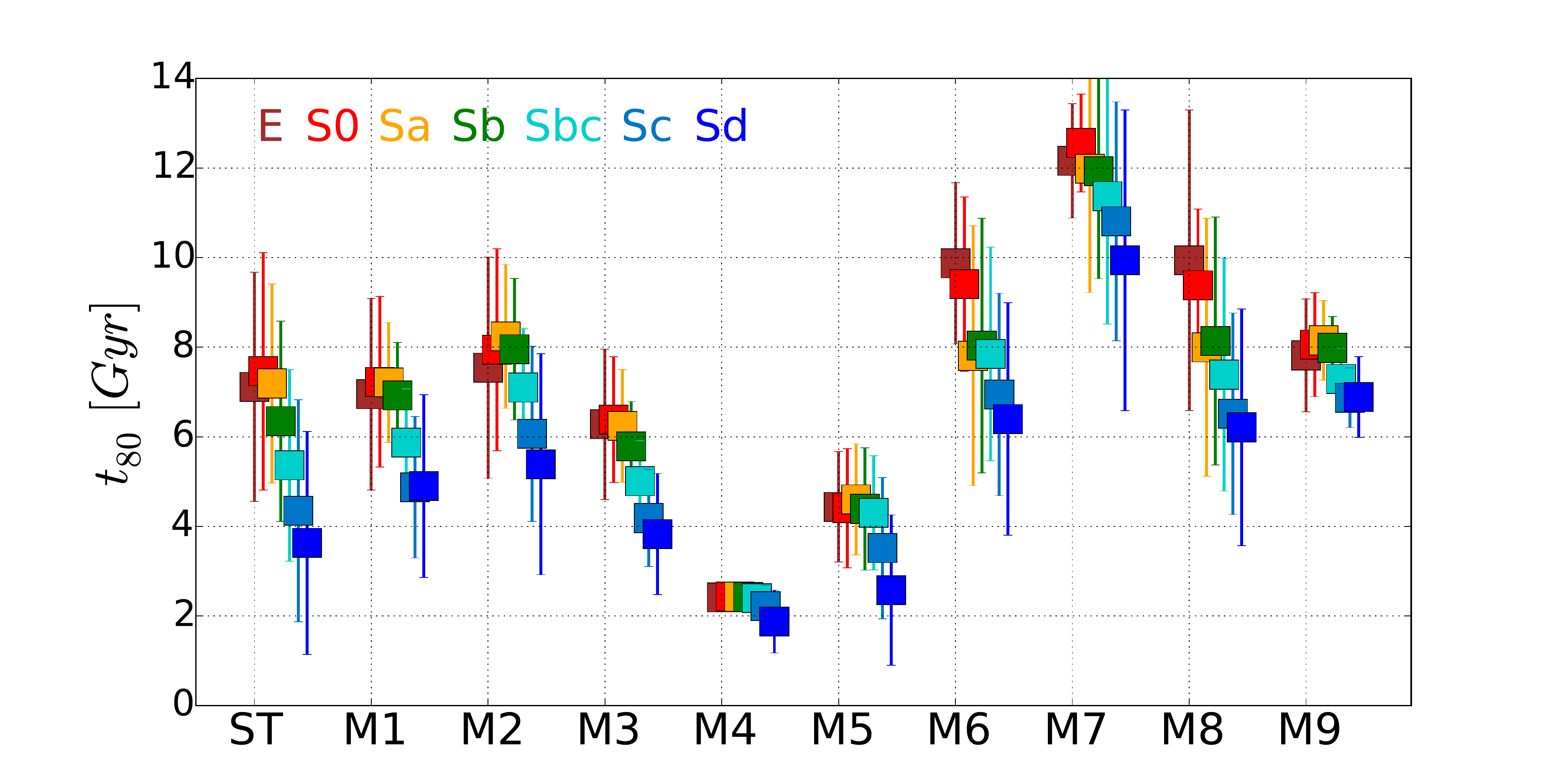}
\caption{Lookback time at which galaxies  assemble  $80 \%$ of their mass,  $t_{80}$, from the different parametric models and  \starlight. The average values are color coded as a function of morphology. 
}
\label{fig:t80_allmodels}
\end{figure}

To show further  differences and similarities between the  parametric models and  \starlight, Fig.\ \ref{fig:t80_allmodels} shows  $t_{80}$ calculated for  the whole galaxy ($R < 2$ HLR).  The results are averaged by Hubble type:

\begin{itemize}

{\item Except for M4, the parametric and non-parametric models show that $t_{80}$ decreases with Hubble type. This is a  manifestation of the downsizing scenario since in our sample, on average, the galaxy mass scales with Hubble type. Notice, however, that the range of variation from E to Sd is smaller with M5 and M9.}

{\item With \starlight,  $t_{80}\sim7$ Gyr for Sa and earlier types and decreases for later types, down to
 $t_{80}\sim3.5$ Gyr for Sd. Results with M1 are very similar to \starlight\  for Sa and earlier types, with  $t_{80}\sim7$ Gyr; but for later  types, Sd,  \starlight\ gives smaller $t_{80}$  than  M1.}

{\item  For M2,  $t_{80}\sim8$ Gyr for S0 and Sb, and larger for Sa. For E,  $t_{80}\sim7.5$ Gyr and smaller values are obtained for later spirals, with  $t_{80} \sim 5.5$ Gyr for Sd.}

{\item    M3 gives smaller values than  M1. For Sa and earlier types  $t_{80}\sim6-7$ Gyr,  $t_{80}\sim6$ Gyr for Sb galaxies, and  $t_{80}\sim4$ Gyr for Sc and Sd.}

{\item   Results with M6 and M8 are very similar. In both cases,  $t_{80}$ decreases from 10 Gyr in E galaxies
to   6-7 Gyr for Sc and Sd.}

{\item  With M7,  $t_{80}$ is larger for all the Hubble types. For Sd   $t_{80}\sim10$ Gyr, which is the value
obtained for E galaxies with M6 and M8. For E and S0 galaxies,   $t_{80}\sim12$-13 Gyr. }

{\item  The range of variation of  $t_{80}$ with M9 is lower than with the other models. For Sb and earlier
types,  $t_{80}\sim8$ Gyr, while for Sc and Sd  $t_{80}\sim7$ Gyr.}
 
\end{itemize}

%***FIG***FIG***FIG***FIG***FIG***FIG***FIG***FIG***FIG***FIG***
%\begin{figure*}[!ht]
%***FIG***FIG***FIG***FIG***FIG***FIG***FIG***FIG***FIG***FIG***

\begin{table*}
\centering % used for centering table
\scalebox{1}{
\begin{tabular}{c c c c c c c c c} % centered columns (4 columns)
& & E & S0 & Sa & Sb & Sbc & Sc & Sd\\
\hline
ST &
\begin{tabular}{c}
$m_{t< 1\, Gyr}$\\ $m_{1 < t< 4\, Gyr}$ \\ $m_{4 < t< 9\, Gyr}$ \\ $m_{t> 9\, Gyr}$
\end{tabular} 
& \begin{tabular}{c}
$1 \pm 3$ \\ $11 \pm 11$ \\ $30 \pm 21$ \\ $58 \pm 25$ \end{tabular}
& \begin{tabular}{c}
$1 \pm 1$ \\ $10 \pm 14$ \\ $28 \pm 18$ \\ $61 \pm 23$ \end{tabular}
& \begin{tabular}{c}
$1 \pm 1$ \\ $9 \pm 8$ \\ $29 \pm 16$ \\ $61 \pm 20$ \end{tabular}
& \begin{tabular}{c}
$1 \pm 1$ \\ $12 \pm 10$ \\ $32 \pm 17$ \\ $54 \pm 22$ \end{tabular}
& \begin{tabular}{c}
$2 \pm 2$ \\ $18 \pm 14$ \\ $35 \pm 12$ \\ $45 \pm 19$ \end{tabular}
& \begin{tabular}{c}
$7 \pm 14$ \\ $19 \pm 16$ \\ $33 \pm 15$ \\ $40 \pm 20$ \end{tabular}
& \begin{tabular}{c}
$13 \pm 18$ \\ $16 \pm 14$ \\ $34 \pm 15$ \\ $36 \pm 19$ \end{tabular} \\

\hline
M1 &
\begin{tabular}{c}
$m_{t< 1\, Gyr}$\\ $m_{1 < t< 4\, Gyr}$ \\ $m_{4 < t< 9\, Gyr}$ \\ $m_{t> 9\, Gyr}$
\end{tabular} 
& \begin{tabular}{c}
$1 \pm 1$ \\ $17 \pm 31$ \\ $51 \pm 30$ \\ $32 \pm 28$ \end{tabular}
& \begin{tabular}{c}
$1 \pm 1$ \\ $14 \pm 27$ \\ $51 \pm 28$ \\ $35 \pm 27$ \end{tabular}
& \begin{tabular}{c}
$1 \pm 1$ \\ $10 \pm 13$ \\ $53 \pm 20$ \\ $36 \pm 21$ \end{tabular}
& \begin{tabular}{c}
$2 \pm 2$ \\ $12 \pm 11$ \\ $51 \pm 16$ \\ $34 \pm 17$ \end{tabular}
& \begin{tabular}{c}
$4 \pm 2$ \\ $20 \pm 14$ \\ $49 \pm 10$ \\ $27 \pm 11$ \end{tabular}
& \begin{tabular}{c}
$6 \pm 3$ \\ $26 \pm 17$ \\ $46 \pm 10$ \\ $21 \pm 12$ \end{tabular}
& \begin{tabular}{c}
$7 \pm 4$ \\ $28 \pm 21$ \\ $45 \pm 15$ \\ $21 \pm 17$ \end{tabular} \\

\hline
M2 &
\begin{tabular}{c}
$m_{t< 1\, Gyr}$\\ $m_{1 < t< 4\, Gyr}$ \\ $m_{4 < t< 9\, Gyr}$ \\ $m_{t> 9\, Gyr}$
\end{tabular} 
& \begin{tabular}{c}
$1 \pm 1$ \\ $16 \pm 20$ \\ $47 \pm 26$ \\ $37 \pm 27$ \end{tabular}
& \begin{tabular}{c}
$1 \pm 1$ \\ $11 \pm 26$ \\ $44 \pm 24$ \\ $44 \pm 25$ \end{tabular}
& \begin{tabular}{c}
$1 \pm 1$ \\ $9 \pm 14$ \\ $39 \pm 23$ \\ $51 \pm 26$ \end{tabular}
& \begin{tabular}{c}
$2 \pm 2$ \\ $10 \pm 12$ \\ $37 \pm 16$ \\ $51 \pm 20$ \end{tabular}
& \begin{tabular}{c}
$3 \pm 1$ \\ $16 \pm 16$ \\ $37 \pm 10$ \\ $48 \pm 14$ \end{tabular}
& \begin{tabular}{c}
$8 \pm 18$ \\ $20 \pm 19$ \\ $37 \pm 13$ \\ $35 \pm 17$ \end{tabular}
& \begin{tabular}{c}
$6 \pm 4$ \\ $30 \pm 27$ \\ $39 \pm 23$ \\ $25 \pm 24$ \end{tabular} \\

\hline
M3 &
\begin{tabular}{c}
$m_{t< 1\, Gyr}$\\ $m_{1 < t< 4\, Gyr}$ \\ $m_{4 < t< 9\, Gyr}$ \\ $m_{t> 9\, Gyr}$
\end{tabular} 
& \begin{tabular}{c}
$1 \pm 1$ \\ $18 \pm 28$ \\ $61 \pm 27$ \\ $21 \pm 20$ \end{tabular}
& \begin{tabular}{c}
$1 \pm 1$ \\ $15 \pm 23$ \\ $60 \pm 20$ \\ $24 \pm 16$ \end{tabular}
& \begin{tabular}{c}
$1 \pm 1$ \\ $14 \pm 11$ \\ $62 \pm 15$ \\ $23 \pm 16$ \end{tabular}
& \begin{tabular}{c}
$2 \pm 2$ \\ $18 \pm 15$ \\ $59 \pm 14$ \\ $21 \pm 9$ \end{tabular}
& \begin{tabular}{c}
$5 \pm 2$ \\ $24 \pm 12$ \\ $54 \pm 9$ \\ $17 \pm 6$ \end{tabular}
& \begin{tabular}{c}
$7 \pm 3$ \\ $29 \pm 11$ \\ $50 \pm 8$ \\ $14 \pm 6$ \end{tabular}
& \begin{tabular}{c}
$8 \pm 4$ \\ $33 \pm 17$ \\ $46 \pm 14$ \\ $12 \pm 7$ \end{tabular} \\

\hline
M4 &
\begin{tabular}{c}
$m_{t< 1\, Gyr}$\\ $m_{1 < t< 4\, Gyr}$ \\ $m_{4 < t< 9\, Gyr}$ \\ $m_{t> 9\, Gyr}$
\end{tabular} 
& \begin{tabular}{c}
$15 \pm 1$ \\ $38 \pm 2$ \\ $38 \pm 2$ \\ $9 \pm 1$ \end{tabular}
& \begin{tabular}{c}
$15 \pm 1$ \\ $38 \pm 1$ \\ $38 \pm 1$ \\ $9 \pm 1$ \end{tabular}
& \begin{tabular}{c}
$15 \pm 1$ \\ $38 \pm 1$ \\ $38 \pm 1$ \\ $9 \pm 1$ \end{tabular}
& \begin{tabular}{c}
$15 \pm 1$ \\ $38 \pm 1$ \\ $38 \pm 1$ \\ $9 \pm 1$ \end{tabular}
& \begin{tabular}{c}
$15 \pm 2$ \\ $38 \pm 3$ \\ $38 \pm 3$ \\ $9 \pm 1$ \end{tabular}
& \begin{tabular}{c}
$18 \pm 12$ \\ $40 \pm 8$ \\ $35 \pm 9$ \\ $7 \pm 3$ \end{tabular}
& \begin{tabular}{c}
$24 \pm 19$ \\ $42 \pm 12$ \\ $29 \pm 14$ \\ $5 \pm 4$ \end{tabular} \\

\hline
M5 &
\begin{tabular}{c}
$m_{t< 1\, Gyr}$\\ $m_{1 < t< 4\, Gyr}$ \\ $m_{4 < t< 9\, Gyr}$ \\ $m_{t> 9\, Gyr}$
\end{tabular} 
& \begin{tabular}{c}
$10 \pm 5$ \\ $29 \pm 11$ \\ $39 \pm 10$ \\ $22 \pm 11$ \end{tabular}
& \begin{tabular}{c}
$14 \pm 20$ \\ $25 \pm 7$ \\ $38 \pm 11$ \\ $23 \pm 10$ \end{tabular}
& \begin{tabular}{c}
$13 \pm 18$ \\ $25 \pm 8$ \\ $38 \pm 10$ \\ $25 \pm 8$ \end{tabular}
& \begin{tabular}{c}
$14 \pm 21$ \\ $25 \pm 8$ \\ $38 \pm 10$ \\ $23 \pm 10$ \end{tabular}
& \begin{tabular}{c}
$11 \pm 12$ \\ $28 \pm 10$ \\ $39 \pm 10$ \\ $21 \pm 11$ \end{tabular}
& \begin{tabular}{c}
$16 \pm 18$ \\ $32 \pm 12$ \\ $36 \pm 11$ \\ $15 \pm 12$ \end{tabular}
& \begin{tabular}{c}
$25 \pm 24$ \\ $379 \pm 16$ \\ $29 \pm 16$ \\ $9 \pm 11$ \end{tabular} \\

\hline
M6 &
\begin{tabular}{c}
$m_{t< 1\, Gyr}$\\ $m_{1 < t< 4\, Gyr}$ \\ $m_{4 < t< 9\, Gyr}$ \\ $m_{t> 9\, Gyr}$
\end{tabular} 
& \begin{tabular}{c}
$1 \pm 1$ \\ $11 \pm 17$ \\ $20 \pm 21$ \\ $69 \pm 30$ \end{tabular}
& \begin{tabular}{c}
$1 \pm 1$ \\ $9 \pm 12$ \\ $35 \pm 26$ \\ $55 \pm 28$ \end{tabular}
& \begin{tabular}{c}
$1 \pm 4$ \\ $14 \pm 22$ \\ $49 \pm 20$ \\ $36 \pm 21$ \end{tabular}
& \begin{tabular}{c}
$2 \pm 5$ \\ $11 \pm 21$ \\ $47 \pm 20$ \\ $40 \pm 21$ \end{tabular}
& \begin{tabular}{c}
$2 \pm 3$ \\ $8 \pm 21$ \\ $62 \pm 21$ \\ $28 \pm 20$ \end{tabular}
& \begin{tabular}{c}
$2 \pm 3$ \\ $8 \pm 20$ \\ $71 \pm 25$ \\ $19 \pm 24$ \end{tabular}
& \begin{tabular}{c}
$3 \pm 7$ \\ $14 \pm 23$ \\ $67 \pm 26$ \\ $16 \pm 22$ \end{tabular} \\

\hline
M7 &
\begin{tabular}{c}
$m_{t< 1\, Gyr}$\\ $m_{1 < t< 4\, Gyr}$ \\ $m_{4 < t< 9\, Gyr}$ \\ $m_{t> 9\, Gyr}$
\end{tabular} 
& \begin{tabular}{c}
$1 \pm 1$ \\ $8 \pm 8$ \\ $5 \pm 7$ \\ $86 \pm 11$ \end{tabular}
& \begin{tabular}{c}
$1 \pm 1$ \\ $5 \pm 9$ \\ $4 \pm 6$ \\ $90 \pm 10$ \end{tabular}
& \begin{tabular}{c}
$1 \pm 2$ \\ $12 \pm 20$ \\ $2 \pm 4$ \\ $85 \pm 21$ \end{tabular}
& \begin{tabular}{c}
$1 \pm 4$ \\ $8 \pm 15$ \\ $4 \pm 6$ \\ $87 \pm 19$ \end{tabular}
& \begin{tabular}{c}
$1 \pm 3$ \\ $7 \pm 16$ \\ $6 \pm 7$ \\ $86 \pm 20$ \end{tabular}
& \begin{tabular}{c}
$2 \pm 4$ \\ $6 \pm 14$ \\ $10 \pm 10$ \\ $82 \pm 21$ \end{tabular}
& \begin{tabular}{c}
$3 \pm 8$ \\ $8 \pm 16$ \\ $12 \pm 10$ \\ $76 \pm 24$ \end{tabular} \\

\hline
M8 &
\begin{tabular}{c}
$m_{t< 1\, Gyr}$\\ $m_{1 < t< 4\, Gyr}$ \\ $m_{4 < t< 9\, Gyr}$ \\ $m_{t> 9\, Gyr}$
\end{tabular} 
& \begin{tabular}{c}
$1 \pm 1$ \\ $7 \pm 8$ \\ $39 \pm 25$ \\ $53 \pm 26$ \end{tabular}
& \begin{tabular}{c}
$2 \pm 6$ \\ $3 \pm 7$ \\ $36 \pm 27$ \\ $59 \pm 28$ \end{tabular}
& \begin{tabular}{c}
$1 \pm 3$ \\ $13 \pm 25$ \\ $48 \pm 29$ \\ $38 \pm 23$ \end{tabular}
& \begin{tabular}{c}
$1 \pm 3$ \\ $9 \pm 20$ \\ $50 \pm 23$ \\ $39 \pm 25$ \end{tabular}
& \begin{tabular}{c}
$1 \pm 2$ \\ $9 \pm 22$ \\ $65 \pm 30$ \\ $25 \pm 29$ \end{tabular}
& \begin{tabular}{c}
$2 \pm 3$ \\ $12 \pm 25$ \\ $70 \pm 25$ \\ $16 \pm 21$ \end{tabular}
& \begin{tabular}{c}
$3 \pm 7$ \\ $20 \pm 22$ \\ $60 \pm 28$ \\ $17 \pm 22$ \end{tabular} \\

\hline
M9 &
\begin{tabular}{c}
$m_{t< 1\, Gyr}$\\ $m_{1 < t< 4\, Gyr}$ \\ $m_{4 < t< 9\, Gyr}$ \\ $m_{t> 9\, Gyr}$
\end{tabular} 
& \begin{tabular}{c}
$1 \pm 1$ \\ $9 \pm 8$ \\ $40 \pm 7$ \\ $50 \pm 12$ \end{tabular}
& \begin{tabular}{c}
$1 \pm 1$ \\ $8 \pm 7$ \\ $39 \pm 7$ \\ $52 \pm 11$ \end{tabular}
& \begin{tabular}{c}
$1 \pm 1$ \\ $8 \pm 5$ \\ $39 \pm 5$ \\ $52 \pm 9$ \end{tabular}
& \begin{tabular}{c}
$2 \pm 1$ \\ $9 \pm 4$ \\ $39 \pm 4$ \\ $50 \pm 7$ \end{tabular}
& \begin{tabular}{c}
$3 \pm 1$ \\ $14 \pm 4$ \\ $40 \pm 1$ \\ $43 \pm 6$ \end{tabular}
& \begin{tabular}{c}
$4 \pm 1$ \\ $16 \pm 4$ \\ $40 \pm 1$ \\ $40 \pm 5$ \end{tabular}
& \begin{tabular}{c}
$4 \pm 2$ \\ $16 \pm 5$ \\ $40 \pm 2$ \\ $40 \pm 7$ \end{tabular} \\

\hline %inserts single line
\end{tabular}}
\centering
\caption{Average mass fraction, corresponding to the whole galaxy, due to stars in different age ranges as a function of  Hubble type, obtained with {\sc starlight} and with the parametric models. 
}
\label{table:mass_fraction_all} 
\end{table*}

\subsection{Fossil cosmology vs galaxy surveys: $\rho_{\rm SFR}$, sSFR, and $\rho_\star$ }

%***FIG***FIG***FIG***FIG***FIG***FIG***FIG***FIG***FIG***FIG***
\begin{figure*}[!ht]
%\begin{figure*}
\includegraphics[width=0.79\textwidth]{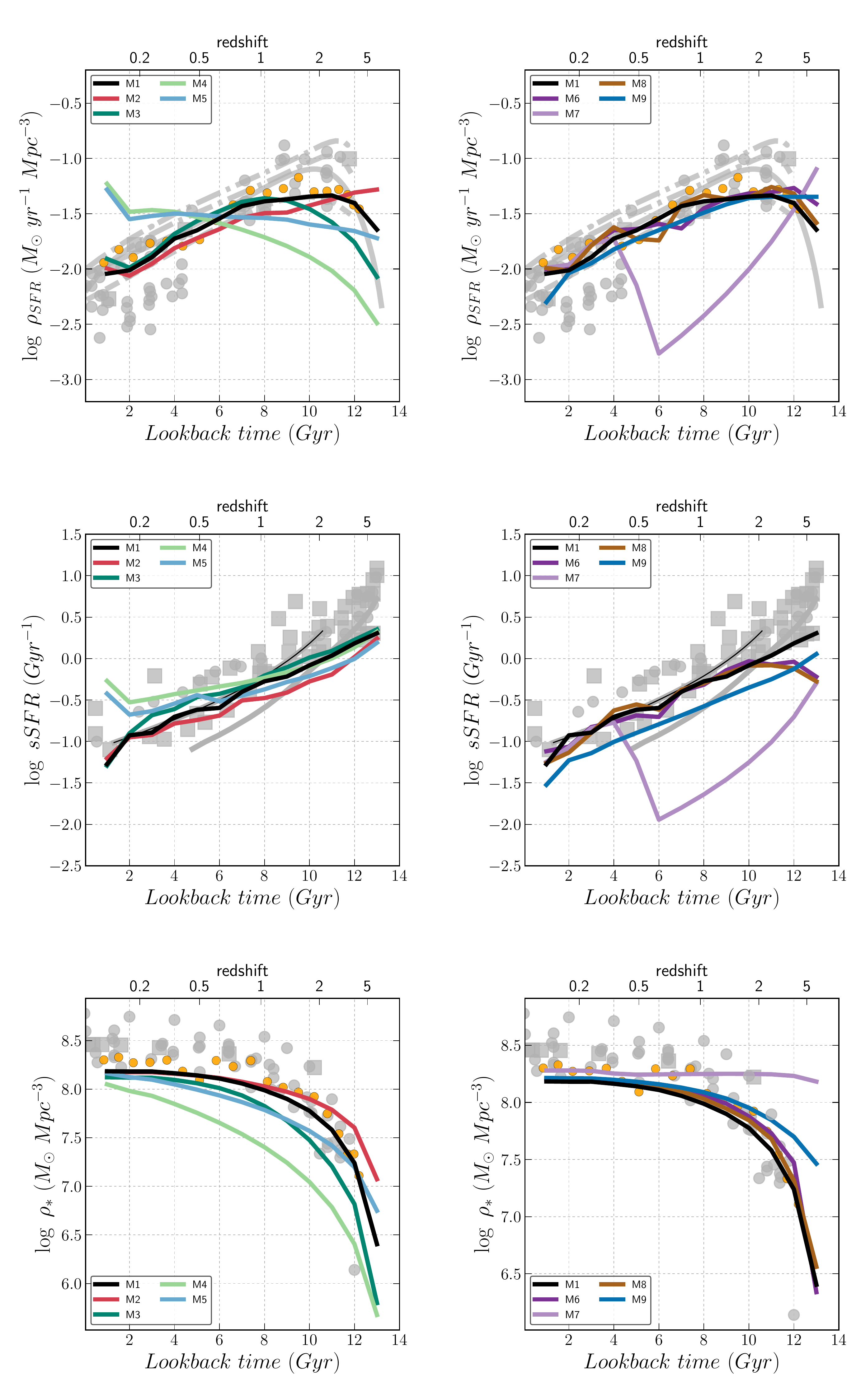}
\caption{ Cosmic evolution of SFR ($\rho_{\rm SFR}$, upper panels), sSFR (middle panels), and stellar mass ($\rho_\star$, bottom panels), for one single parametric SFH (left panels) and a combination of two parametric SFH (right panels). The reference delayed-$\tau$ model (M1, black line) is plotted in all the panels. Gray and yellow points, and gray curves are taken from the literature as explained in the text and in the captions of Fig.\ \ref{fig:cosmicSFR},  Fig.\ \ref{fig:sSFR_all}, and  Fig.\ \ref{fig:Massdensitycosmic}. }
\label{fig:csfr_all}
\end{figure*}
%***FIG***FIG***FIG***FIG***FIG***FIG***FIG***FIG***FIG***FIG***

Finally we compare the results from the different models with fossil cosmology with respect to galaxy surveys,   calculating $\rho_{\rm SFR}$, sSFR, and $\rho_\star$ as a function of redshift. In Section 5 we compared these properties for model M1 with results from galaxy surveys from the literature. Here we extend the comparisons to all the parametric models presented in this Appendix. From  Fig.\ \ref{fig:csfr_all}, we conclude:

\begin{itemize}

\item {\bf M2}: $\rho_{\rm SFR}$ increases with redshift, and it does not show a maximum (or plateau) at $z \sim 2$ as the cosmological galaxy surveys do. $\rho_\star$ increases with time more rapidly than any other single parametric SFH, and the galaxy surveys. sSFR($t$) for $z > 0.5$ are below the lower envelope of points from the cosmological galaxy surveys.

\item {\bf  M3}: $\rho_{\rm SFR}$ has the maximum at intermediate redshift $z\sim 1$. The evolution of sSFR($t$) is very similar to model M1 and the individual measurements from galaxy surveys. However, $\rho_\star$ increases more slowly than M1 and galaxy surveys, although it gets a similar value than M1 at $z = 0$. 

\item  {\bf M4 and M5}:  they give very unrealistic results for the SFH of nearby galaxies, and are not able to fit well the observational constrains (see Fig.\ \ref{fig:qualityfits}). Moreover, $\rho_{\rm SFR}$, $\rho_\star$, and sSFR($t$) do not match the results from galaxy surveys.  At z = 0, with these models $\rho_{\rm SFR}$ is significantly higher than in cosmological galaxy surveys, and at $z\geq$ 1 significantly lower.  sSFR($t$)  evolves with time more smoothly than galaxy surveys do. M4 gives a cosmic evolution of  $\rho_\star$ that does not match at all the points from galaxy surveys; and M5 gives values of $\rho_\star$ at intermediate redshift that are not in agreement with results from galaxy surveys.

\item  {\bf M6 and M8}:  $\rho_{\rm SFR}$, sSFR($t$), and  $\rho_\star$  are similar to galaxy surveys, except that sSFR($t$) at $z\geq$2 is lower than the values of the literature. These models not provide $m(t)$ that is very different to that obtained by \citet{madau14} (see Fig.\ \ref{fig:massfraction_mean}). 

\item {\bf M7}: the fraction in stars older than 9 Gyr is significantly higher than for the other models, and thus the galaxies grow their mass very rapidly and very early on. As a consequence, the fraction of stellar mass in components of intermediate ages are low. Thus, although at $z= 0$ and $z\sim 4$, $\rho_{\rm SFR}$ is similar to cosmological galaxy surveys, its evolution  is quite different to the results from these works, since  a minimum hapens  at intermediate redshifts. Similarly, sSFR($t$) for $z > 0.5$ is quite low in comparison with galaxy surveys. In contrast, $\rho_\star$ does not evolve with time, being almost constant since $z \leq 5$.

\item {\bf M9}: the evolution of  $\rho_{\rm SFR}$  up to $z = 2$ is similar to galaxy surveys, but it does not show a change of  slope at $z > 2$ and it continues increasing with redshift. In absolute values $\rho_{\rm SFR}\sim0.2 - 0.3$ dex below the \citet{madau14} curve. sSFR($t$) is significantly below the \cite{elbaz11} curve, this is also a consequence of the rapid growth of $\rho_\star$.

\end{itemize}

%************************************************************************************
\end{document}